\documentclass[fleqn,usenatbib]{mnras}

\usepackage[T1]{fontenc}
\usepackage{ae,aecompl}

\usepackage{graphicx}	
\usepackage{amsmath}	
\usepackage{amssymb}	
\usepackage{bm}
\usepackage[percent]{overpic}
\usepackage[normalem]{ulem}
\usepackage{tikz}
\usetikzlibrary{arrows,positioning}
\usepackage{xcolor}

\usepackage{soul}
\setstcolor{red}

\usepackage{newtxtext,newtxmath}

\title[Real-Time Anomaly Detection]{Real-Time Detection of Anomalies in Large-Scale Transient Surveys}

\author[D. Muthukrishna et al.]{
Daniel Muthukrishna$^{1,2}$\thanks{E-mail: \href{mailto:danmuth@mit.edu}{danmuth@mit.edu}},
Kaisey S. Mandel$^{1,3,4}$,
Michelle Lochner$^{5,6,7}$,
Sara Webb$^{8}$,
and Gautham Narayan$^{9}$
\\
$^{1}$Institute of Astronomy, University of Cambridge, Madingley Road, Cambridge CB3 0HA, United Kingdom\\
$^{2}$Kavli Institute for Astrophysics and Space Research, Massachusetts Institute of Technology, Cambridge, MA 02139, USA\\
$^{3}$Statistical Laboratory, DPMMS, University of Cambridge, Wilberforce Road, Cambridge, CB3 0WB, United Kingdom \\
$^{4}$The Alan Turing Institute, Euston Road, London NW1 2DB, United Kingdom\\
$^{5}$Department of Physics and Astronomy, University of the Western Cape, Bellville, Cape Town, 7535, South Africa \\
$^{6}$South African Radio Astronomy Observatory (SARAO), 2 Fir Street, Observatory, Cape Town, 7925, South Africa \\
$^{7}$African Institute for Mathematical Sciences, 6 Melrose Road, Muizenberg, 7945, South Africa \\
$^{8}$Centre for Astrophysics and Supercomputing, Swinburne University of Technology, John St, Hawthorn VIC 3122, Australia\\
$^{9}$Department of Astronomy, University of Illinois at Urbana-Champaign, Urbana, IL 61801, USA\\
}

\date{Accepted XXX. Received YYY; in original form ZZZ}

\pubyear{2020}

\begin{document}
\label{firstpage}
\pagerange{\pageref{firstpage}--\pageref{lastpage}}
\maketitle

\begin{abstract}
New time-domain surveys, such as the Vera C. Rubin Observatory Legacy Survey of Space and Time (LSST), will observe millions of transient alerts each night, making standard approaches of visually identifying new and interesting transients infeasible. We present two novel methods of automatically detecting anomalous transient light curves in real-time. Both methods are based on the simple idea that if the light curves from a known population of transients can be accurately modelled, any deviations from model predictions are likely anomalies. The first modelling approach is a probabilistic neural network built using Temporal Convolutional Networks (TCNs) and the second is an interpretable Bayesian parametric model of a transient. We demonstrate our methods' ability to provide anomaly scores as a function of time on light curves from the Zwicky Transient Facility. We show that the flexibility of neural networks, the attribute that makes them such a powerful tool for many regression tasks, is what makes them less suitable for anomaly detection when compared with our parametric model. The parametric model is able to identify anomalies with respect to common supernova classes with high precision and recall scores, achieving area under the precision-recall curves (AUCPR) above 0.79 for most rare classes such as kilonovae, tidal disruption events, intermediate luminosity transients, and pair-instability supernovae. Our ability to identify anomalies improves over the lifetime of the light curves. Our framework, used in conjunction with transient classifiers, will enable fast and prioritised followup of unusual transients from new large-scale surveys.  

\end{abstract}




\section{Introduction}
\label{sec:Introduction}
Astronomy is reaching an unprecedented era of big data, where astronomers are observing more transient events than they can possibly visually examine. Upcoming large-scale surveys of the transient universe such as the Vera C. Rubin Observatory Legacy Survey of Space and Time (LSST) will observe transient alerts at a rate more than an order of magnitude larger than any survey to date \citep{Ivezic2009LSST:Products}. LSST is expected to observe over 10 million transient alerts each night, making it infeasible to visually examine or follow up any significant fraction of transient candidates. However, for a long time, discovery in astronomy has been driven by serendipity and by identifying anomalies in data sets. To this end, identifying anomalous objects and prioritising which of the millions of alerts are most suitable for spectroscopic followup is a challenge that needs to be automated. In this paper, we develop a novel framework for identifying anomalous transients in real-time.

There have been several efforts to automate the identification of astronomical transients in large-scale surveys (e.g. \citealt{Lochner2016,Narayan2018MachineStream,PELICANPasquet2019, Webb2020}). These efforts are useful for dealing with the big datasets in recent surveys, but they require the full phase coverage of each light curve for classification. While retrospective classification after the full light curve of an event has been observed is useful, it also limits the scientific questions that can be answered about these events, many of which exhibit interesting physics at early times. To prioritise followup, the type of transient and its phase of evolution are most important. 

Obtaining detailed follow-up observations shortly after a transient's explosion provides insights into the progenitor systems that power the event and hence improves our understanding of the object's physical mechanism. While the mechanism of some transients are reasonably well-understood, the central engine of various exotic classes such as calcium-rich gap transients, super-luminous supernovae, and some newly discovered fast blue optical transients (FBOTs) are poorly understood \citep[e.g.][]{CoppejansMargutti2020FBOTs}. Moreover, even though Type Ia Supernovae (SNe Ia) have been well studied, their progenitor system remains mysterious \citep[e.g.][for reviews of the current state of SNIa progenitor origin]{Livio2018,Ruiter2020SNIaProgenitors}. Furthermore, the discovery of the electromagnetic counterpart from the binary neutron star merger gravitational wave event, GW170817 \citep{Abbott2017GW170817:Inspiral}, and the considerable human effort that went into the followup, has made it clear that automated photometric classifiers are necessary. The need for rapid identification of these events means automatically sifting through the millions of transient alerts produced each night and identifying candidates at early times.

To this end, recent methods such as \texttt{SuperNNova} \citep{SupernnoovaMoller2019}  and \texttt{RAPID} \citep{Muthukrishna19RAPID} have developed early and real-time classifiers capable of identifying the specific type of transient shortly after explosion. These approaches use state-of-the-art deep recurrent neural networks (RNNs) to model a function that maps real-time photometric information onto a range of different transient classes, and are able to update their prediction as new photometric data along a transient's light curve become available. They enable astronomers to prioritise candidates for follow-up observations.

However, one major caveat of all existing approaches, is that classification is inherently a supervised learning task, and hence, requires either comprehensive labelled data for training an algorithm, or well-understood models that enable simulating a training set. They are unable to classify events that they have not been specifically trained on. But, with the deluge of data coming from upcoming wide-field surveys, that are probing deeper, wider, and faster than ever before, we should be prepared for discovering rare and unexpected classes of transients. LSST will have a point source depth of r$\sim$27.5 \citep{LSST_Book_2009}, and will be able to probe fainter than any other wide-scale survey to date, while the Transiting Exoplanet Science Survey (TESS, \citealt{TESS_Ricker_2015}) will use its wide field-of-view to explore transient phenomena at the minutes to hours timescale which is a region of parameter space that has been relatively unexplored. Consequently, astronomers are in need of methods capable of discovering new and unknown transient phenomena within the context of the huge datasets in modern time-domain astronomy.

Anomaly detection is a data-driven approach to finding such outliers. The goal is to detect outliers that are scientifically interesting, rather than random statistical fluctuations. Within astronomy, anomaly detection algorithms have been used in a range of applications, and recently \citet{Lochner2020Astronomaly} and \citet{Ishida2021_Timeseries} have developed active learning frameworks to make the identification of anomalies in a range of datasets systematic and easily accessible.

However, applying anomaly detection to time-series such as astronomical light curves is a more challenging problem than identifying anomalies in static datasets such as images or spectra. Recently, there have been a few anomaly detection algorithms applied to astronomical light curves \citep[e.g.][]{Rebbapragada2009_Timeseries,Nun2014_Timeseries,Solarz2017,Giles2019_Timeseries,Sadeh2019_Timeseries, Pruzhinskaya2019, Soraism2020Novelties, Webb2020,Villar2021_Anomalydetection, MartinezGalarza2020Waldo, Lochner2020Astronomaly,Ishida2021_Timeseries,Malanchev2021}. These approaches predominantly use unsupervised clustering algorithms such as Density-Based Spatial Clustering of Applications with Noise (DBSCAN) \citep[e.g.][]{Giles2019_Timeseries,Webb2020}, RNN-based autoencoders that identify anomalies in a lower dimensional subspace \citep[e.g.][]{Sadeh2019_Timeseries, Villar2021_Anomalydetection}, or outlier detection algorithms such as Isolation Forests \citep[e.g.][]{Pruzhinskaya2019,Ishida2021_Timeseries,Giles2019_Timeseries,Lochner2020Astronomaly, Malanchev2021} and one-class support vector machines \citep[e.g.][]{Solarz2017, Malanchev2021}. These approaches are effective at identifying anomalies once the full light curve has been observed, but many of them prove problematic for real-time detection in large-scale transient surveys. However, \citet{Soraism2020Novelties} and \citet{Villar2021_Anomalydetection} have recently developed some of the first methods that perform real-time anomaly detection. \citet{Villar2021_Anomalydetection} uses a variational recurrent autoencoder to learn an encoded form of each light curve before obtaining anomaly scores by passing the encoded form into an isolation forest. \citet{Soraism2020Novelties} uses the distribution of magnitude changes over time intervals in a population of light curves and computes the likelihood of a new observation being consistent with the population to identify outliers. In this paper, we employ a unique method that performs regression over light curves to predict future fluxes, and uses the deviation between the predictions and observations to identify anomalies. 

This paper is organised as follows. In \S\ref{sec:Data} we detail the ZTF light curve simulations and preprocessing methods used in this analysis. In \S\ref{sec:Models} we develop two independent autoregressive models, the first being a probabilistic deep neural network that predicts future fluxes in a light curve, and the second being a Bayesian parametric model of a transient class built from the Bazin model of a light curve \citep{Bazin_function}. We present and compare the results of these two models at fitting transients and identifying anomalies in \S\ref{sec:Results}. We then also apply our models to real ZTF observational data taken from the public MSIP data stream in \S\ref{sec:application_to_real_ztf_data}, and finally, in \S\ref{sec:Conclusions}, we present our conclusions and discuss future applications of our work.

\section{Data}
\label{sec:Data}

\subsection{Zwicky Transient Facility}
The Zwicky Transient Facility \citep[ZTF, ][]{Bellm2019ZTF} is the first of the new generation of optical synoptic survey telescopes and is a precursor survey to the upcoming LSST. It builds upon the infrastructure of the Palomar Transient Factory \citep[PTF, ][]{PTFrau2009} using the 48-inch Schmidt telescope yielding an order of magnitude improvement in survey speed. It employs a 47 $\mathrm{deg}^2$ field-of-view camera to scan more than 3750 $\mathrm{deg}^2$ per hour to a depth of 20.5-21 mag \citep{GrahamZTF2019PASP}. Using a prototype of the LSST alert distribution system, it is streaming up to one million transient alerts per night in two passband filters ($gr$). Due to this unprecedented data volume, it is necessary to build automated algorithms capable of processing and sifting through this amount of data. 

ZTF has been producing transient observations since 2017, and has successfully identified several thousand supernovae. In \S\ref{sec:application_to_real_ztf_data}, we detail our collection of over 2000 supernova light curves from the public ZTF Mid Scale Innovations Program (MSIP) survey to illustrate the performance of our method on real ZTF observations.

\subsection{Simulations}
The number of confirmed ZTF SNe is impressive, but the distribution is dominated by SNe Ia (see \S\ref{sec:application_to_real_ztf_data}). Neural network based algorithms notoriously require a large training set before they are able to develop a model that generalises well to new data. Thus, while we may be able to create a good training set for SNe Ia, there are very few observations of many of the other classes, and we are not able to create a dataset of these classes that encompasses the variety of objects we expect to observe, even with significant data augmentation.

To this end, we used simulations that match the observing properties of the ZTF as described in \citet{Muthukrishna19RAPID} for the results shown in \S\ref{sec:Results}. These simulations were created using the \texttt{SNANA} \citep{Kessler2010SNANA:Analysis} software developed for the Photometric LSST Astronomical Time-series Classification Challenge \citep[PLAsTiCC, ][]{plasticcNote, KesslerPlasticcModels}. As described in Section 2 of \citealt{Muthukrishna19RAPID}, the simulations were made using a year's worth of observing logs from the public MSIP survey at the ZTF. The simulated light curves mimic the ZTF observing properties with a median cadence of 3 days in the $g$ and $r$ passbands \footnote{Phase II of ZTF now uses a 2-day cadence for the MSIP survey.}. Each simulated transient consists of a time-series of flux and flux uncertainty measurements in the $g$ and $r$ ZTF passbands, an indicator of whether the flux was a detection or non-detection, sky position, Milky Way dust reddening, and a host galaxy redshift. The models and LSST simulations developed for PLAsTiCC, which these simulations are based from, were validated using numerous techniques as described in \citet{Hlozek2020PLAsTiCCResults}. 

We define the date of \textit{trigger} throughout this paper as the first detection in a light curve, defined as the first observation that exceeds a $5\sigma$ signal-to-noise (S/N) measurement in a difference image. Hence, in the rest of this paper, time $t$ refers to the number of Modified Julian Date (MJD) days since trigger:
\begin{equation}
    t = \mathrm{MJD} - \mathrm{MJD}_{\mathrm{trigger}}
\end{equation}

\subsubsection{Training and testing classes}
We simulated approximately 10,000 events for each of the following classes: SNIa, SNIbc, SNII, superluminous supernovae (SLSN), tidal disruption events (TDE), active galactic nuclei (AGN), kilonovae, pair-instability supernovae (PISN), intermediate luminosity transients (ILOT), calcium-rich gap transients (CART), microlensing from binary star systems (uLens-BSR). Example light curves from each class are illustrated in Figures 1-3 of \citealt{KesslerPlasticcModels}. 

The PISN, ILOT, CART, and uLens-BSR classes were used as the anomaly class in the aforementioned PLAsTiCC challenge. We deemed that these classes along with kilonovae were too poorly understood to be used as a model class in this work (see \citet{Chatterjee2021} for a novel method of detecting kilonovae using photometry and contextual information). We also decided that AGN would be too difficult to model with our approaches because of their wide variability, and thus do not use them as one of the model classes. 

We have trained models for the SNIa, SNIbc, SNII, SLSN, and TDE classes, and identify anomalies with respect to each of these classes throughout this work. We split the total set of transients for each model class into two parts: 80\% for the \textit{training set} and 20\% for the \textit{testing set}, respectively. The \textit{training set} is used to train the model that predicts future fluxes, while the \textit{testing set} is used to test the performance of the model. We also apply our methods to transients from each of the anomaly classes (kilonova, AGN, PISN, ILOT, CART, and uLens-BSR) and evaluate how well we can identify these as anomalies. While AGN are not rare, we still include them as one of the anomaly classes in order to understand how our methods will respond to any AGN that are not quickly identified by other methods.

\subsection{Preprocessing}
\label{sec:preprocessing}
One of the most important aspects in an effective learning algorithm is the quality of the training set. We ensured that the data were processed in a uniform and systematic way before training the model. We perform `sigma clipping' to reject photometric points with flux uncertainties that are more than $3\sigma$ from the mean uncertainty in each passband, and iteratively repeat this clipping $5$ times. Next, we correct the light curves for Milky Way extinction using the reddening function of \citet{Fitzpatrick1998}. We assume an extinction law, $R_V = 3.1$, use the central wavelength of each ZTF filter ($g$: 4767 \AA,  $r$: 6215 \AA) and the sky position to compute the line-of-sight reddening caused by the Milky Way and de-redden each light curve\footnote{We use the extinction code: \url{https://extinction.readthedocs.io}}.

As we are most interested in parts of the light curve near trigger, we ignored any observations more than 70 days before trigger and removed any data more than 150 days from the first observation after $t=-70$ days\footnote{Some late-time phenomena such as late-time CSM interactions, that the Rubin Observatory will be particularly good at finding, will be missed \citep[e.g.][]{Soderberg2005GCN,Graham2019LateTimeCSM}. However, these late-time observations will not be well modelled by our methods and are outside of the scope of this work.}.

\subsubsection{Training set preparation}
The ZTF observations are irregularly sampled due to intranight cadence choices and seasonal constraints that lead to naturally arising temporal gaps. However, our neural network framework requires regular time-sampling of the input data. Thus, despite the ZTF observations having a roughly 3-day cadence, we interpolate our observed data onto a grid with a cadence of exactly 3 days. This interpolation is not necessary for our Bazin method, but for the sake of comparing the results between the two methods, we use the interpolated data as the input for both models described in \S\ref{sec:Models}.

Gaussian process (GP) regression \citep{RasmussenGPsBook} has been shown to be effective for astronomical light curve modelling and interpolation \citep{Lochner2016, Boone2019Avacodo}. However, typically when GPs are used for preprocessing light curves to interpolate irregularly sampled data to a regular grid \citep[e.g.][]{Boone2019Avacodo,Villar2021_Anomalydetection}, the GP is conditioned on the entire light curve and makes use of long-range covariance kernels (e.g. squared-exponential). For the purposes of retrospective analyses using the full light curve, this interpolation method is effective; however, for real-time usage this approach unrealistically uses future observations that would not be available at a particular prediction time. Instead we use linear interpolation which respects causality as follows. The linearly interpolated value at a given grid time depends only on the two neighboring observations.  Hence, relative to a prediction  time $T_{\mathrm{pred}}$ between two observation times $t_{\mathrm{obs}, i} < T_{\mathrm{pred}} < t_{\mathrm{obs}, i+1}$, the interpolated values at all grid points $T_{\mathrm{grid},j}$ before $t_{\mathrm{obs}, i} <  T_{\mathrm{pred}}$ depend only on past observations at times earlier or equal to $t_{\mathrm{obs}, i} <  T_{\mathrm{pred}}$. In contrast, the GP conditioned on the full light curve produces interpolated values at earlier grid times $T_{\mathrm{grid},j} < T_{\mathrm{pred}}$ that depend, through the covariance kernel, on $t_{\mathrm{obs}, i+1} > T_{\mathrm{pred}}$ and all future data points and, therefore, does not respect causality in real-time applications. Linear interpolations also have the added benefit over GPs of not over-smoothing and being less computationally intensive. With few and noisy observations characteristic of early real-time data, the GP may also be more prone to overfitting the light curve than linear interpolation.
 
Our method for obtaining linear interpolations with uncertainties is detailed as follows. For each data point of transient $s$ in passband $p$ at time $t$ since trigger in a light curve, we used the observed flux $\hat{F}_{spt}$ and uncertainty $\hat{\sigma}_{F,{spt}}$ as the mean and standard deviation of a normal distribution to draw $100$ random fluxes indexed by $i$,
\begin{equation}
    \hat{F}_{spt, i} \sim \mathcal{N}(\hat{F}_{spt}, \hat{\sigma}_{F,{spt}}).
\end{equation}
This gives $100$ different replications of the observed light curve. We linearly interpolate each of these $100$ generated light curves at $3$-day intervals. We use $D_{spt, i}$ to denote the flux of each of these interpolations, where the subscript $t$ now refers to the interpolated times instead of the observed times. We obtain the interpolated flux and flux uncertainty by computing the mean and standard deviation of these light curves draws, respectively:
\begin{equation}
    D_{spt} = \frac{1}{100}\sum_{i=1}^{100} D_{spt, i},
\end{equation}
\begin{equation}
    \sigma_{D,{spt}} = \sqrt{\frac{1}{100}\sum_{i=1}^{100} \left(D_{spt, i} - D_{spt} \right)^2}
\end{equation}

We removed any data more than 150 days from the first observation after $t=-70$, and hence with the 3-day interpolations make a matrix of length $N_t=50$, with each point for transient $s$, passband $p$, and interpolated time $t$ having a flux and uncertainty as follows,
\begin{equation}
    \bm{X}_{spt} = \left[D_{spt}, \sigma_{D,{spt}} \right] 
\label{eq:X_input}
\end{equation}
We similarly define the output flux predictions of our models described in \S\ref{sec:Models} as a vector of the predictive flux and uncertainty,
\begin{equation}
    \bm{Y}_{spt} = \left[y_{spt}, \sigma_{y,{spt}} \right] 
\label{eq:Y_predictions}
\end{equation}
While not strictly necessary in our architecture, the fixed length vectors are useful for passing the data to the neural network framework. Many light curves do not have observations that cover the full range, $-70 < t < 80$, required by the input matrix. We set the data in the input matrix at times before the first observation and after the last observation to zero. We ensure that the neural network skips over the time-steps where observations don’t exist and hence doesn’t use the zeroed entries by using a \textit{Masking Layer} as discussed in \S\ref{sec:NN_Model_architecture}. The final input matrix $\bm{X}_s$ for each transient $s$ is a matrix of shape $N_t \times 2N_p$ where the rows are composed of the interpolated flux and flux uncertainty in each of the $N_p$ passbands across the $N_t$ time-steps.

If contextual information such as the redshift or host galaxy properties were known, we could include this information as a extra columns in the input matrix. However, as this information is not always available without additional host spectra, we have not included it in this work, but note that our framework allows for an easy incorporation of contextual data. Future work should aim to use host galaxy information to improve transient identification. Studies by \citealt{Foley2013ClassifyingData} and \citealt{Gagliano2021} have shown that using only host galaxy properties without any photometric data can achieve $\sim 70\%$ accuracy when classifying SNe Ia and CC SNe. 

\section{Models}
\label{sec:Models}

Our methods for anomaly detection involve first developing an autoregressive sequence model of a transient class, and then using the model's ability to predict future fluxes as an anomaly score (or conversely, a goodness of fit score). We develop two methods for regressing over a transient. The first is a probabilistic deep neural network (DNN) approach using Temporal Convolutional Networks (TCNs) (described in \S\ref{sec:Model_NN}), and the second is a Bayesian parametric approach using the flexible Bazin function \cite{Bazin_function} of transients (described in \S\ref{sec:Model_Bazin}). 

Each model aims to do real-time detection, and is hence causal, using only past values to predict future values. Specifically, our model is a function that predicts future fluxes in a time-series as well as the uncertainty of that prediction; it then compares the prediction with the observed data to obtain an anomaly score. 

In the following two subsections (\S \ref{sec:Model_NN}, \S \ref{sec:Model_Bazin}), we describe our two approaches of developing a function that maps the interpolated fluxes up to time $T$ onto flux predictions $y_{sp(T+3)}$ and predictive uncertainties $\sigma_{y,{sp(T+3)}}$ three days after a given set of observations:
\begin{tikzpicture}[node distance = 6mm and 12mm, align=center]
\node (adc) [draw,minimum size=14mm] {Model \\(DNN or Bazin)};

\coordinate[above left = 3mm and 12mm of adc.west]   (a1);
\coordinate[below = of a1]              (a2);
\coordinate[below = of a2]              (a3);

\coordinate[above right= 3mm and 12mm of adc.east]  (b1);
\coordinate[below=of b1]  (b2);

\draw[-latex']  (a1) node[left] {$\bm{D}_{sp(t \le T)}$} -- (a1-| adc.west);
\draw[-latex']  (a2) node[left] {$\bm{\sigma}_{D,{sp(t \le T)}}$} -- (a2-| adc.west);

\draw[-latex'] (adc.east |- b1) -- (b1) node[right] {$y_{sp(T+3)}$};
\draw[-latex'] (adc.east |- b2) -- (b2) node[right] {$\sigma_{y,{sp(T+3)}}$};

\end{tikzpicture}

In \S\ref{sec:Anomaly_score_definition}, we define an anomaly score metric that uses the discrepancy between the fluxes $D$ and predictions $y$ to quantify anomalies.

The DNN approach builds a neural network that effectively performs regression over past data in order to predict the flux 3 days in the future. On the other hand, the Bazin approach performs regression over time to predict the flux at any time. We then feed in partial light curves into the Bazin model and infer a prediction 3 days after given data to obtain anomaly scores comparable with the DNN.

\begin{figure*} 
    \centering
    \includegraphics[width=1.0\linewidth]{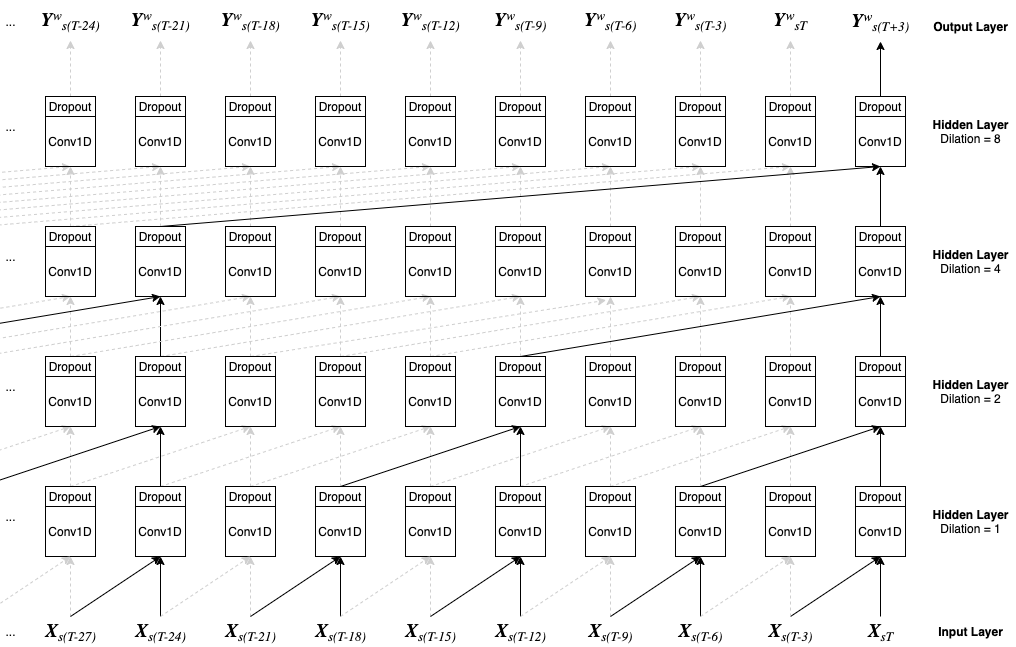}
    \caption[Temporal Convolutional Network architecture used to predict light curve fluxes for anomaly detection.]{Temporal Convolutional Neural Network architecture used in this work. Each column in the diagram is a subsequent time-step from left to right, and each time-step is 3-days after the previous one. The bottom row is the input light curve (from equation \ref{eq:X_input}) where each input is a vector of the interpolated flux and flux uncertainty in all passbands for a transient $s$ at a time $t$. The input fluxes and uncertainties of two adjacent time-steps are passed into a residual block consisting of a 1D convolutional neural network layer (Conv1D) with dropout. While not shown in the figure, the residual block also contains a second Conv1D layer with dropout. The outputs of these are then convolved with the outputs from some previous time-steps in the above hidden layers as shown in the diagram, until the final Output Layer is the predicted light curve at the following time-step (from equation \ref{eq:Y_predictions}). The solid arrows show how the prediction $\bm{Y}^w_{s(T+3)}$ is made, and the gray dashed arrows show the neural network layers that lead to all other predictions. The network is causal, whereby new predictions only use information from previous time-steps in the light curve. We set the dropout rate to $20\%$ for all layers in the network. We build this model using the \texttt{Keras} and \texttt{TensorFlow Probability} libraries after adapting the TCN model from \citet{Bai2018} and their code in \url{https://github.com/philipperemy/keras-tcn}.}
    \label{fig:tcn_architecture}
\end{figure*}

\subsection{Probabilistic Neural Network}
\label{sec:Model_NN}

\subsubsection{Model definition}
The DNN is an autoregressive mapping function that aims to map an input multi-passband light curve matrix, $\bm{X}_{s(t \le T)}$, for transient $s$ up to an interpolated time $T$, onto an output multi-passband flux vector at the next time-step $T+3$ (where we recall that each time-step is 3 days after the previous interpolated time),
\begin{equation}
    \bm{Y}^w_{s(T+3)} = \bm{f}_T(\bm{X}_{s(t\le T)}; \bm{w})
    \label{eq:learned_function}
\end{equation}
where $\bm{w}$ are the parameters (i.e. weights and biases) of the network. We define $\bm{X}_{s(t \le T)}$ as the matrix $\bm{X}_s$ but up to a time $T$ in each of its $N_p$ passbands. The model output prediction $\bm{Y}^w_{s(T+3)}$ is a $1 \times 2N_p$ vector consisting of the predicted mean flux $\tilde{y}_{sp(T+3)} (\bm{w})$ and intrinsic uncertainty $\tilde{\sigma}_{\mathrm{int},sp(T+3)}(\bm{w})$ in the $g$ and $r$ passbands at the next time-step for a particular set of network weights (these outputs are explained in \S\ref{sec:NN_capturing_uncertatinties}). The model $\bm{f}_T(\bm{X}_{s(t \le T)}; \bm{w})$ in equation \ref{eq:learned_function} is represented by the complex DNN architecture illustrated in Fig.~\ref{fig:tcn_architecture}, and the details of the architecture are described in the following subsection.

\subsubsection{Model architecture}
\label{sec:NN_Model_architecture}
We developed a deep neural network architecture as our first approach for the autoregressive sequence model that learns the function described in equation \ref{eq:learned_function}. The problem of time-series prediction falls in the wider machine-learning area of \textit{sequence learning}. Recurrent Neural Networks (RNNs) such as Long Short Term Memory (LSTM, \citealt{LSTM}) Networks and Gated Recurrent Units (GRU, \citealt{GRU}) are considered the default starting point for sequence modelling tasks in the machine learning community after they were shown to achieve state-of-the-art performance in many benchmark time-series and sequential data applications \citep[e.g.][]{Pascanu2014,Chung2014,Sutskever2014SequenceNetworks,Jozefowicz2015,Zhang2015,Bahdanau2014NeuralTranslate,Che2018RecurrentValues}. RNN's ability to retain an internal memory of long-term temporal dependencies of variable length observations made it well suited for time-series applications, and it has been shown to be successful in light curve classification \citep[e.g.][]{Charnock2016,Moss2018,Muthukrishna19RAPID,SupernnoovaMoller2019,Martinez2020,JamalBloom2020}. 

However, RNNs suffer from a few drawbacks not present in Convolutional Neural Network (CNN) approaches that have been so successful in image analysis and a range of other groundbreaking problems. Most notably, RNNs are notoriously slow and difficult to train using standard stochastic gradient descent (SGD) algorithms \citep{Pascanu2014,Bai2018}. In the past couple of years, Temporal Convolutional Networks (TCNs, first proposed in \citealt{Lea2016}) have risen as a powerful alternative to RNNs. A thorough systematic empirical evaluation of RNNs and TCNs conducted by \citet{Bai2018} suggest that TCNs are able to convincingly outperform LSTMs and GRUs across a broad range of sequence modelling tasks. In particular, \citet{Bai2018} demonstrates that TCNs exhibit a substantially longer memory of sequential data (being able to capture a longer history of data in the model), have a more flexible receptive field size (being able to control how many historic data points to remember), are much faster to train because of their parallelism (where RNNs need to wait for preceding blocks to complete but convolutions can be done in parallel since the same filter is used in each layer), are less memory intensive, and are able to capture local information through convolutions along with temporal information \citep{Kalchbrenner2016,Lea2016,Bai2018}. Furthermore, as TCNs are much simpler and clearer than RNNs, we have used this architecture in favour of RNNs in this paper. In practice, we found that the TCNs were much faster to train, but we did not notice significant differences in the performance of our TCN architecture when we compared it to a similar LSTM/GRU architecture that was used in \citet{Muthukrishna19RAPID}.

The TCN architecture used in this work is illustrated in Figure \ref{fig:tcn_architecture}, and is based on the model developed by \citet{Bai2018}\footnote{The TCN code was adapted from  \url{https://github.com/philipperemy/keras-tcn}}. We used the high level \texttt{Python} API, \texttt{Keras} \citep{Keras}, and the \texttt{TensorFlow Probability} library that are built on the efficient \texttt{TensorFlow} machine learning system \citep{Abadi2015} to develop our deep probabilistic neural network model. We describe the architecture in detail here.
\begin{description}
    \item[\textit{Input}:] The input at each time-step is the vector $\bm{X}_{st}$ for transient $s$ at time $t$. Each $\bm{X}_{st}$ input has shape $1 \times 2N_p$ containing the interpolated flux and flux uncertainty for each of the $N_p$ passbands.
    \item[\textit{Residual block:}] Each residual block performs a 1D convolution on two vector inputs using a sigmoid activation function on the neurons, and then applies dropout to each layer. While not shown in Figure \ref{fig:tcn_architecture}, each residual block also contains a second 1D convolution and dropout layer. We ensure that the convolutions are dilated and causal, whereby each output only uses information from preceding time-steps. We set the dilations to 1, 2, 4, and 8 such that the total receptive field includes $2\times8 = 16$ time-steps (equivalent to 48 days as we've set each time-step to be 3 days apart). Such a receptive field was considered a sufficient light curve history to make a prediction.
    \item[\textit{Conv1D:}] A 1D convolution is applied to the two vector inputs that each have a shape $1 \times 2N_p$. The convolutional kernel size is 2, applied to the two preceding time-steps in the input layer, and dilated to other time-steps in the hidden layers as illustrated in the Figure.
    \item[\textit{Dropout:}] We also implement dropout regularization to each layer of the neural network to reduce overfitting during training. This is an important step that effectively ignores randomly selected neurons such that their contribution to the network is temporarily removed. This process causes other neurons to more robustly handle the representation required to make predictions for the missing neurons, making the network less sensitive to the specific weights of any individual neuron. We set the dropout rate to 20\% of the neurons present in the previous layer. This effectively means that each neuron's weight has a 20\% probability of being set to zero. While this approach is ubiquitously used for regularisation during training, we also apply dropout during test time to obtain model uncertainties (see \S\ref{sec:NN_capturing_uncertatinties}).
    \item[\textit{Neurons:}]
    The output of each neuron in a neural network layer is expressed as a function of the weighted sum of the connections to it from the previous layer.
    \item[\textit{Activation function:}] 
    As with any neural network, each neuron applies an activation function to bring non-linearity to the network and hence help it to learn complex patterns in the data. We use a sigmoid activation function for the 1D convolutional layers as, after some testing, it appeared to have more stability while training the network when compared to a ReLU function. This bounded the outputs of the neurons to values between $0$ and $1$ and ensured that the weights did not become too large.
    \item[\textit{Masking Layer:}] TCNs have the advantage over standard CNNs of allowing variable length input sequences. This is particularly useful for light curves, where our observations do not always span the full $-70 < t < 80$ days used in the input matrix. However, the \texttt{Python} API requires a fixed length input for ease of computation. To employ the TCN's flexibility, we make an $N_t=50$ length input matrix for each light curve, but set the time-steps where data does not exist to an arbritrary value (or zero). We then use \texttt{Keras}'s Masking Layer to mask this value and hence ensure that time-steps where data are not available are not used in the model.

\end{description}

\subsubsection{Capturing uncertainties in the model}
\label{sec:NN_capturing_uncertatinties}
Standard deep learning tools for regression and classification do not capture model uncertainty. Nevertheless, the power and success of neural networks at a wide range of benchmark problems has led to their widespread use in science. They are particularly useful when the underlying physical processes that generated the data is not well-understood. However, gaining an intuitive understanding of the neural network's high-dimensional model is difficult and often impossible. In fact, a common and significant issue in deep learning is its over-confident predictions on unseen data \citep[e.g.][]{Guo2018CalibrationofNN}. Getting a neural network to say that it ``does not know'' and to state its confidence in a prediction is imperative for its use in science. The softmax probability output in many neural network classification problems is often erroneously interpreted as model confidence in spite of it being infamously falsely overconfident \citep[e.g.][]{Szegedy2014AdversarialAttacks,Goodfellow2015AdversarialAttacks,Gal2015}. Obtaining uncertainties on predictions is important to help overcome these issues and for the continued use of deep learning in science. See \citet{Caldeira_Nord2020} for a comparison of different uncertainty quantification methods for deep learning.

To characterise the intrinsic uncertainty of our network's ability to represent a light curve, we build a probabilistic neural network using the \texttt{Tensorflow Probability} \texttt{Python} library. While typical neural networks model the output as a point estimate, a probabilistic neural network allows us to parameterise a probability distribution with the output of a neural network. In this work, our DNN parameterises a Normal distribution and outputs a predictive mean $\tilde{y}_{spt}(\bm{w})$ and standard deviation $\tilde{\sigma}_{\mathrm{int},spt}(\bm{w})$ for a particular set of network weights $w$. The predictions $\tilde{y}_{spt}(\bm{w})$ and $\tilde{\sigma}_{\mathrm{int},spt}(\bm{w})$ are components of the vector $\bm{Y}^w_{spt}$ from equation \ref{eq:learned_function}. We include the learned uncertainty $\tilde{\sigma}_{\mathrm{int},spt} (\bm{w})$ because we know that our DNN model is not a perfect representation of a light curve, and even if we had no measurement error and had an infinite training set, there would still be some discrepancy between our DNN predictions and the observed light curves. 

A Bayesian neural network enables us to also quantify the uncertainty in our model's predictions of the outputs, $\tilde{y}_{spt}(\bm{w})$ and $\tilde{\sigma}_{\mathrm{int},spt}(\bm{w})$. The key advantage of a Bayesian neural network over a standard neural network, is that we are able to sample over a posterior distribution of network parameters (i.e. the weights and biases). The ideal way to perform a Bayesian inference over the neural network model would be to sample over the model parameters with a method such as Markov Chain Monte Carlo (MCMC). However, the huge number of parameters in a deep neural network make this a computationally intractable problem. 

Instead, an approach called Monte Carlo (MC) dropout sampling that places a Bernoulli distribution over the network weights using the commonly used dropout regularisation technique has become the popular approach for implementing approximate Bayesian neural networks (see \citet{Gal2015} for an explanation of how MC dropout approximates a Bayesian NN). The method is significantly simpler to implement than standard neural network Variational Inference (VI) approaches (such as \textit{Bayes by Backprop} \citep{Blundell2015BayesbyBackprop}, the \textit{Flipout estimator} \citep{Wen2018Flipout}, and the \textit{Reparameterization estimator} \citep{Kingma2013DenseReparemerisation}) - simply requiring dropout to be applied to all the network weights during validation (instead of just training). It has the further advantage over standard approaches to VI in neural networks of not increasing training time or reducing test accuracy. Throughout this work, we use MC dropout with our probabilistic neural network to estimate the predictive uncertainty. We do this by collecting the results of stochastic forward passes through the network as approximate posterior draws of $\bm{Y}^w_{spt}$, and use the mean and standard deviation of these draws as our marginal predictive mean $y_{spt}$ and predictive uncertainty $\sigma_{y,{spt}}$.

\subsubsection{Model loss function}
\label{sec:NN_loss_function}
Before defining the loss function, we first develop a generative model of the latent flux of a transient $s$ in passband $p$ 3 days in the future at time $T+3$. We aim to model the underlying latent flux with the neural network as follows,
\begin{equation}
   F_{sp(T+3)}(\bm{w}) = \tilde{y}_{sp(T+3)}(\bm{w}) + \epsilon_{\mathrm{int},sp(T+3)}(\bm{w}),
\label{eq:DNN_latent_flux}
\end{equation}
where the error $\epsilon_{\mathrm{int},sp(T+3)}(\bm{w}) \sim \mathcal{N}\left(0, \tilde{\sigma}^2_{\mathrm{int},sp(T+3)}(\bm{w})\right)$ is a zero-mean Gaussian random variable with variance $\tilde{\sigma}^2_{\mathrm{int},sp(T+3)}(\bm{w})$.
Thus, we write the predictive distribution of the latent flux as follows,
\begin{equation}
    \begin{split}
      &\mathcal{P}(\bm{F}_{s(T+3)}|\bm{X}_{s(t \le T)}, \bm{w}) \\
      &= \prod^{N_p}_{p=1} \mathcal{N} \left( F_{sp(T+3)}(\bm{w}) \mid \tilde{y}_{sp(T+3)}(\bm{w}), \tilde{\sigma}^2_{\mathrm{int},sp(T+3)}(\bm{w}) \right).
    \end{split}
\label{eq:posterior_predictive}
\end{equation}
Next, a generative model of the observed flux is derived by adding a measurement error to the latent flux as follows,
\begin{equation}
    D_{sp(T+3)} = F_{sp(T+3)}(\bm{w}) + \epsilon_{D, sp(T+3)},
\label{eq:generative_model_data_DNN}
\end{equation}
where we assume that the measurement error $\epsilon_{D, sp(T+3)} \sim \mathcal{N}(0,\sigma^2_{D,{sp(T+3)}})$ is a zero-mean Gaussian random variable with variance $\sigma^2_{D,{sp(T+3)}}$.

Typically, researchers will not use the uncertainty in the data within the loss function (e.g. \citealt{JamalBloom2020,Villar2021_Anomalydetection}, however, work by \citealt{Naul2018AStars} included data uncertainty without model uncertainty). In this work, we construct our loss function to include both predictive uncertainties $\tilde{\sigma}_{\mathrm{int},sp(T+3)}(\bm{w})$ and flux uncertainties $\sigma_{D,{sp(T+3)}}$.
Given equations \ref{eq:DNN_latent_flux} and \ref{eq:generative_model_data_DNN}, we write the likelihood function of the probabilistic DNN as follows,
\begin{equation}
    \begin{split}
      &\mathcal{P}(\bm{D}_{s(T+3)}|\bm{X}_{s(t \le T)}, \bm{w}) \\
      &= \prod^{N_p}_{p=1} \mathcal{N} \left( D_{sp(T+3)} \mid \tilde{y}_{sp(T+3)}(\bm{w}), \tilde{\sigma}^2_{\mathrm{int},sp(T+3)}(\bm{w}) + \sigma_{D,{sp(T+3)}}^2  \right) \\
            &= \prod^{N_p}_{p=1} \left( 2\pi(\tilde{\sigma}^2_{\mathrm{int},sp(T+3)}(\bm{w}) + \sigma_{D,{sp(T+3)}}^2) \right)^{-0.5} \\
            &\times \exp{\left( -0.5 \frac{(\tilde{y}_{sp(T+3)}(\bm{w})-D_{sp(T+3)})^2}{\tilde{\sigma}^2_{\mathrm{int},sp(T+3)}(\bm{w}) + \sigma_{D,{sp(T+3)}}^2} \right) }.
    \end{split}
\label{eq:DNN_likelihood}
\end{equation}

Following \citet{Gal2015}, we define the prior over the weights as a zero-mean Normal distribution,
\begin{equation}
    \mathcal{P}(\bm{w}) = \mathcal{N}(\bm{w} \mid 0, \bm{I}/l^2),
\label{eq:DNN_prior}
\end{equation}
where $\bm{I}$ is the identity matrix and $l$ is the prior length-scale that regularises how large the weights can be. The posterior over the weights is given by the product of the prior distribution and the likelihood function over all $N_s$ transients at all $N_t$ time-steps,
\begin{equation}
     \mathcal{P}(\bm{w} |\bm{X}) \propto \mathcal{P}(\bm{w}) \prod\limits_{s=1}^{N_s}\prod\limits_{T=-70}^{80} \mathcal{P}(\bm{D}_{s(T+3)}|\bm{X}_{s(t \le T)}, \bm{w}),
\label{eq:DNN_posterior}
\end{equation}
where we ignore the Bayesian evidence as a scaling constant that is unnecessary for this work. We would ideally like to sample the negative log posterior while training our DNN, and so we derive the log prior from equation \ref{eq:DNN_prior} as $\log{\mathcal{P}(\bm{w})} = \mathrm{constant} - l^2 ||\bm{w}||_2^2 / 2$. We can ignore the additive constant not necessary for our optimisation and follow \citet{Gal2015} to implement the log prior by including an $L_2$ regularisation term $\lambda ||\bm{w}||_2^2$ weighted by some weight decay that averages over the number of transients $N_s$ and time-steps $N_t$,
\begin{equation}
    \lambda = \frac{l^2 (1-d)}{2 N_s N_t},
\end{equation}
where $d$ is the dropout rate (set to $0.2$ in this work), and we set $l=0.2$ consistent with work by \citet{Gal2015}\footnote{See Section 4.2 of \citet{Gal2015Appendix} for a detailed explanation of this prior and \url{https://github.com/yaringal/DropoutUncertaintyExps/blob/master/net/net.py} for an example implementation of this $L_2$ regularisation by Yarin Gal.}. Here, we have included the $(1-d)$ term to account for the dropout regularisation used in our work. With this term, the $L_2$ regularisation term $\lambda ||\bm{w}||_2^2$ matches the log prior, $\log{\mathcal{P}(\bm{w})}$, averaged over the number of transients and time-steps. However, we point out that our $\lambda$ slightly differs from equation 18 of \citet{Gal2015Appendix} because we use the negative log-likelihood instead of a squared loss as the cost function of our DNN. Furthermore, we also add the caveat that because of this difference of loss functions and our inclusion of a probabilistic neural network that outputs a predictive mean and standard deviation instead of a point estimate, it is not clear that the demonstration of MC dropout as an approximation to Bayesian neural networks in \citet{Gal2015} necessarily holds true in our work. Future machine learning research should check the validity of MC dropout as a Bayesian approximation in a broader range of neural network architectures.

Since we use dropout regularisation, we define a dropout objective function over all time-steps and over all transients that we aim to minimise while training the neural network model as follows,
\begin{equation}
     \mathrm{obj}(\bm{w}) = \sum\limits_{s=1}^{N_s}\sum\limits_{T=-70}^{80} 
     \left[ -\log \mathcal{P}(\bm{D}_{s(T+3)}|\bm{X}_{s(t \le T)}, \bm{w}) + \lambda ||\bm{w}||_2^2 \right]
    \label{eq:objective}
\end{equation}
where we sum the log-likelihood and $L_2$ regularisation term over all $N_t$ time-steps (between -70 and 80 days) and $N_s$ transients in the training set. To train the DNN and determine optimal values of its parameters $\bm{\hat{w}}$, we minimise the dropout objective function with the sophisticated and commonly used \texttt{Adam} gradient descent optimiser \citep{Kingma2014}. 

To make predictions, we evaluate the predictive distribution of the latent flux defined as follows,
\begin{equation}
    \mathcal{P}(\bm{F}_{s(T+3)} | \bm{X}_{s(t \le T)}) = \int \mathcal{P}(\bm{F}_{s(T+3)} | \bm{X}_{s(t \le T)}, \bm{w}) \mathcal{P}(\bm{w} |\bm{X}) d\bm{w},
    \label{eq:DNN_predicitve_distribution}
\end{equation}
where we are marginalising over the weights of the network by integrating the product of the predictive distribution of the latent flux given the network weights (first term in the integrand and defined in equation \ref{eq:posterior_predictive}) and the posterior distribution over the network weights (second term in the integrand and defined in equation \ref{eq:DNN_posterior}). The integral is intractable, and so we approximate it by using Monte Carlo dropout at inference time to sample the posterior distribution, as described in \cite{Gal2015}. We draw 100 samples from the posterior $\mathcal{P}(\bm{w} |\bm{X})$ by running 100 forward passes of the neural network for a given input. Each run of the neural network outputs both a mean $\tilde{y}_{sp(T+3)}(\bm{w}_\mathrm{draw})$ and standard deviation $\tilde{\sigma}_{\mathrm{int},sp(T+3)}(\bm{w}_\mathrm{draw})$ because of our probabilistic neural network architecture. To include the variance of each draw in the marginal predictive uncertainty $\sigma_{y,{sp(T+3)}}$, we compute $F_{sp(T+3)}(\bm{w}_\mathrm{draw}) \sim \mathcal{N} \left( \tilde{y}_{sp(T+3)}(\bm{w}_\mathrm{draw}), \tilde{\sigma}^2_{\mathrm{int},sp(T+3)}(\bm{w}_\mathrm{draw}) \right)$. We estimate the marginal predictive mean and uncertainty as the sample mean and standard deviation of the 100 values of $F_{sp(T+3)}(\bm{w}_\mathrm{draw})$ taken from the 100 forward passes of the neural network, respectively:
\begin{equation}
    y_{sp(T+3)} = \frac{1}{100} \sum_{\mathrm{draw=1}}^{100} F_{sp(T+3)}(\bm{w}_\mathrm{draw}),
\end{equation}
\begin{equation}
    \sigma_{y,{sp(T+3)}}  = \sqrt{\frac{1}{100} \sum_{\mathrm{draw=1}}^{100} \left(F_{sp(T+3)}(\bm{w}_\mathrm{draw}) - y_{sp(T+3)} \right)^2}.
\end{equation}
We use these to compute the anomaly scores discussed in \S\ref{sec:Anomaly_score_definition}.

\begin{figure*}
\centering
    {\includegraphics[width=0.49\linewidth]{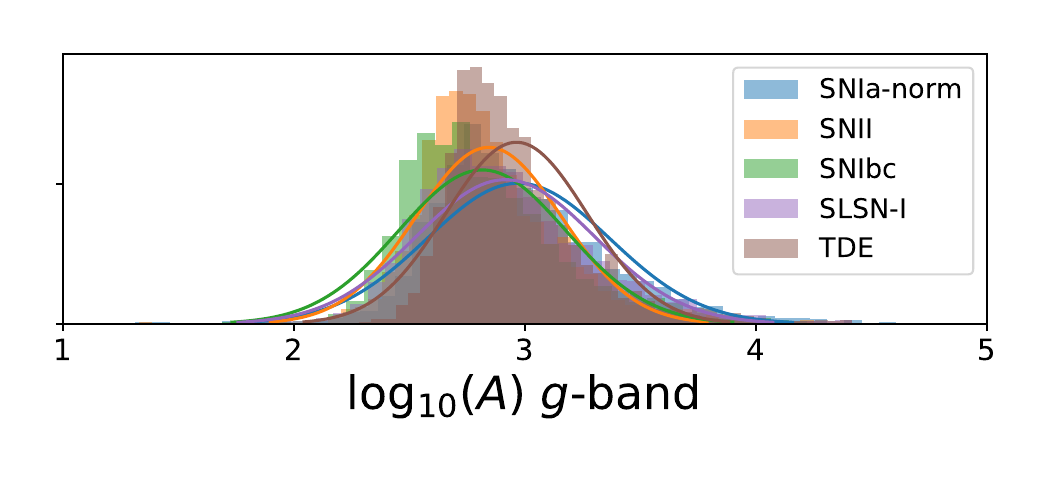}}
    \vspace{-1.5em}
    {\includegraphics[width=0.49\linewidth]{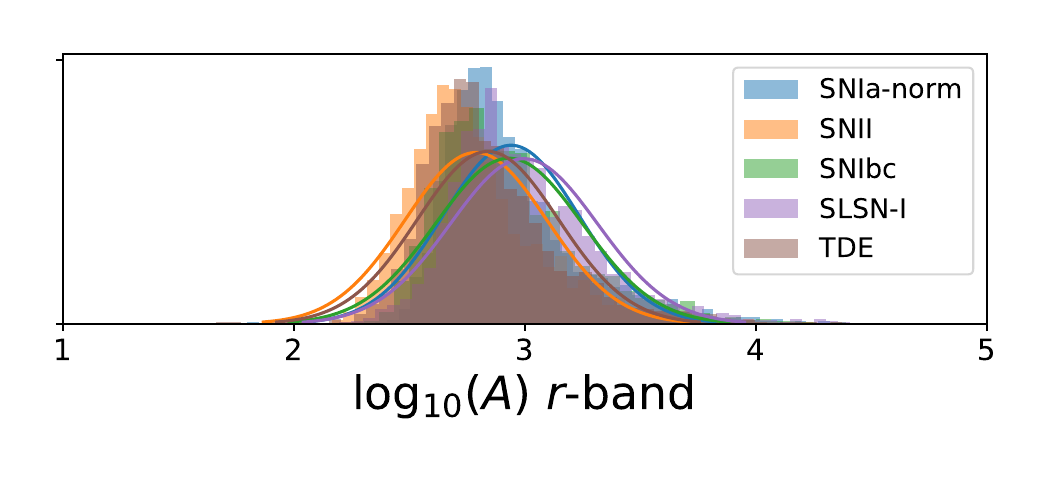}}
    {\includegraphics[width=0.49\linewidth]{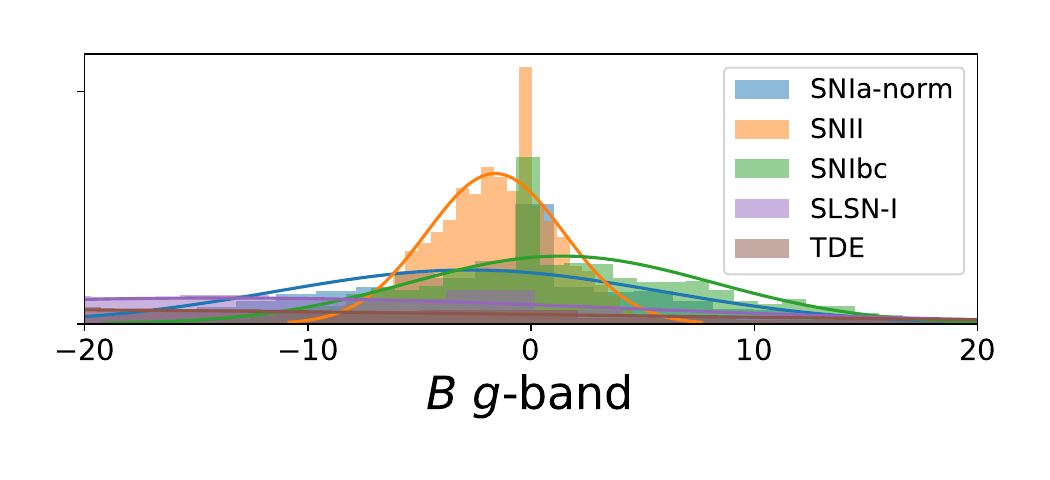}}
    \vspace{-1.5em}
    {\includegraphics[width=0.49\linewidth]{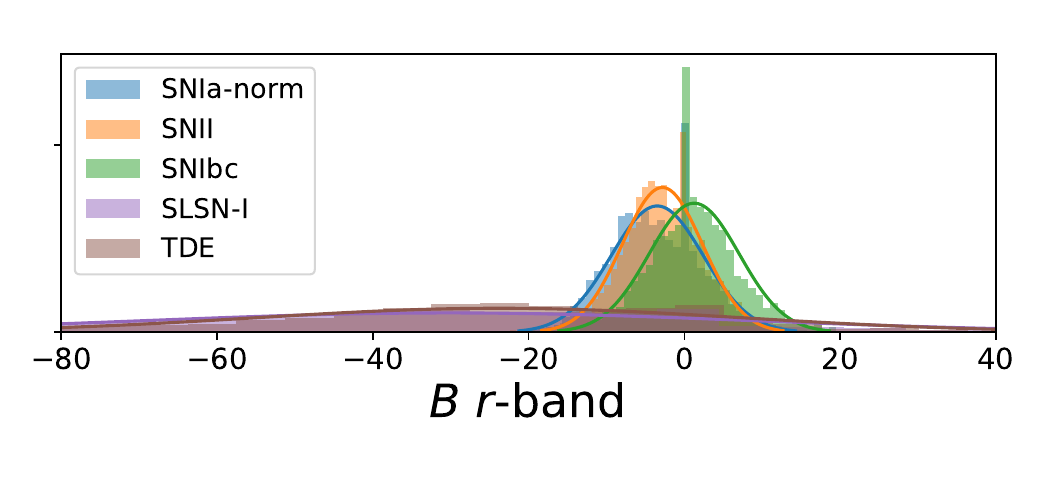}}
    {\includegraphics[width=0.49\linewidth]{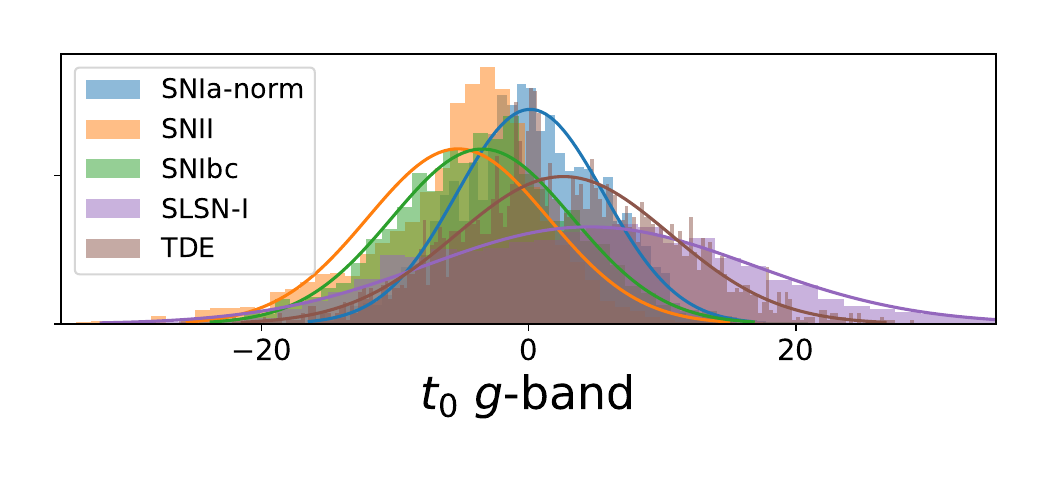}}
    \vspace{-1.5em}
    {\includegraphics[width=0.49\linewidth]{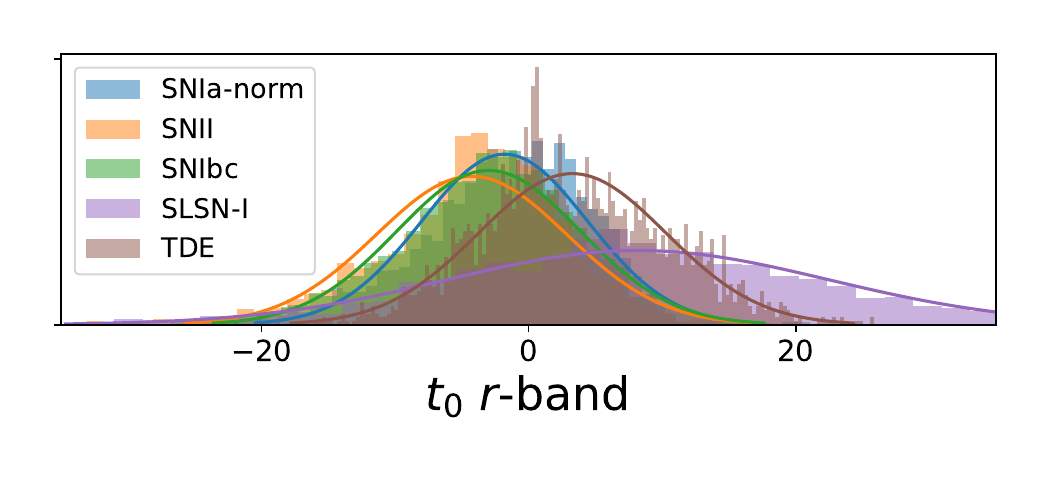}}
    {\includegraphics[width=0.49\linewidth]{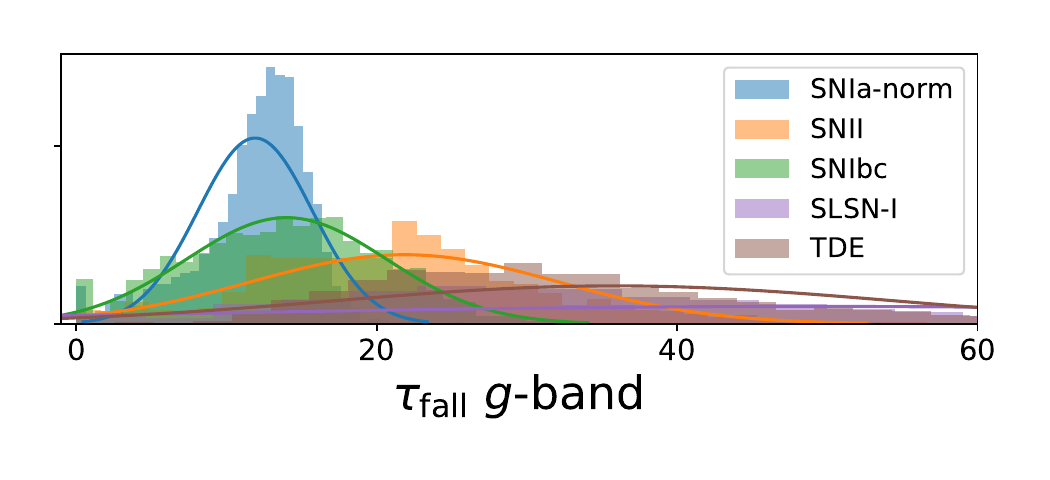}}
    \vspace{-1.5em}
    {\includegraphics[width=0.49\linewidth]{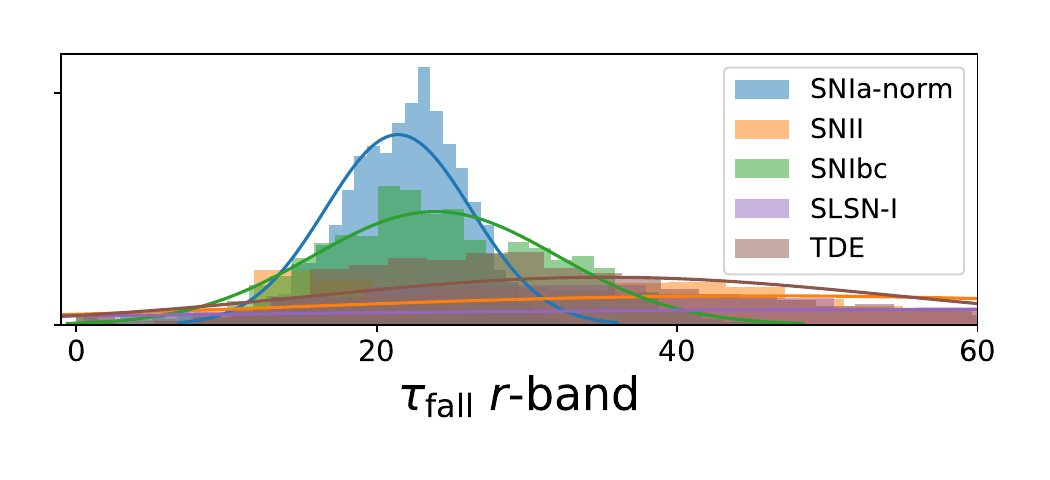}}
    {\includegraphics[width=0.49\linewidth]{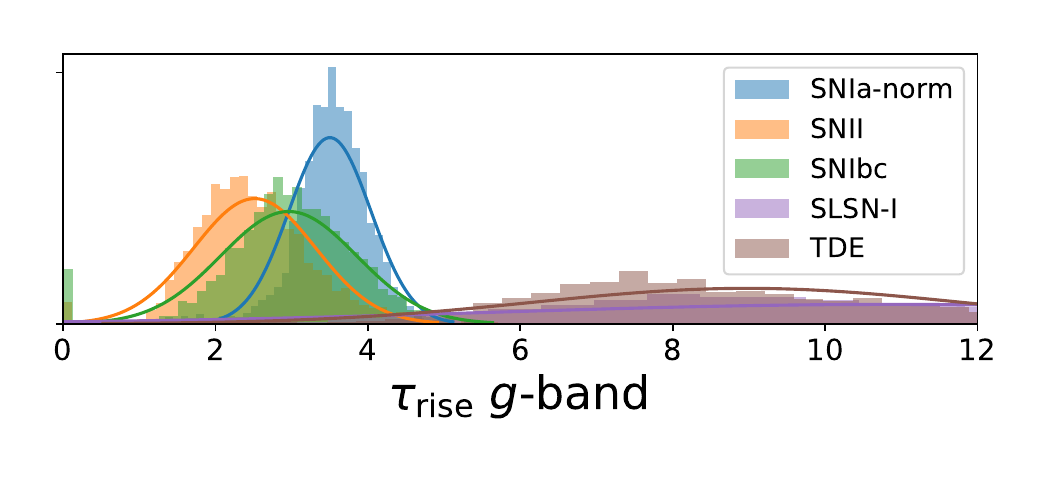}}
    \vspace{-1.5em}
    {\includegraphics[width=0.49\linewidth]{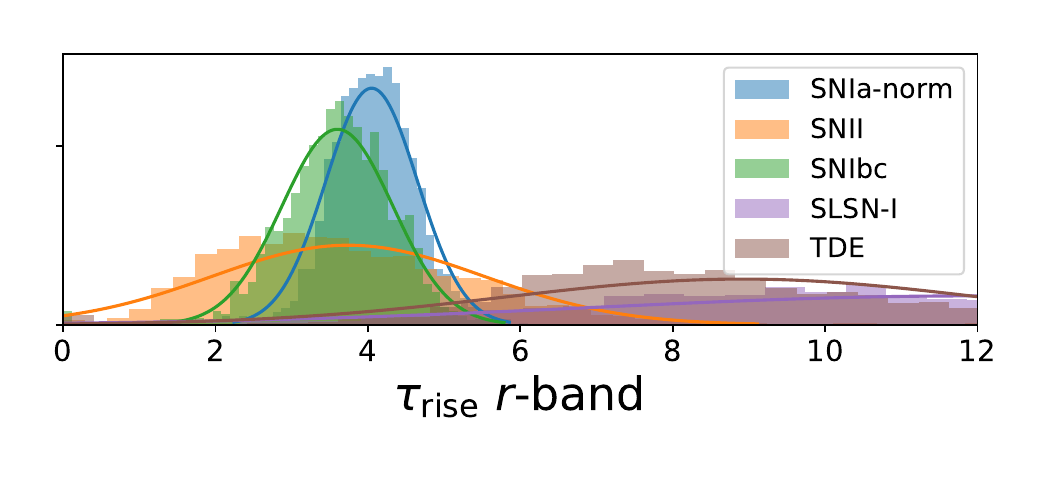}}
    {\includegraphics[width=0.49\linewidth]{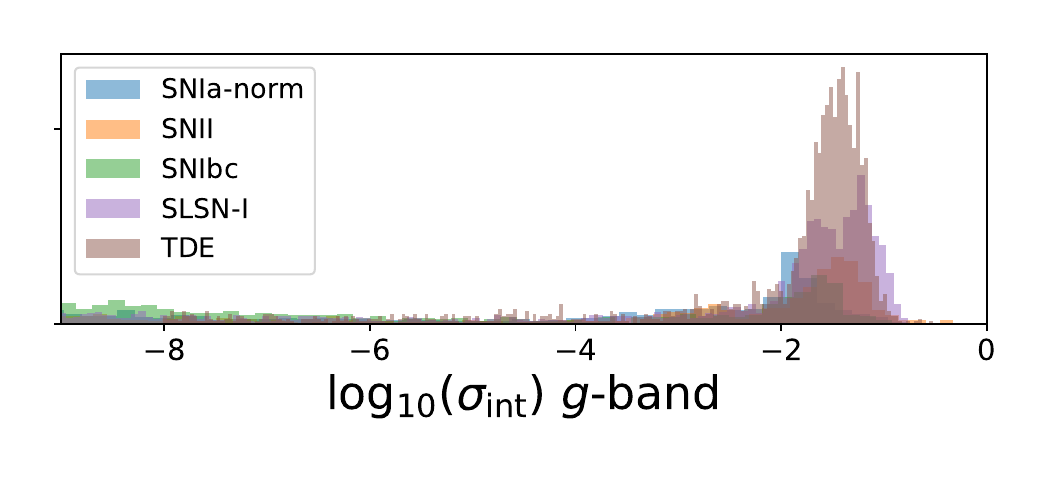}}
    {\includegraphics[width=0.49\linewidth]{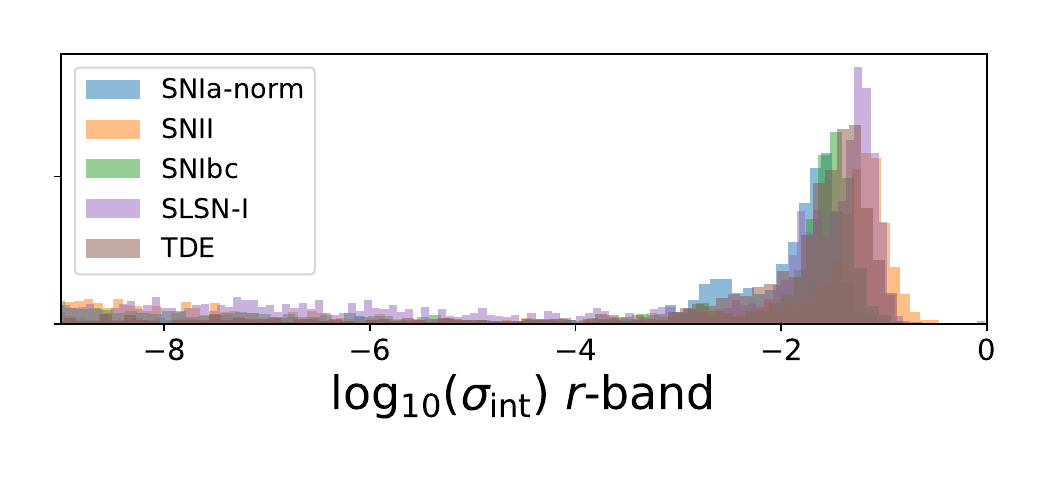}}

\caption{The histograms of the best fit Bazin parameters for the population of simulated light curves in each transient class and passband. We ignored any transient light curves that did not have 10 data points on each side of trigger. We modelled the population distribution as a multivariate Gaussian, and show the one-dimensional slices as the solid lines. We used this multivariate Gaussian as the priors for the Bazin parametric model defined in equation \ref{eq:Gaussian_prior}.}
    \label{fig:Bazin_parameter_distribution}
\end{figure*}

\subsection{Parametric Bayesian Bazin function}

\label{sec:Model_Bazin}
\subsubsection{Model Definition}
The fast inference speed of the DNN model makes it scaleable for the enormous data streams expected from surveys such as LSST. However, the large number of parameters in the model makes it difficult to ascertain how the model makes decisions. Thus, as a comparison, we have built a Bayesian model of each transient light curve based on the widely used phenomenological Bazin function from \citet{Bazin_function}. This method has the advantage of not requiring regularly sampled input data. However, to make it comparable to the DNN method, we use the interpolated fluxes $\bm{D}_{spt}$ and uncertainties $\bm{\sigma}_{D,{spt}}$ as the input data.

We begin by augmenting the standard Bazin function with an additional error term, $\epsilon_{\mathrm{int}}(t)$. We include this error because we know that the Bazin function is not a perfect representation of a light curve, and even if we had no measurement error, there would still be some discrepancy between the Bazin model and the observed light curve. Hence, we define a generative model of a transient's luminosity as follows,
\begin{equation}
    L(t) = L_0 \left ( \frac{e^{-(t-t_0)/\tau_{\mathrm{fall}}}}{1 + e^{-(t-t_0)/\tau_{\mathrm{rise}}}} + \epsilon_{\mathrm{int}}(t) \right)
    \label{eq:luminosity_bazin_function}
\end{equation}
where $L(t)$ is the luminosity as a function of time $t$ in days since trigger, $L_0$, $t_0$, $\tau_{\mathrm{fall}}$, $\tau_{\mathrm{rise}}$ are free parameters of the model, and $\epsilon_{\mathrm{int}}(t) \sim \mathcal{N}(0,\sigma_{\mathrm{int}}^2)$ is a zero-mean Gaussian random variable with intrinsic variance $\sigma_{\mathrm{int}}^2$. 

The Bazin model has the advantage of being much more interpretable than a DNN as the parameters can be intuitively understood with respect to the shape of a light curve, whereby, $L_0$ can be interpreted as the intrinsic luminosity of the transient, $\tau_{\mathrm{fall}}$, $\tau_{\mathrm{rise}}$ relate to the sloping rise and fall times of the light curve, and $t_0$ relates to the time of peak brightness.

We then model the latent flux of a transient $s$ in passband $p$ by first dividing equation \ref{eq:luminosity_bazin_function} by $4\pi d^2$, where $d$ is the distance to the transient object, and then adding a term for the measured background flux $B$,
\begin{equation}
    F_{spt}(\bm{\theta}) = A \frac{e^{-(t-t_0)/\tau_{\mathrm{fall}}}}{1 + e^{-(t-t_0)/\tau_{\mathrm{rise}}}} + B + A\epsilon_{\mathrm{int}}(t),
\label{eq:bazin_function}
\end{equation}
where $A = L_0/4\pi d^2$. Then, a generative model for the observed flux is derived by adding a measurement error $\epsilon_{D,{spt}}$, as follows, 
\begin{equation}
    D_{spt} = A \frac{e^{-(t-t_0)/\tau_{\mathrm{fall}}}}{1 + e^{-(t-t_0)/\tau_{\mathrm{rise}}}} + B + A\epsilon_{\mathrm{int}}(t) + \epsilon_{D,{spt}},
\label{eq:bazin_function_with_error}
\end{equation}
where we assume that the measurement error $\epsilon_{D_{spt}} \sim \mathcal{N}(0,\sigma_{D,{spt}}^2)$ is a zero-mean Gaussian random variable with variance $\sigma_{D,{spt}}^2$. The mean of the model in equation \ref{eq:bazin_function_with_error} is the Bazin function described in \citet{Bazin_function},
\begin{equation}
    f_{spt}(\bm{\theta}) = A \frac{e^{-(t-t_0)/\tau_{\mathrm{fall}}}}{1 + e^{-(t-t_0)/\tau_{\mathrm{rise}}}} + B,
\label{eq:bazin_function_no_error}
\end{equation}
with free parameters $\bm{\theta} = [\log_{10}{(A)}, B, t_0, \tau_{\mathrm{fall}}, \tau_{\mathrm{rise}}, \log_{10}{(\sigma_{\mathrm{int}})}]$. 

To model a partial light curve from -70 days before trigger or when observations begin up to time $T$, and given equations \ref{eq:bazin_function} and \ref{eq:bazin_function_with_error}, we write the likelihood function as follows,
\begin{equation}
\begin{split}
    &\mathcal{P}(\bm{D}_{sp(t \le T)}|\bm{t}, \bm{\theta}) =  \prod^{T}_{t=-70} \mathcal{N} \left(  D_{spt}  \mid f_{spt}(\bm{\theta}), A^2\sigma_{\mathrm{int}}^2 +  \sigma_{D,{spt}}^2\right) \\
      &= \prod^{T}_{t=-70} \left( 2\pi(A^2\sigma_{\mathrm{int}}^2 +\sigma_{D,{spt}}^2) \right)^{-0.5} \exp{\left( -0.5 \frac{(f_{spt}(\bm{\theta})-D_{spt})^2}{A^2\sigma_{\mathrm{int}}^2 + \sigma_{D,{spt}}^2} \right) }
\end{split}
\label{eq:bazin_likelihood}
\end{equation}

\subsubsection{Bayesian model and prior}
We define a Bayesian model to fit each transient light curve in a particular passband as follows,
\begin{equation}
    \mathcal{P}(\bm{\theta}|\bm{D}_{sp},\bm{t}) \propto \mathcal{P}(\bm{D}_{sp}|\bm{t}, \bm{\theta}) \mathcal{P}(\bm{\theta})
\end{equation}
where $\mathcal{P}(\bm{\theta}|\bm{D}_{sp},\bm{t})$ is the posterior distribution, $\mathcal{P}(\bm{D}_{sp}|\bm{t}, \bm{\theta})$ is the likelihood function from equation \ref{eq:bazin_likelihood}, and $\mathcal{P}(\bm{\theta})$ is the prior distribution of each transient class.

We have chosen to base our prior on the distribution of fits of the transient population in each passband and class. This is used to pass the transient class information into the Bazin model. To construct this population distribution, we ignored light curves that did not have at least ten data points before and after trigger. We deemed that the $\tau_{\mathrm{fall}}$ and $\tau_{\mathrm{rise}}$ parameters were not realistic for light curves that did not cover the full transient phase. We then fit each of the remaining light curves in the population with the Bazin function in equation \ref{eq:bazin_function} by minimising the negative log-likelihood function in equation \ref{eq:bazin_likelihood}. The one-dimensional histograms of the best fit parameters are shown in Figure \ref{fig:Bazin_parameter_distribution}. We computed the mean $\bm{\mu}_{\mathrm{pop}}$ and covariance $\bm{\Sigma}_{\mathrm{pop}}$ of the set of best fit parameters and hence modelled the population distributions as multivariate Gaussians for each passband and each transient class. We use this multivariate Gaussian as the prior distribution,
\begin{equation}
    \mathcal{P}(\bm{\theta}) = \mathcal{N} \left(\bm{\theta} | \bm{\mu}_{\mathrm{pop}}, \bm{\Sigma}_{\mathrm{pop}}  \right).
\label{eq:Gaussian_prior}
\end{equation}
The distribution of the parameters $A$ and $\sigma_{\mathrm{int}}$ have distributions that appear right-skewed. To make these distributions more Gaussian, we instead optimised over $\log_{10}{(A)}$ and $\log_{10}{(\sigma_{\mathrm{int}})}$ and use these reparameterisations in the multivariate Gaussian prior. The one-dimensional slices of the multivariate Gaussian priors are plotted in Figure \ref{fig:Bazin_parameter_distribution}. Some of the distributions look slightly different from the overplotted Gaussian fits. In future work we could consider modelling the distributions with a Gaussian mixture model to better represent the population data. However, we think that our Gaussian fits are reasonable approximations of the population parameter distributions, and will act as reasonable prior distributions for this work.

\subsubsection{Optimisation and fitting routine}
To fit each light curve, we ideally want to first sample the posterior $\mathcal{P}(\bm{\theta}|\bm{D}_{sp},\bm{t})$. Initially, we fit and sampled  the posterior of the 6-parameter model with an MCMC routine. However, this proved far too computationally slow to run over the thousands of light curves in our training set. Instead, we used the Laplace approximation to approximate the posterior as a multivariate Gaussian centred on the mode. To apply this, we first optimised the objective function with the Nelder-Mead optimisation routine (available in the \texttt{scipy} optimisation library) after setting the starting parameter values to the median of the histograms shown in Figure \ref{fig:Bazin_parameter_distribution}. We then computed the Hessian matrix of the negative log posterior, using the autodifferentiation package \texttt{autograd}, and evaluated it at the optimal parameter values,

\begin{equation}
  \mathrm{H}_{qr} = -\frac{\partial^{2} \log{\mathcal{P}(\bm{\theta}|\bm{D}_{sp},\bm{t})}}{\partial \theta_{q} \partial \theta_{r} } \biggr\rvert_{\bm{\theta} = \bm{\hat{\theta}}}
\label{eq:Bazin_Hessian}
\end{equation}
where $\bm{\hat{\theta}}$ is the optimal parameter values, and $q$ and $r$ run across the six parameters in the model. The inverse of this Hessian is the covariance matrix,
\begin{equation}
    \bm{\Sigma} = \bm{\mathrm{H}}^{-1}
\end{equation}
To ensure that the Laplace approximation was a good approximation of the posterior, we compare it with a fit using MCMC. We show the covariance contours from the Hessian matrix and the MCMC samples of an example SNIa $g$-band light curve fit in Appendix Figure \ref{fig:example_MCMC_Bazin_contours}. While the Hessian approximates most parameter covariances well, it is a poor approximation of $\log_{10}{(\sigma_{\mathrm{int}})}$. The MCMC samples show that the posterior is non-Gaussian over $\log_{10}{(\sigma_{\mathrm{int}})}$ and prefers negative values. Since the Laplace method approximates the posterior as a Gaussian, it does not approximate this behaviour well. The large values of $\log_{10}{(\sigma_{\mathrm{int}})}$ can lead to unphysically large estimations of the predicted flux. 

To account for this poor approximation, we define $\bm{\theta'} = [\log_{10}{(A)}, B, t_0, \tau_{\mathrm{fall}}, \tau_{\mathrm{rise}}]$ and $\bm{\Sigma'}$ to not include  $\log_{10}{(\sigma_{\mathrm{int}})}$. Then, to obtain posterior fits, we drew 100 samples from a multivariate Gaussian with mean $\bm{\hat{\theta}'}$ and covariance $\bm{\Sigma'}$ and set $\log_{10}{(\sigma_{\mathrm{int}})}$ to the optimal fit for all draws. We evaluated equation \ref{eq:bazin_function} with these sets of parameters to obtain posterior fits. We noticed that some posterior samples in the Bazin parameter space produced unrealistic light curves that had wildly large flux values that did not fit past data well. These unrealistic parameter values are an artefact of the Laplace Approximation not being a good enough approximation to the true posterior (we discuss this in detail Appendix \ref{sec:Appendix_Fisher_matrix_approximation}). To account for this behaviour, we ignored posterior fits that deviated from data previous to the present time $T$ by $\chi^2 > 10$, leaving $K$ posterior samples. We note that usually, less than 5\% of the samples resulted in spurious fits, and so $K$ was not much less than 100.

Unlike the DNN that performs regression over past data to give a flux predictions at a single time-step, our Bazin method performs regression over time to give flux predictions at all times. Therefore, to compare the predictive power of this method with the DNN, we use the interpolated fluxes as input, use data up to time $T$, and record the prediction 3 days later $(T+3)$. We do this for each time-step, to obtain a sequence of predictions one time-step in the future from given data.

In practice, for anomaly detection, we need to make predictions of $D$ at new times $T$ (where this $T$ corresponds to the interpolated times used in the DNN for easy comparison). This requires that we evaluate the predictive distribution defined by
\begin{equation}
\begin{split}
    &\mathcal{P}(F_{sp(T+3)} | \bm{D}_{sp(t \le T)}, \bm{t}) \\ 
    & \qquad = \int \mathcal{P}(F_{sp(T+3)}|\bm{\theta}, \bm{t}) \mathcal{P}(\bm{\theta}|\bm{D}_{sp(t \le T)}, \bm{t}) d\bm{\theta}.
\end{split}
\label{eq:Bazin_predictive_distribution}
\end{equation}
This distribution can be compared to the predictive distribution for the DNN described in equation \ref{eq:DNN_predicitve_distribution}. The integral on the RHS cannot be computed analytically, and so we approximate it by sampling. We draw $K$ sample parameters $\bm{\theta}_{\mathrm{draw}}$ of the posterior (second term in the integrand) and compute the flux predictions $F_{sp(T+3)}(\bm{\theta}_{\mathrm{draw}})$ for each set of parameters (first term in the integrand) with equation \ref{eq:bazin_function}. The LHS of equation \ref{eq:Bazin_predictive_distribution} is the sampled probability density function, and we estimate the marginal predictive mean and uncertainty as the sample mean and standard deviation of the fluxes computed from the posterior draws, respectively: 
\begin{equation}
    y_{sp(T+3)} = \frac{1}{K} \sum_{\mathrm{draw=1}}^{K} F_{sp(T+3)}(\bm{\theta}_{\mathrm{draw}})
\end{equation}
\begin{equation}
    \sigma_{y,{sp(T+3)}}  = \sqrt{\frac{1}{K} \sum_{\mathrm{draw=1}}^{K} \left( F_{sp(T+3)}(\bm{\theta}_{\mathrm{draw}}) - y_{sp(T+3)} \right)^2}.
\end{equation}
We use these to compute the anomaly scores discussed in \S\ref{sec:Anomaly_score_definition}.  We also plot the $K \approx 100$ posterior fits (excluding the unrealistic spurious fits) and the median of these in the respective plots throughout this paper. 

\begin{figure*}
\centering
\textbf{Bazin}\par\medskip
\vspace{-0.5em}
    {\includegraphics[width=0.33\linewidth]{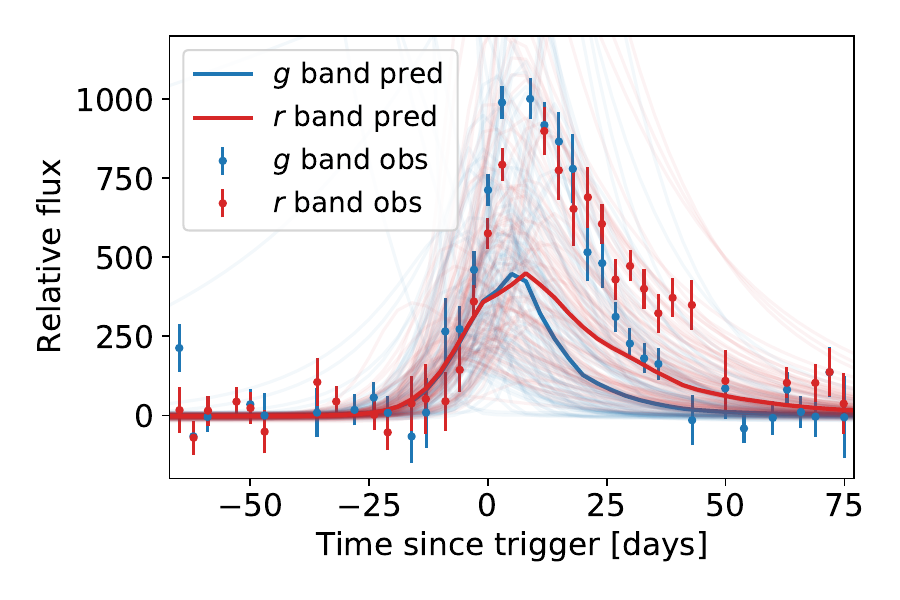}}
    {\includegraphics[width=0.33\linewidth]{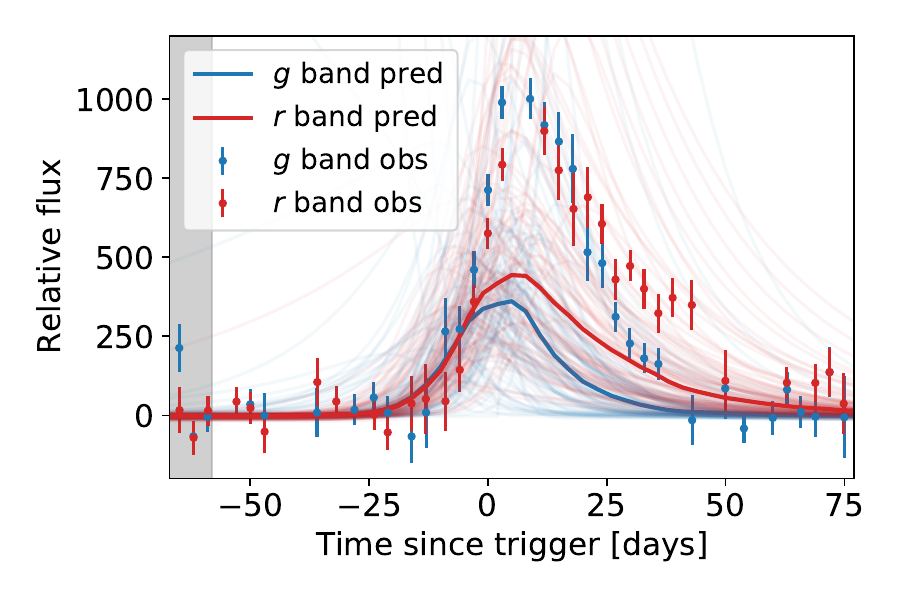}}
    {\includegraphics[width=0.33\linewidth]{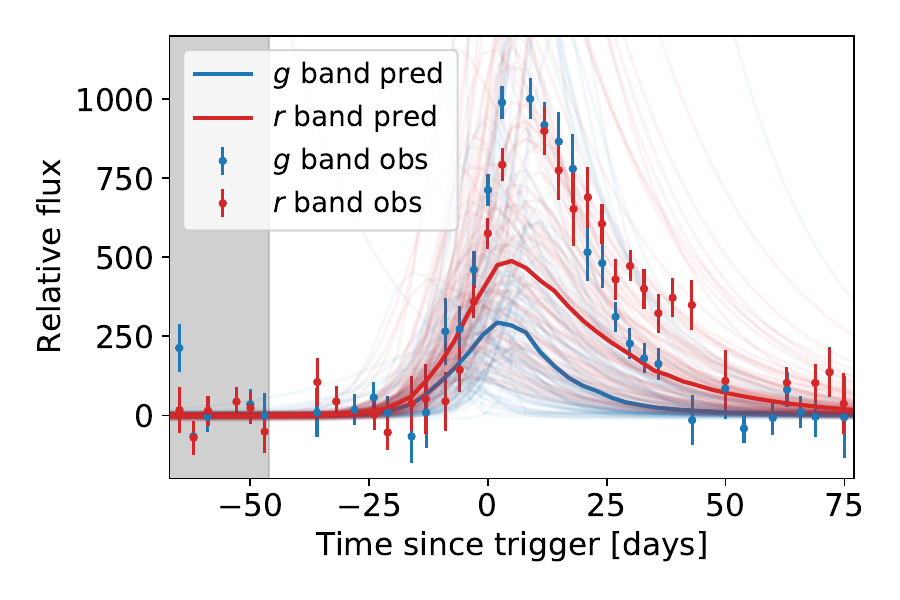}}
    {\includegraphics[width=0.33\linewidth]{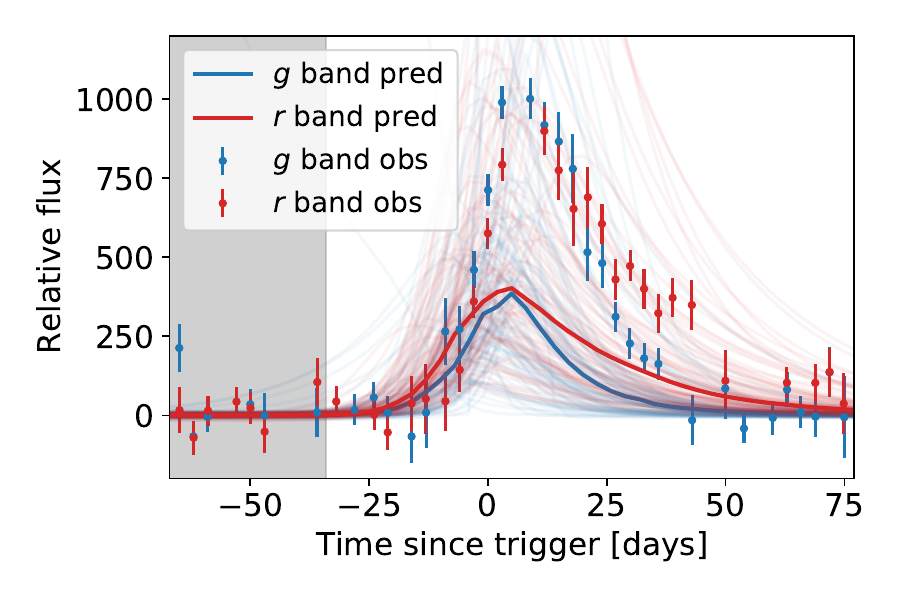}}
    {\includegraphics[width=0.33\linewidth]{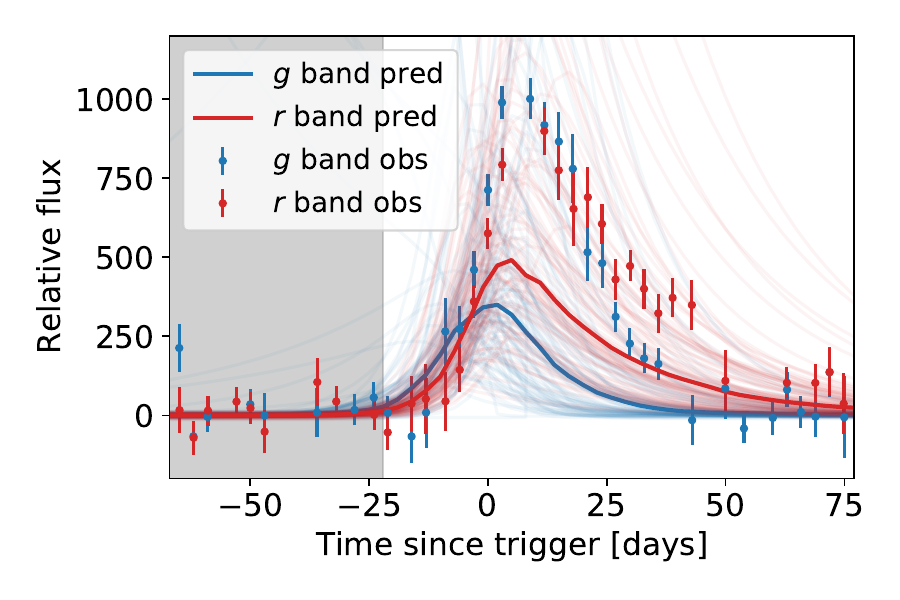}}
    {\includegraphics[width=0.33\linewidth]{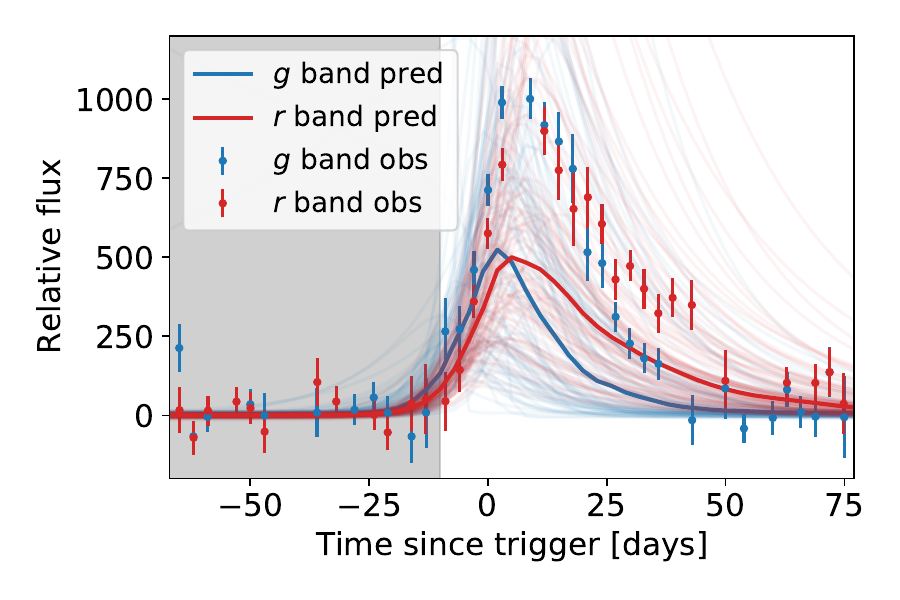}}
    {\includegraphics[width=0.33\linewidth]{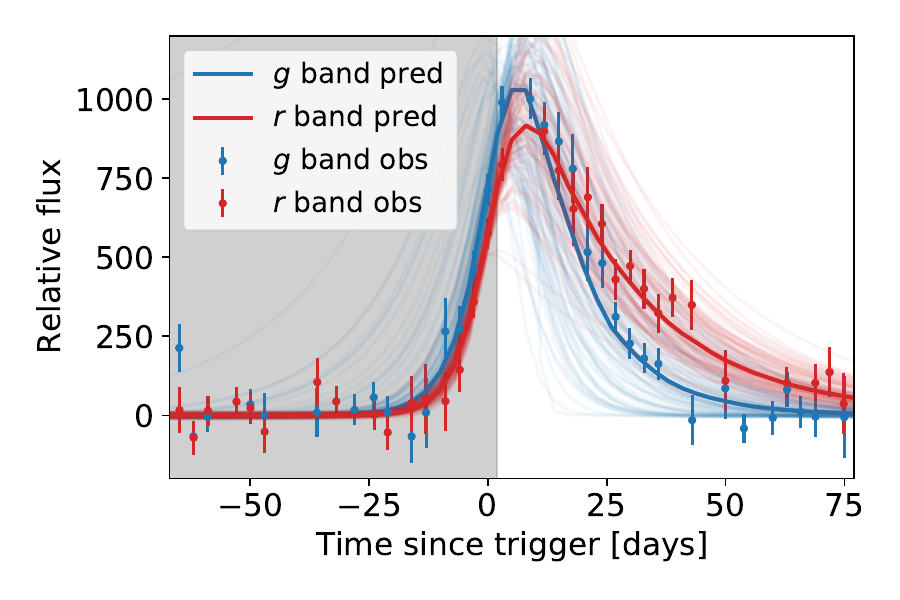}}
    {\includegraphics[width=0.33\linewidth]{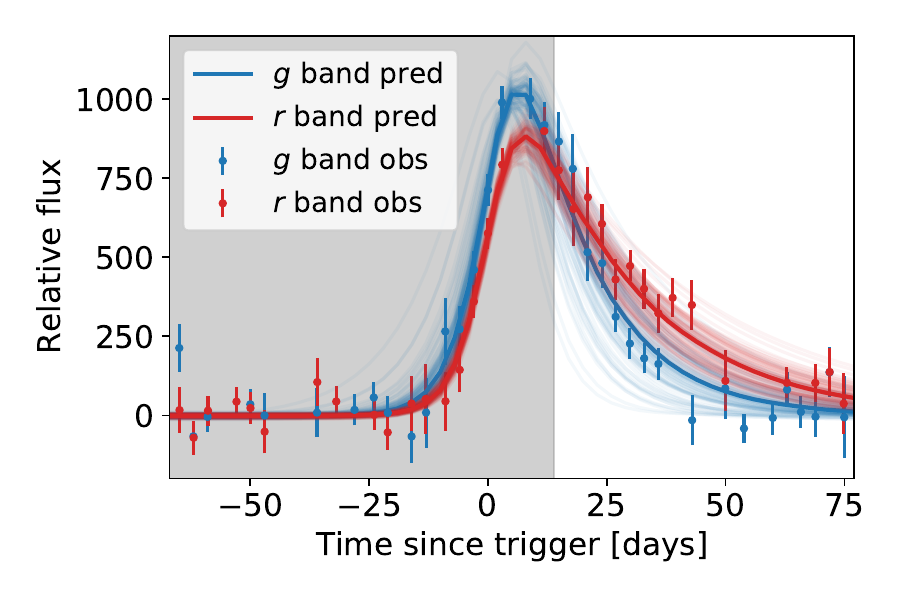}}  {\includegraphics[width=0.33\linewidth]{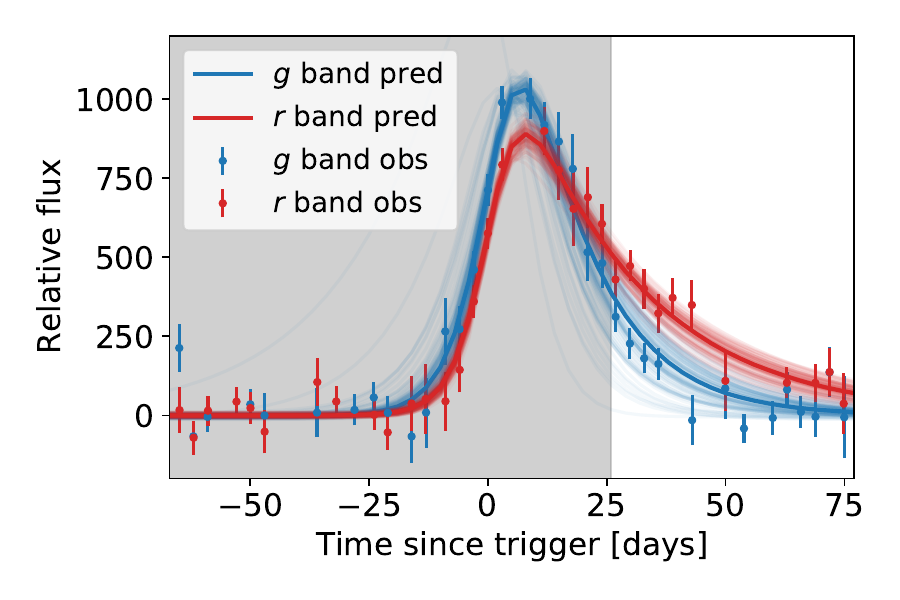}}
    {\includegraphics[width=0.33\linewidth]{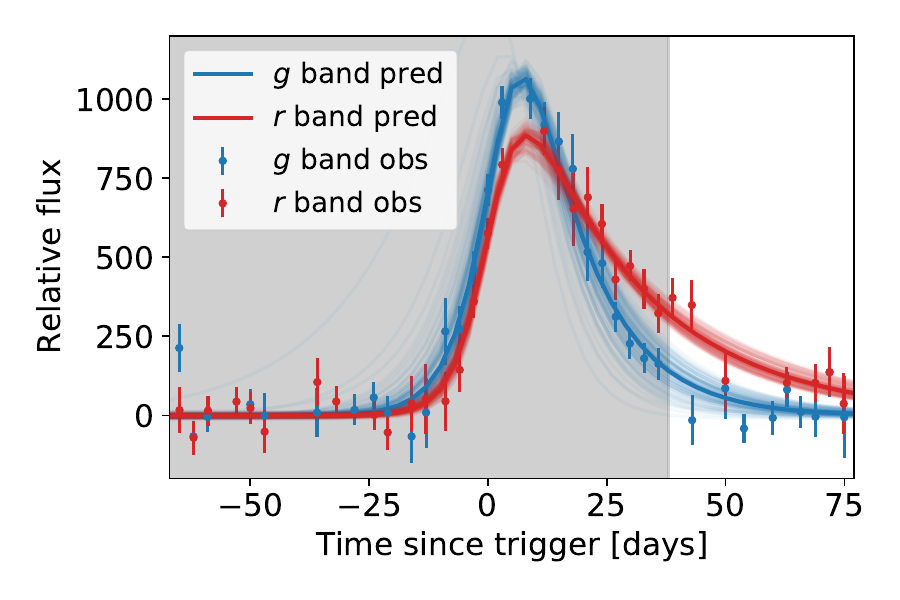}}
    {\includegraphics[width=0.33\linewidth]{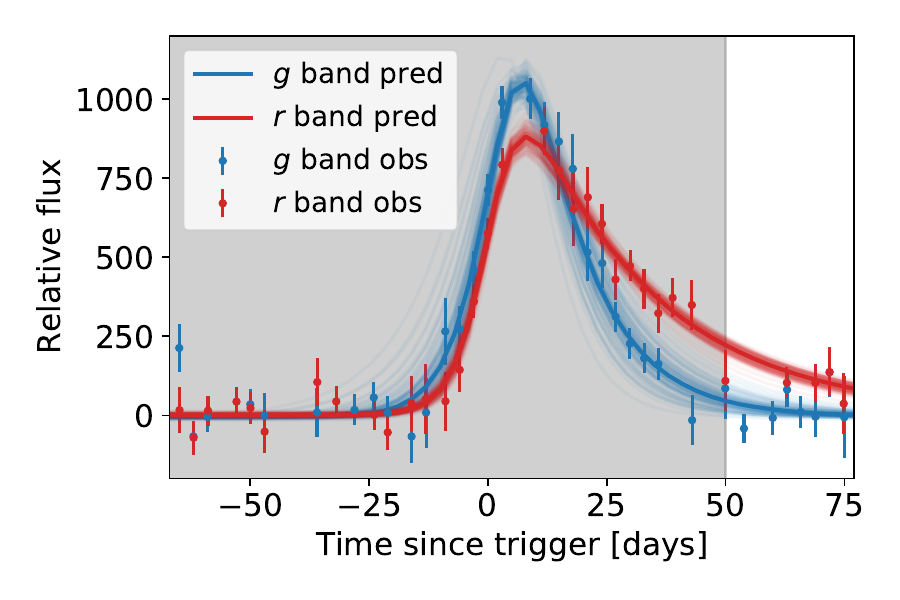}}
    {\includegraphics[width=0.33\linewidth]{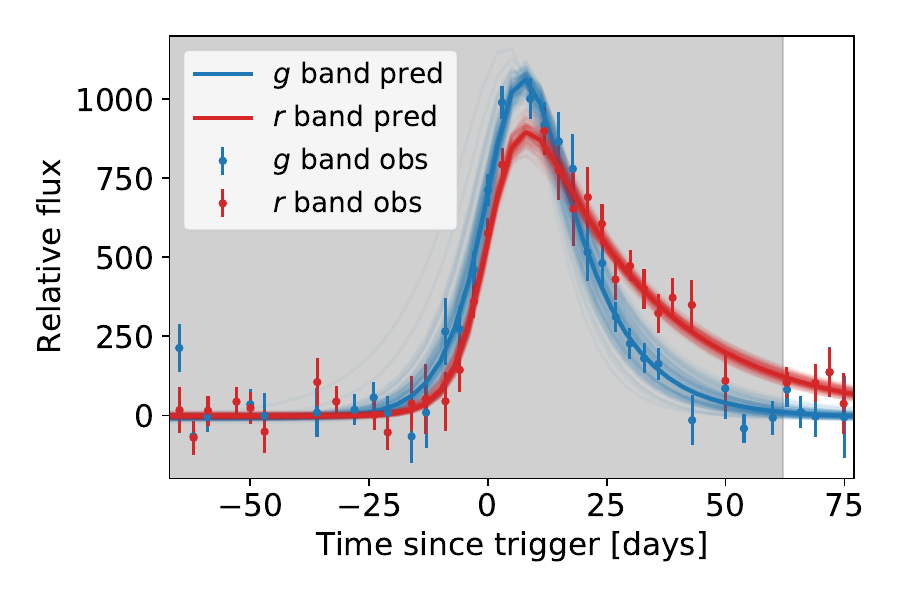}}

\caption{The Bazin parametric method being used as a generative model of a SNIa given a partial light curve. The grey shaded region is the region of data that the model was fit with, while the observations in the white region was not used to fit the model. The trace lines show the normally distributed posterior sample fits from the Laplace approximation. The bold solid line is the median of the posterior fits. The first panel does not use any data from the light curve and thus illustrates the fits from the prior distribution. The following panels use data up to times -58, -46, -34, -22, -10, 2, 14, 26, 38, 50, and 62 days from trigger, respectively. The plots show a fit to an example simulated SNIa.}
    \label{fig:Bazin_generative_plots}
\end{figure*}

\begin{figure*}
\centering
\textbf{DNN}\par\medskip
\vspace{-0.5em}
    {\includegraphics[width=0.33\linewidth]{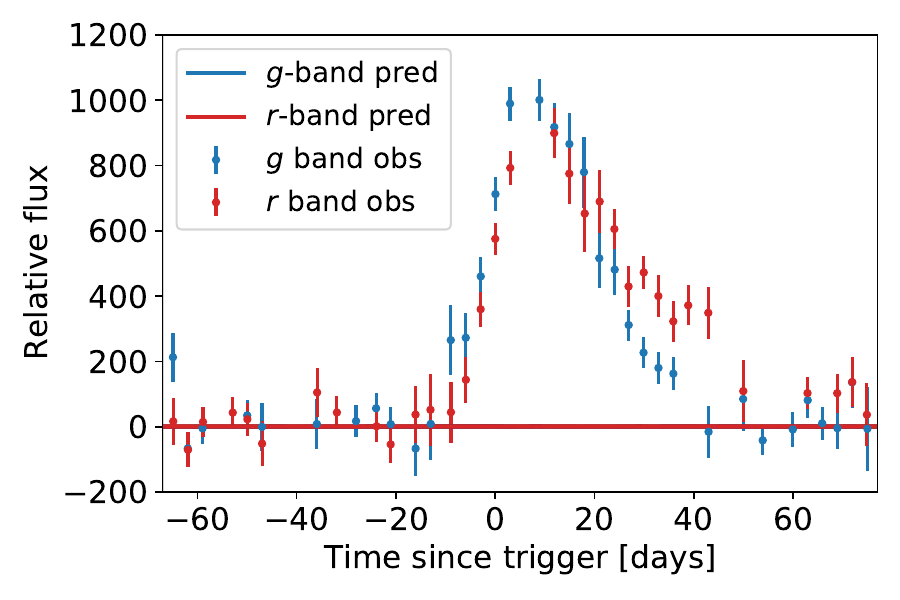}}
    {\includegraphics[width=0.33\linewidth]{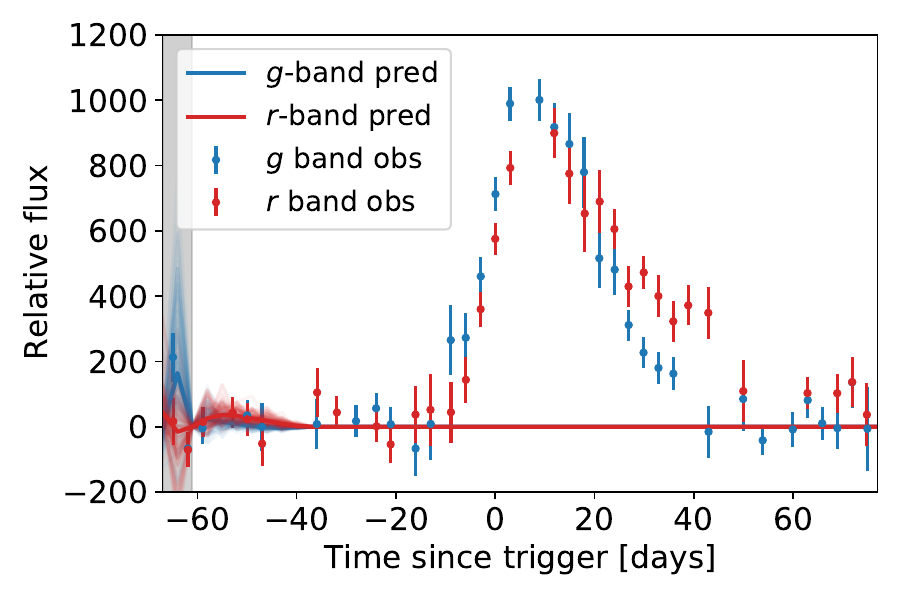}}
    {\includegraphics[width=0.33\linewidth]{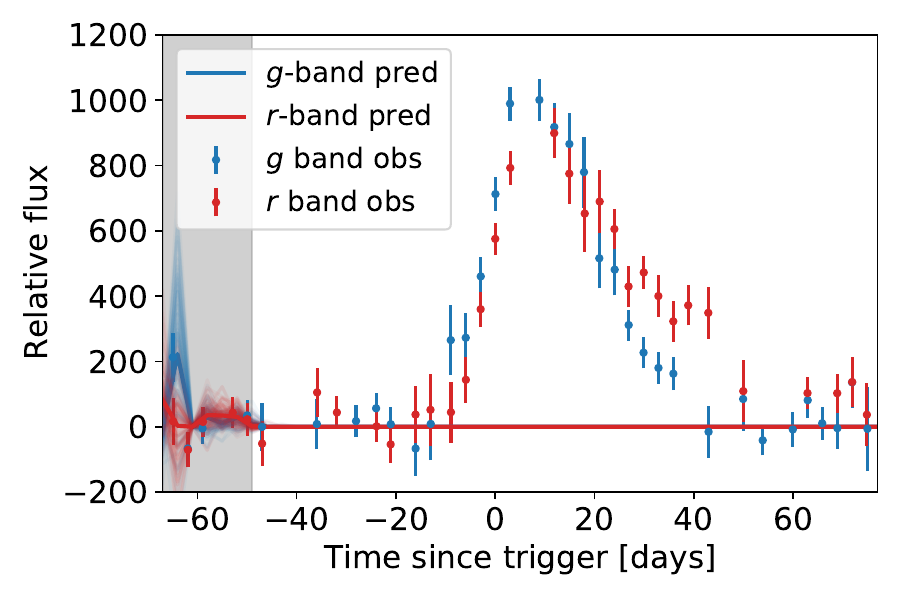}}
    {\includegraphics[width=0.33\linewidth]{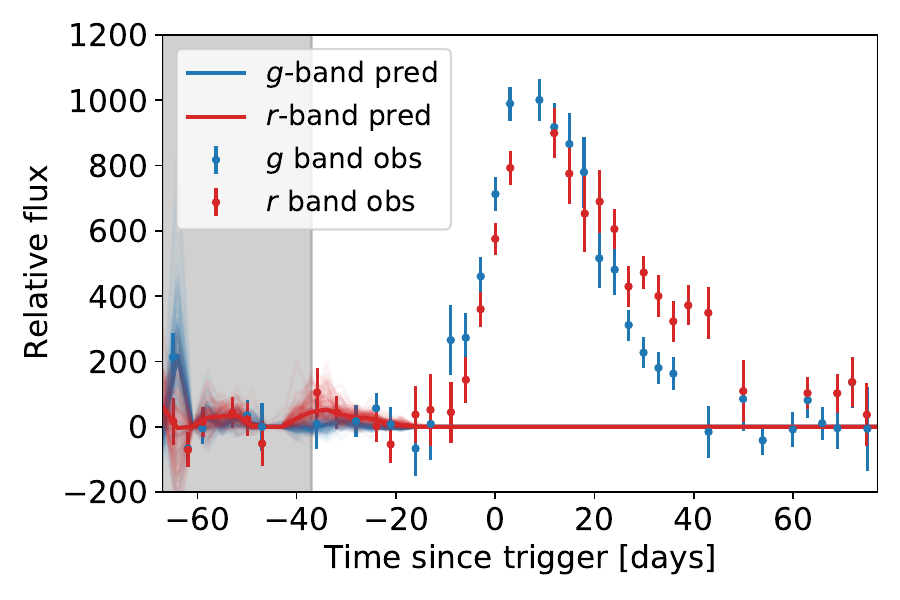}}
    {\includegraphics[width=0.33\linewidth]{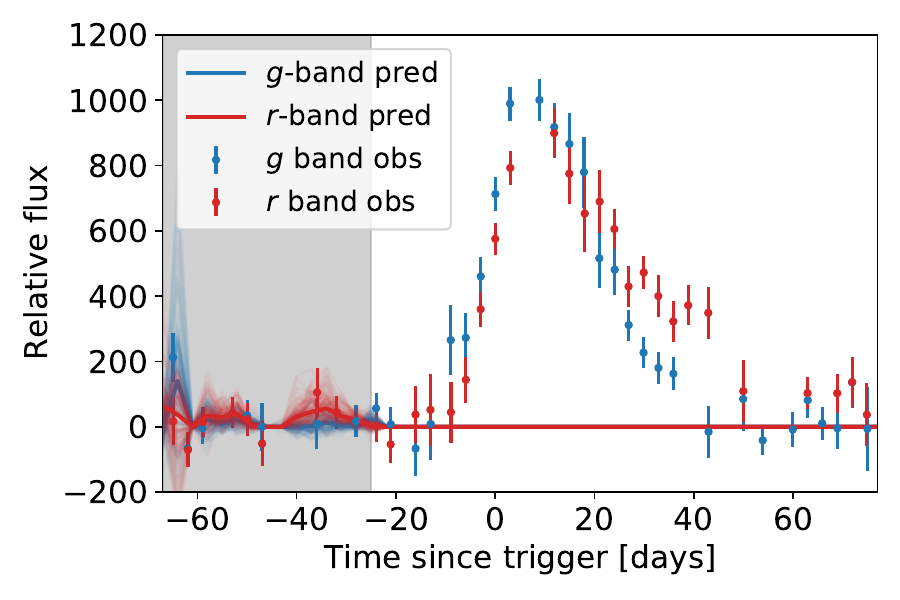}}
    {\includegraphics[width=0.33\linewidth]{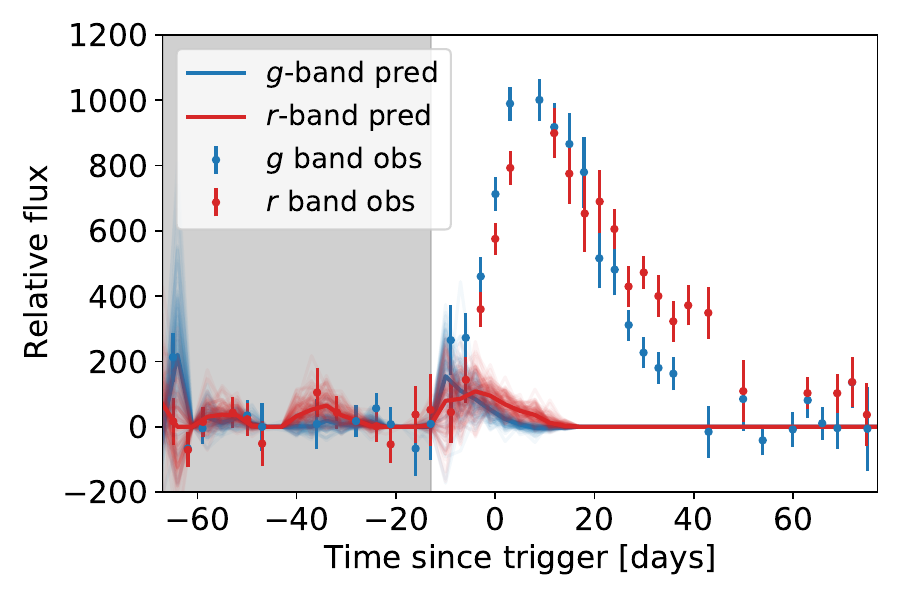}}
    {\includegraphics[width=0.33\linewidth]{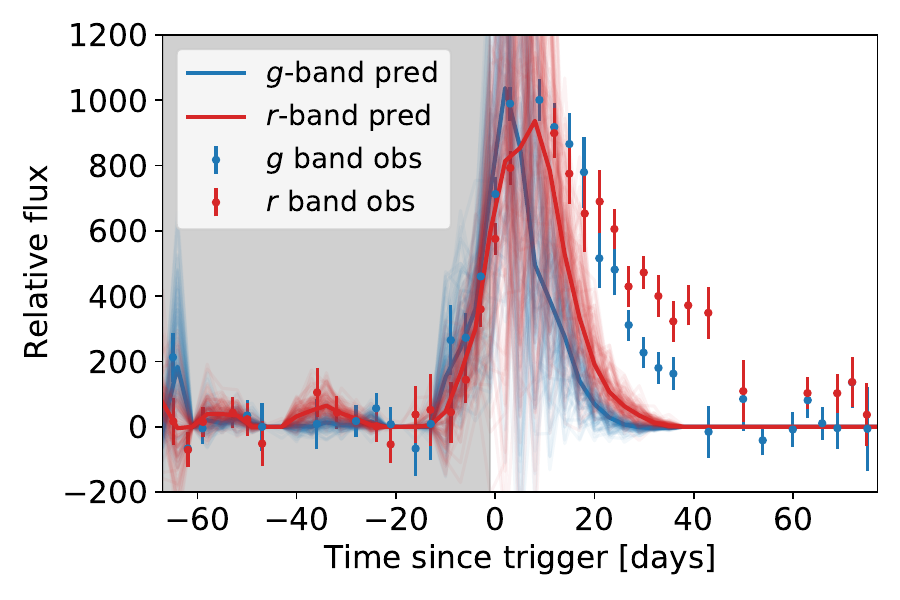}}
    {\includegraphics[width=0.33\linewidth]{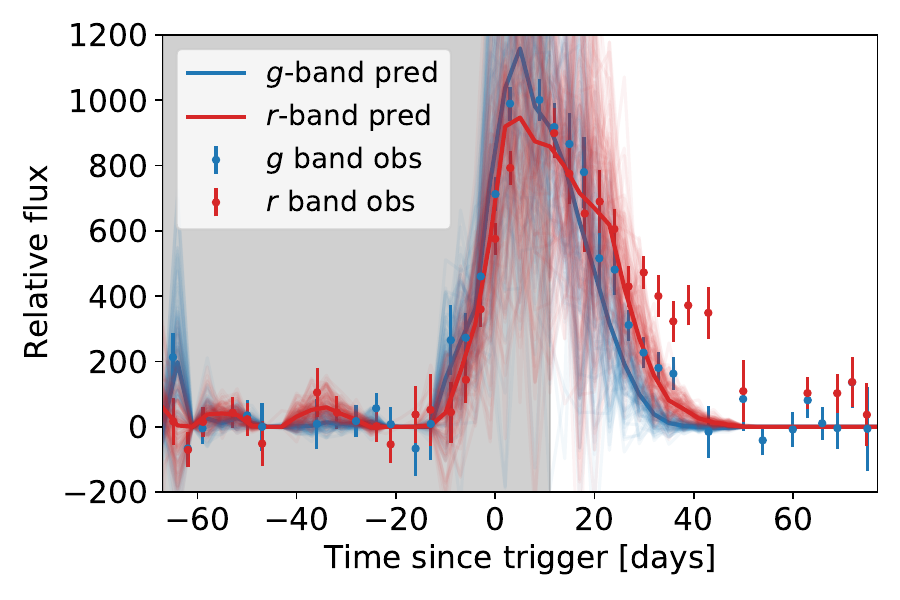}}
    {\includegraphics[width=0.33\linewidth]{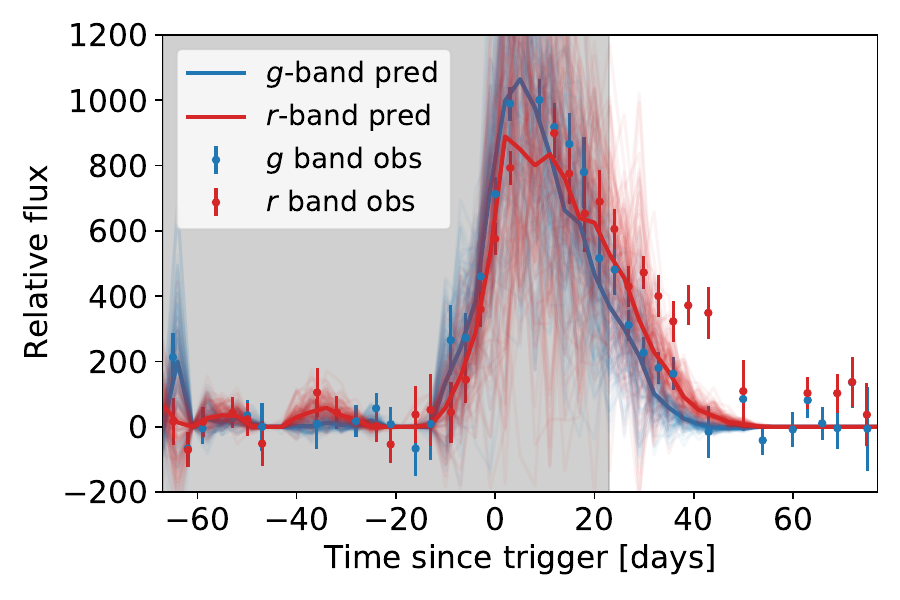}}
    {\includegraphics[width=0.33\linewidth]{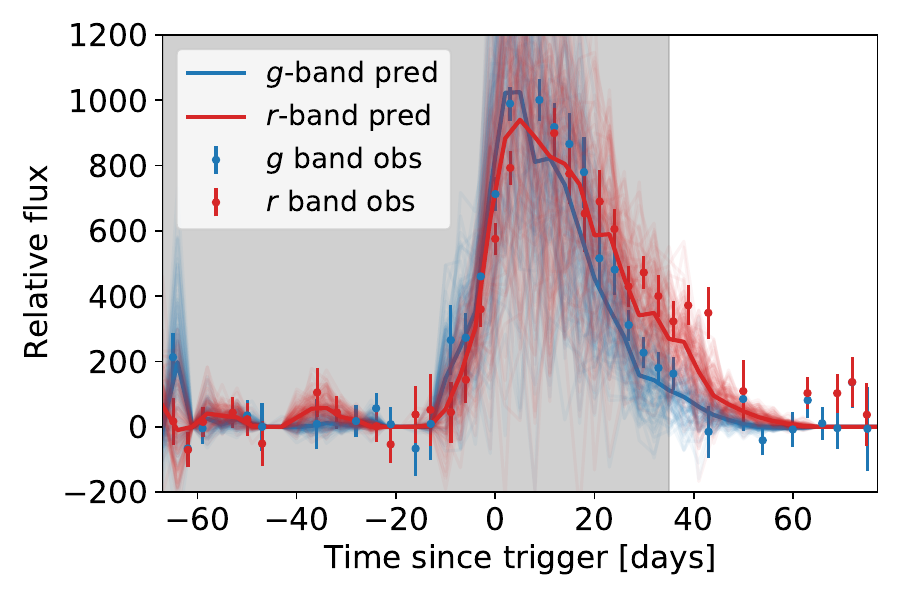}}
    {\includegraphics[width=0.33\linewidth]{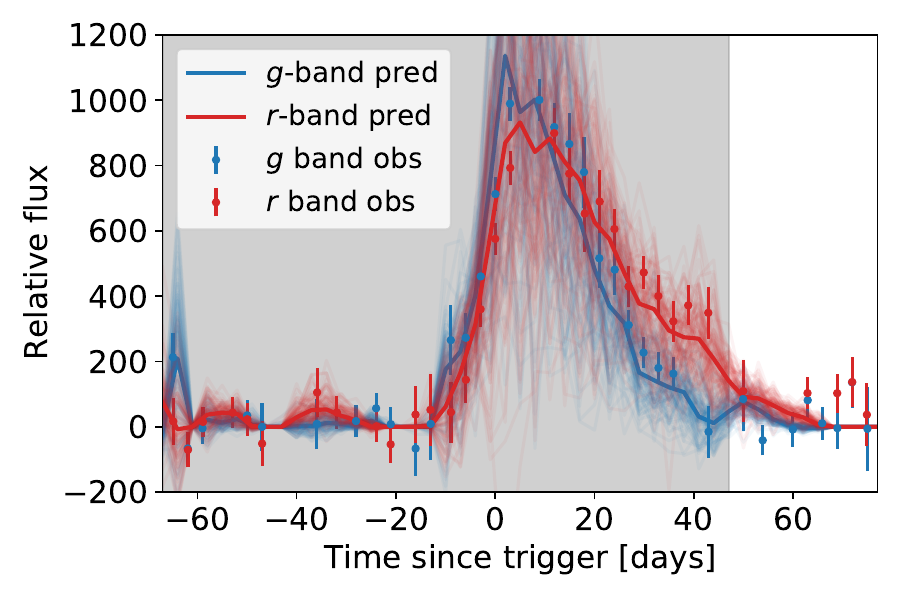}}
    {\includegraphics[width=0.33\linewidth]{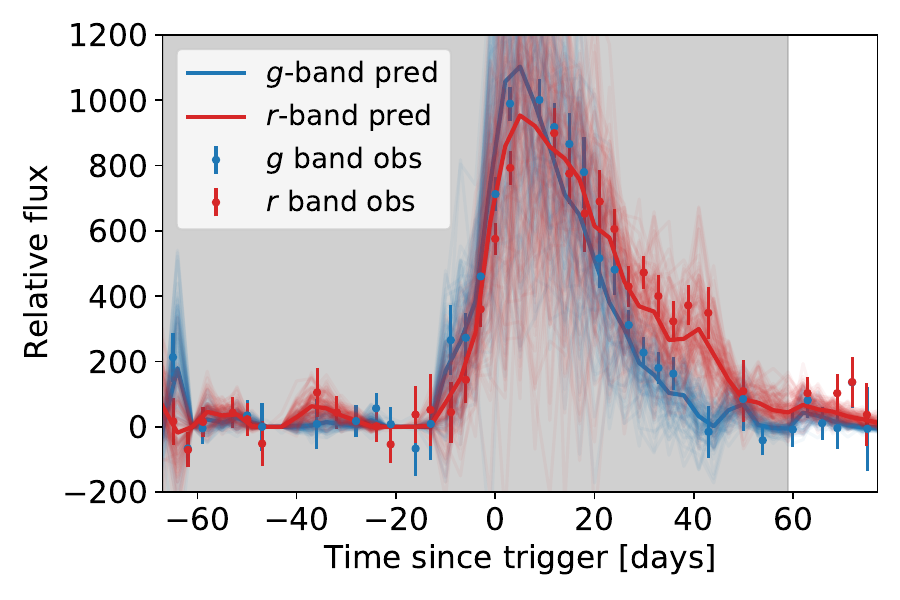}}

\caption[The DNN being used as a generative model of a SNIa given partial light curves.]{The DNN being used as a generative model of a SNIa given a partial light curve. The grey shaded region is the region of data that was used to make a prediction, while the observations in the white region was not used to make predictions. The DNN is data-driven and learns about the light curve as new observations arrive. Initially the DNN sensibly predicts that the light curve will stay flat, and it is only after the light curve begins to rise that the DNN revises its prediction. As the DNN was specifically designed to only predict one time-step in the future and was not designed to generate an entire light curve, it cannot be expected to perform well, but acts as a good comparison to Figure \ref{fig:Bazin_generative_plots}. To obtain a sequence of predictions, we feed in the predicted values back into the DNN as if they were part of the observations. The trace lines illustrate the posterior sample predictions and the bold solid line is the median of the posterior predictions. The panels sequentially show the predicted light curve given increasing amounts of observational data, and each panel uses observations up to times -58, -46, -34, -22, -10, 2, 14, 26, 38, 50, and 62 days from trigger. The plots show predictions on an example simulated SNIa.}
    \label{fig:DNN_generative_plots}
\end{figure*}

\begin{figure*}
    \textbf{DNN}\par\medskip
    \vspace{-0.5em}
    \centering
        {\includegraphics[width=0.3\linewidth]{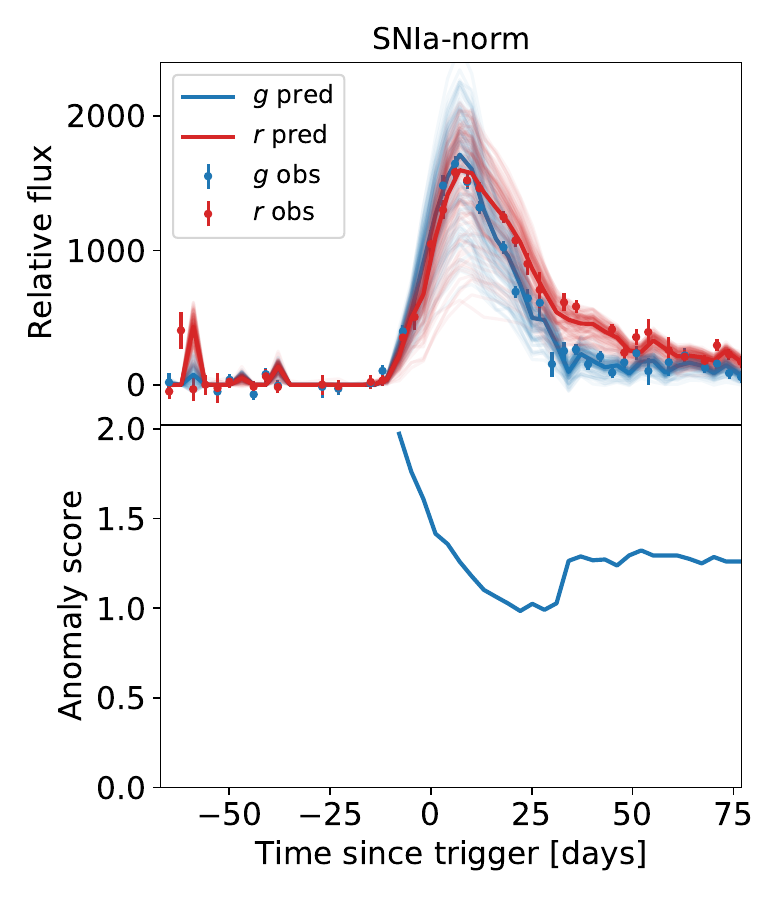}}
        {\includegraphics[width=0.3\linewidth]{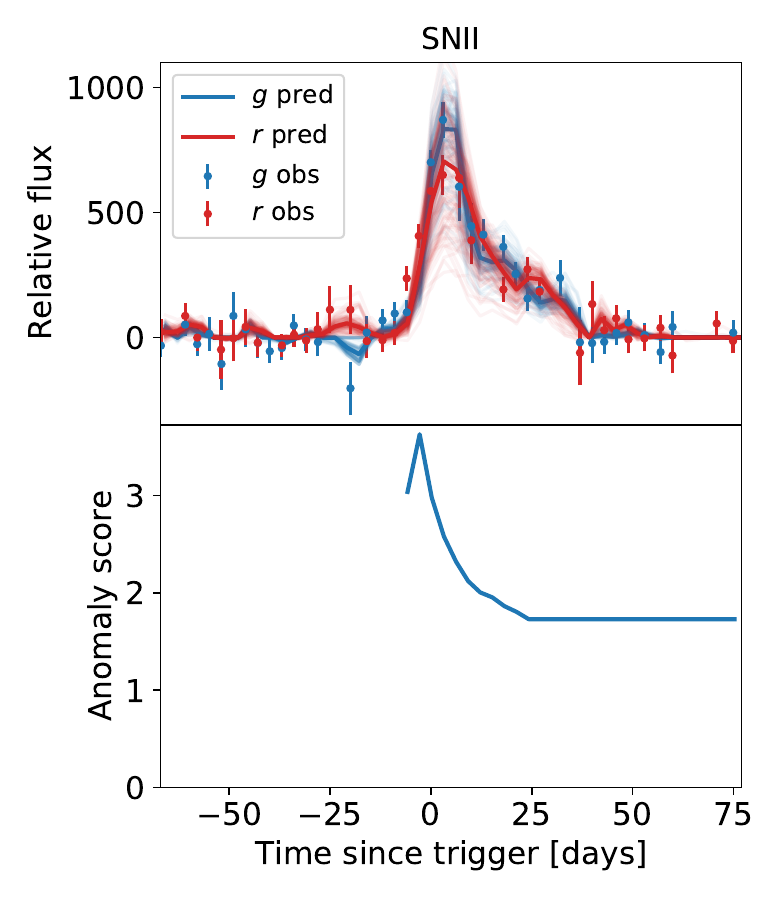}}
        {\includegraphics[width=0.3\linewidth]{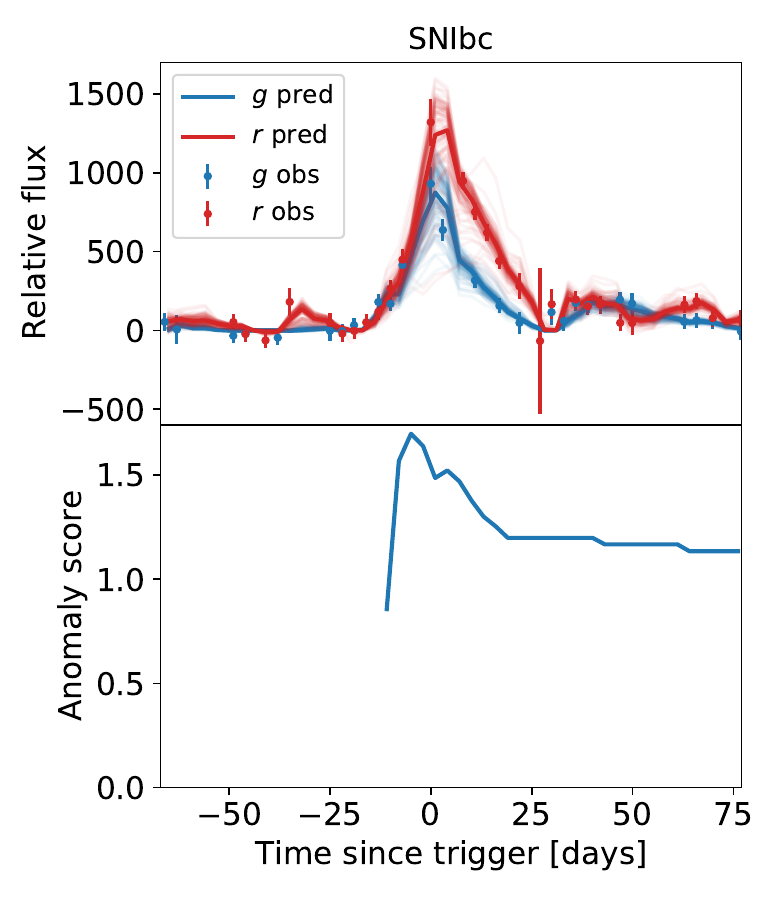}}
        {\includegraphics[width=0.3\linewidth]{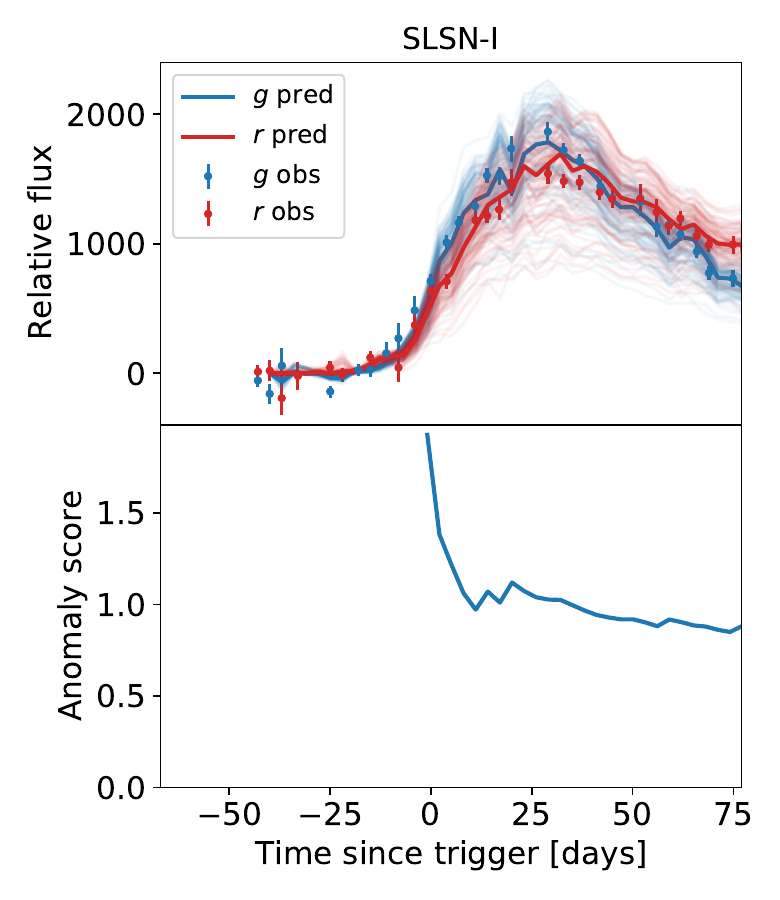}}
        {\includegraphics[width=0.3\linewidth]{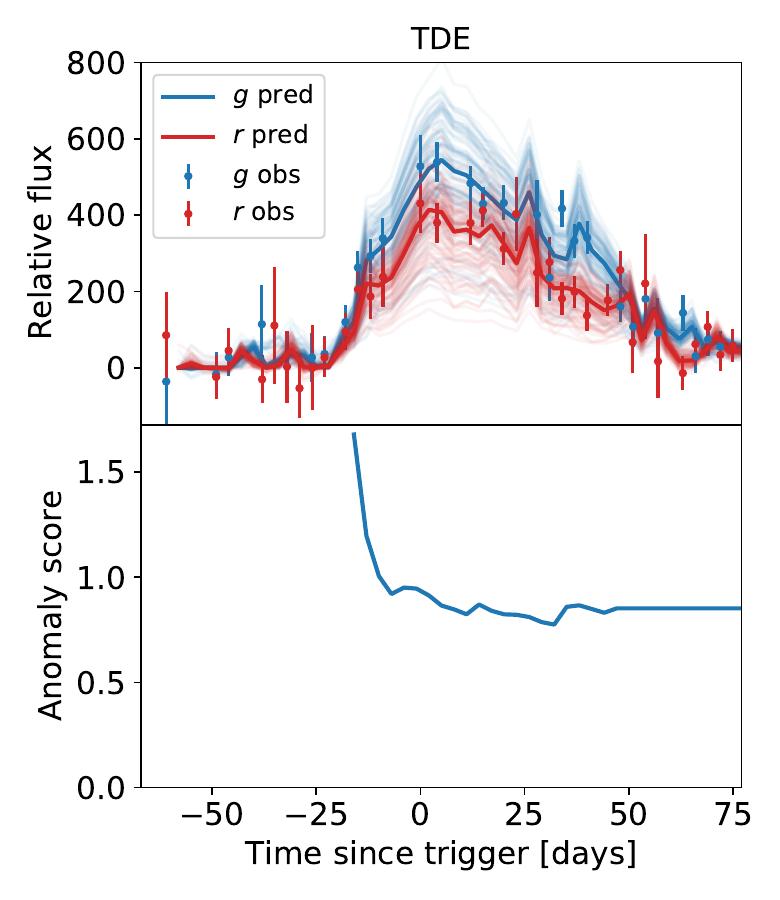}}
    \caption{The DNN being used as a predictive sequence model. Each plot uses a different one of the five trained models and applies it to an example simulated transient from the same class. The trace lines show the posterior predictions and the bold solid line is the median of the posterior predictions. The SNIa plot (first plot) is made up of each of the predictions 3 days after the grey shaded regions in Figure \ref{fig:DNN_generative_plots}. The bottom panels in each plot show the anomaly scores (computed using equation \ref{eq:Anomaly_score}) as a function of time. We expect the anomaly scores to be low since these plots show example objects from the same class the models were trained on.}
        \label{fig:DNN_predictive_plots}
\end{figure*}
\begin{figure*}
    \centering
    \textbf{Bazin}\par\medskip
    \vspace{-0.5em}
        {\includegraphics[width=0.3\linewidth]{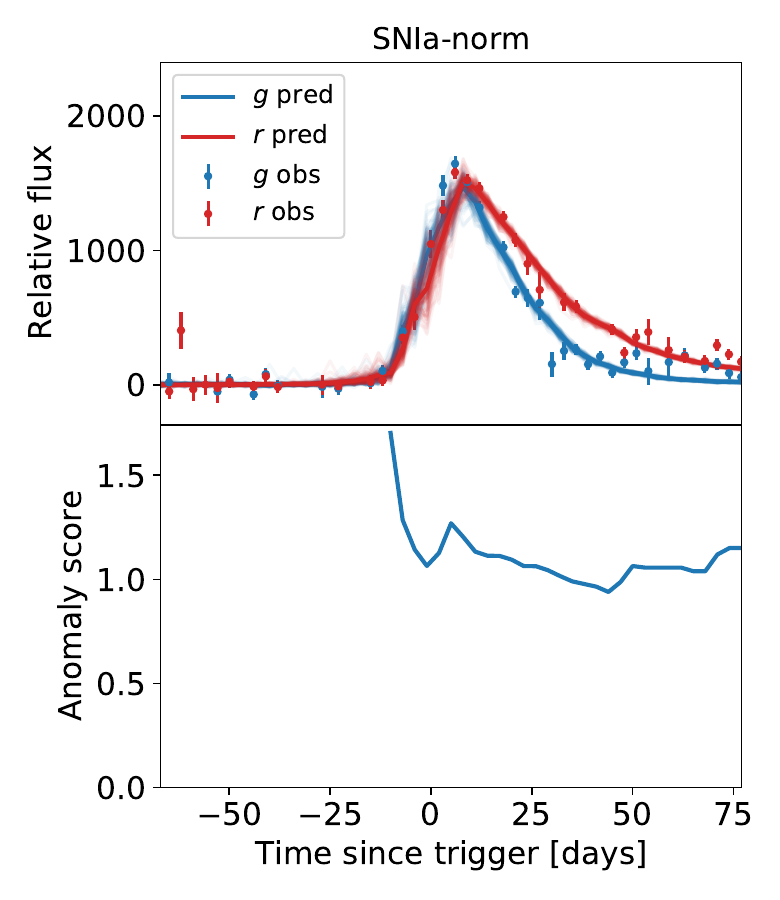}}
        {\includegraphics[width=0.3\linewidth]{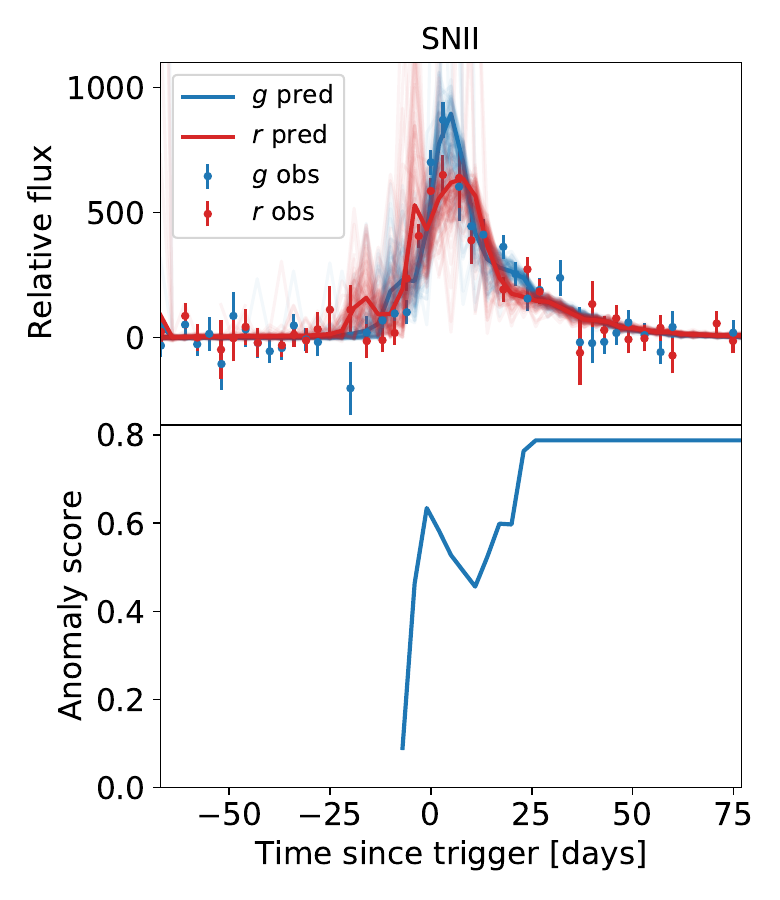}}
        {\includegraphics[width=0.3\linewidth]{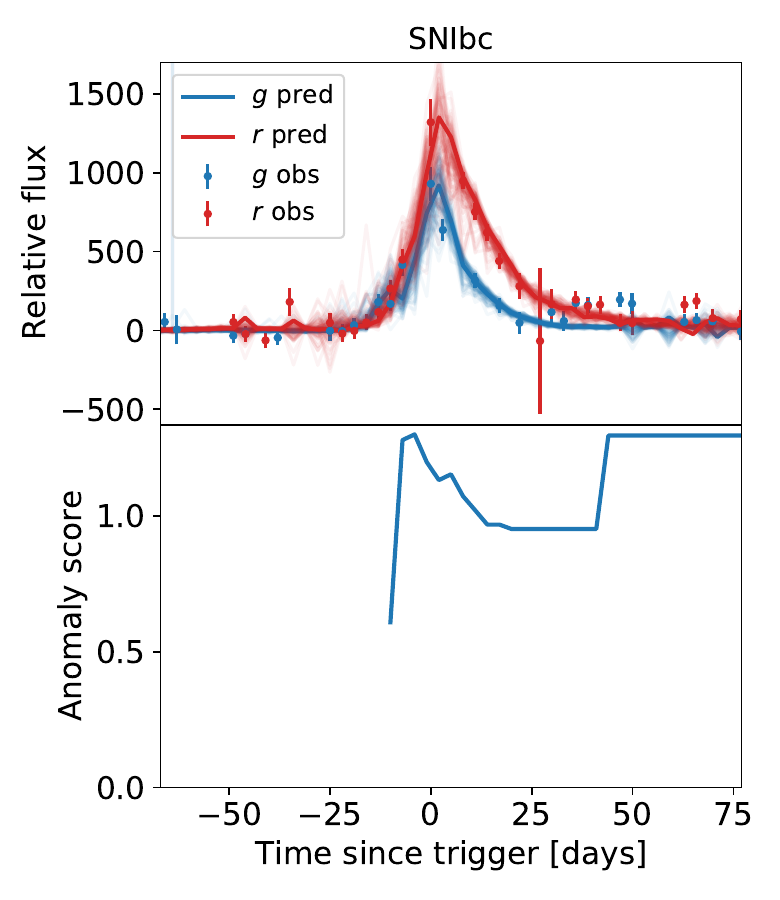}}
        {\includegraphics[width=0.3\linewidth]{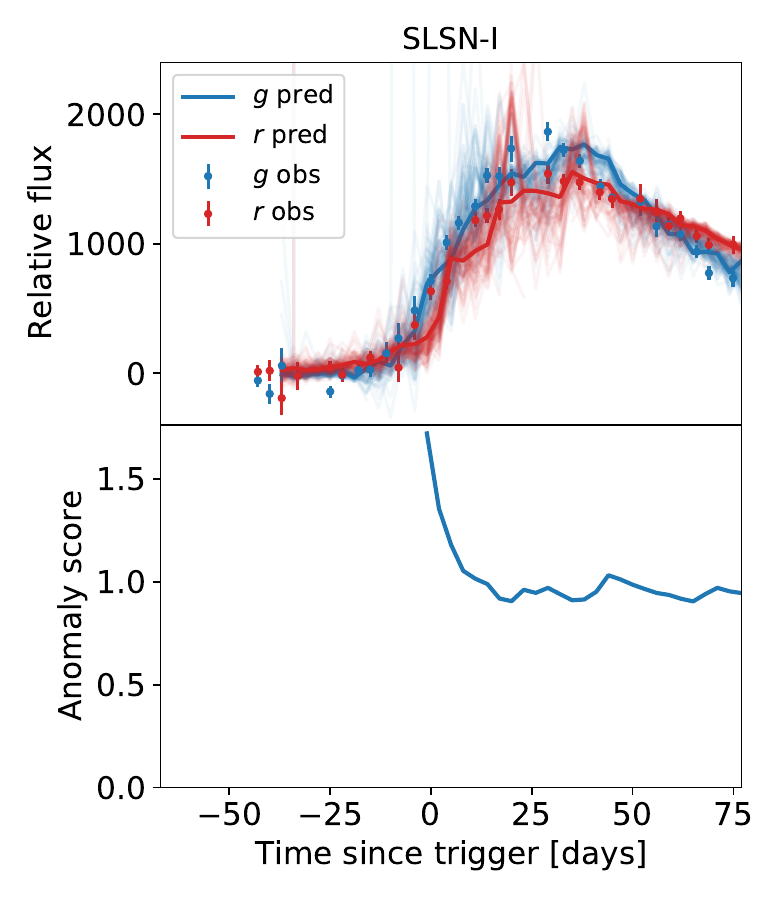}}
        {\includegraphics[width=0.3\linewidth]{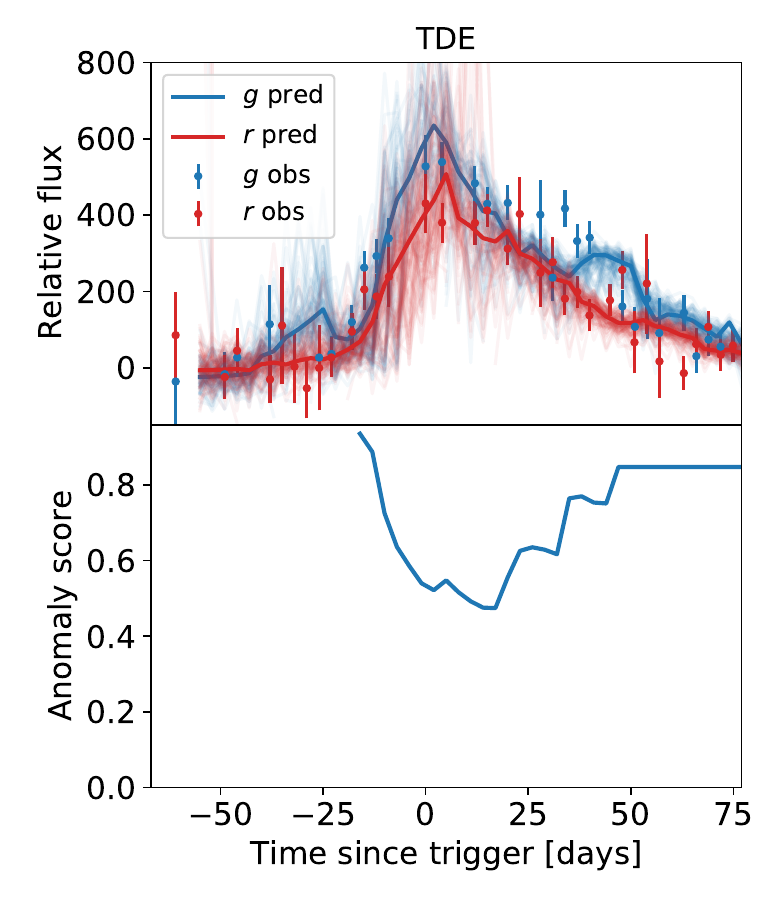}}
    \caption{A sequence of predictions only 3 days in the future of a given partial light curve made using the Bazin function. Each plot uses a different one of the five trained models and applies it to an example simulated transient from the same class. The trace lines show the posterior predictions and the bold solid line is the median of the posterior predictions. The SNIa plot (first plot) is made up of each of the predictions 3 days after the grey shaded regions in Figure \ref{fig:Bazin_generative_plots}. The bottom panels in each plot show the anomaly scores (computed using equation \ref{eq:Anomaly_score}) as a function of time. We expect the anomaly scores to be low since these plots show example objects from the same class the models were trained on.}
    \label{fig:Bazin_predictive_plots}
\end{figure*}

  \begin{figure*}
    \begin{flushleft}
        \hspace{0.33\linewidth} \textbf{DNN} \hspace{0.31\linewidth} \textbf{Bazin}\par\medskip
        \vspace{-1.5em}
    \end{flushleft}
	\centering
	\includegraphics[width=0.35\linewidth]{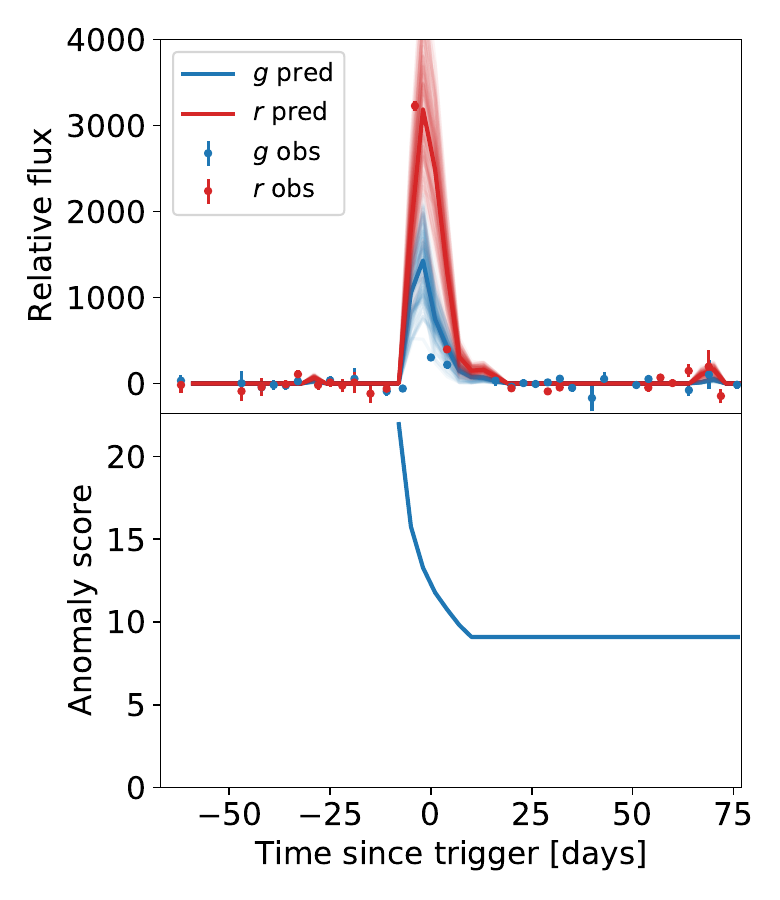}
	\includegraphics[width=0.35\linewidth]{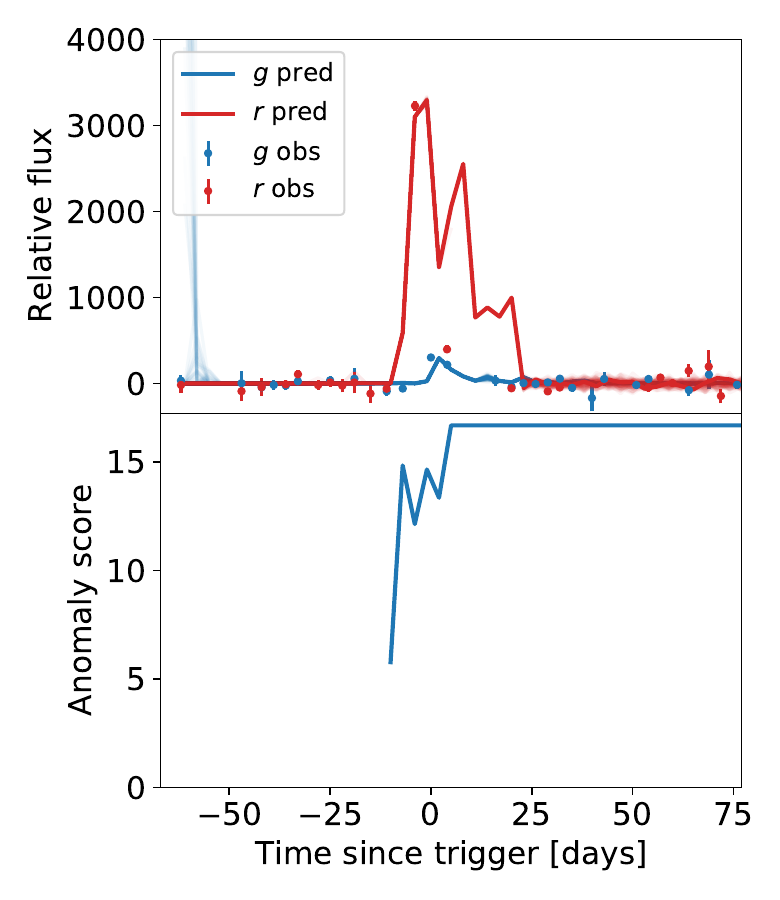}
	\caption{Predictions of an example simulated kilonova modelled with the DNN (left) and Bazin (right) SNIa models.The plots show a sequence of predictions only 3 days in the future of the partial light curves. The trace lines show the posterior predictions and the bold solid line is the median of the posterior predictions. The bottom panels in each plot show the anomaly scores (computed using equation \ref{eq:Anomaly_score}) as a function of time.}
	\label{fig:anomaly_example}
  \end{figure*}

\subsection{Anomaly score definition}
\label{sec:Anomaly_score_definition}
To quantify a potential anomaly, we first define the instantaneous anomaly score as a $\chi^2$ metric to compute the discrepancy between the observed flux at time $t$ and the predictions of a model based on previous data. This $\chi^2$ is weighted by the total variance including the predictive uncertainty and measurement error.
\begin{equation}
    \begin{split}
    \chi^2_{st} = & \frac{1}{N_p} \sum^{N_p}_{p=1} { \frac{\left( y_{spt}-D_{spt} \right)^2}{ c^2 \sigma_{y,{spt}}^2 + \sigma_{D,{spt}}^2} }.
    \end{split}
\label{eq:instantaneous_anomaly_score}
\end{equation}
Next, we define the \textit{Anomaly score}, $\tilde{\chi}$, used throughout this paper, as the square root of the time-averaged $\chi^2$ up to the present time $T$,
\begin{align}
    \tilde{\chi}_{sT} = \sqrt{ \frac{1}{N_\mathrm{avg}} \sum_{\{t \le T : \mathrm{(S/N)}_t > 5 \}} {\chi^2_{st}}}, && N_\mathrm{avg} = \lvert \{ \mathrm{(S/N)_{sT}} > 5 \} \rvert
    \label{eq:Anomaly_score}
\end{align}
where $N_\mathrm{avg}$ is the number of time-steps with signal-to-noise greater than 5 up to the time $T$, and $k$ runs across that index. This metric is effectively the time-averaged reduced $\chi^2$ up to time $T$. After some analysis (see the detailed discussion in Appendix \ref{sec:Appendix_analysis_of_predictive_uncertainty}), we identified that the DNN overestimated the predictive uncertainty. To account for this, we scale the predictive uncertainty with a factor $c=0.2$ for the DNN. We found that the Bazin model's predictive uncertainties were already well-calibrated to actual predictive performance and so use use $c=1$ for the Bazin model. 

This metric is used as our real-time anomaly score. Higher values indicate that the regressive model was less able to predict future data given past data, while lower scores indicate that the model was able to effectively predict future data.

\begin{figure*}
    \begin{flushleft}
        \hspace{0.26\linewidth} \textbf{DNN} \hspace{0.45\linewidth} \textbf{Bazin}\par\medskip
        \vspace{-7em}
    \end{flushleft}
    \centering
    \includegraphics[width=0.495\linewidth]{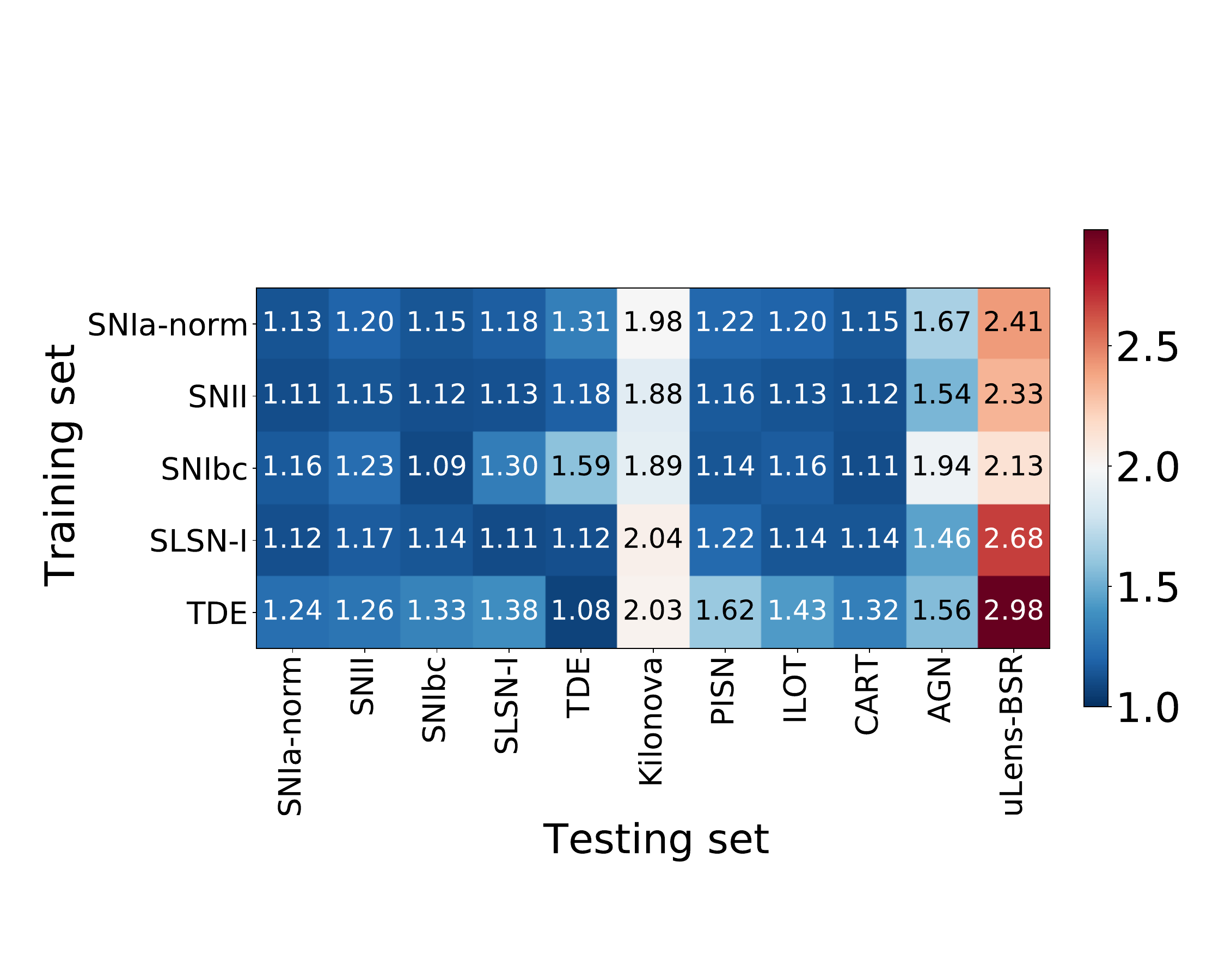}
    \includegraphics[width=0.495\linewidth]{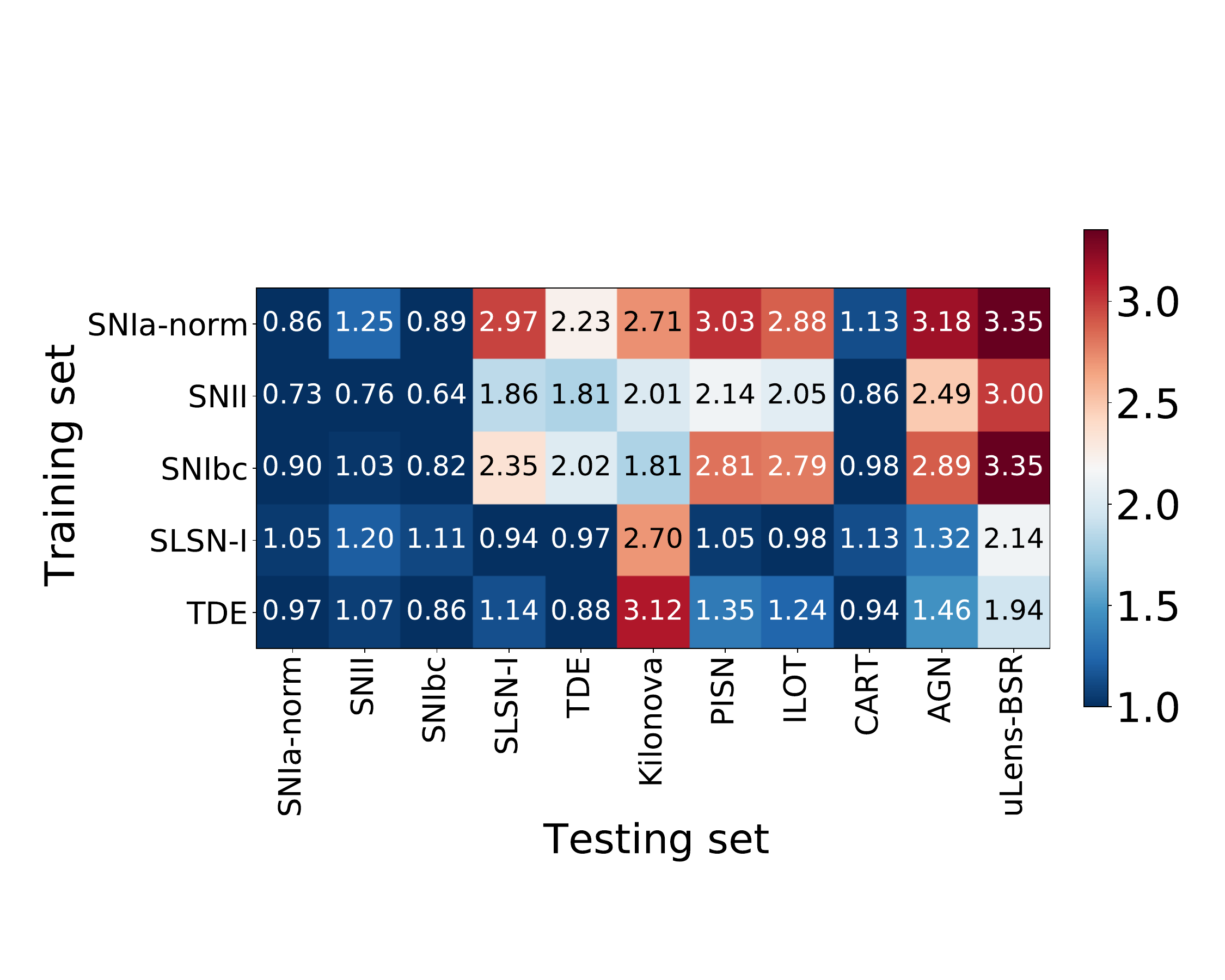}
    \vspace{-3em}
    \caption{The matrix illustrates the similarity of different transient classes, with lower numbers being more similar (less anomalous), and higher numbers being less similar (more anomalous). The vertical axis shows five trained models, and the horizontal axis are transients from a range of test classes. Each transient in our simulated dataset is fit with the five models, and the anomaly score over the full light curve is recorded. The median of the distribution of anomaly scores for each class are the numbers shown.}
    \label{fig:similarity_matrix}
\end{figure*}

\begin{figure*}
    \begin{flushleft}
        \hspace{0.27\linewidth} \textbf{DNN} \hspace{0.45\linewidth} \textbf{Bazin}\par\medskip
        \vspace{-0.1em}
    \end{flushleft}
    \centering
    \includegraphics[width=0.495\linewidth]{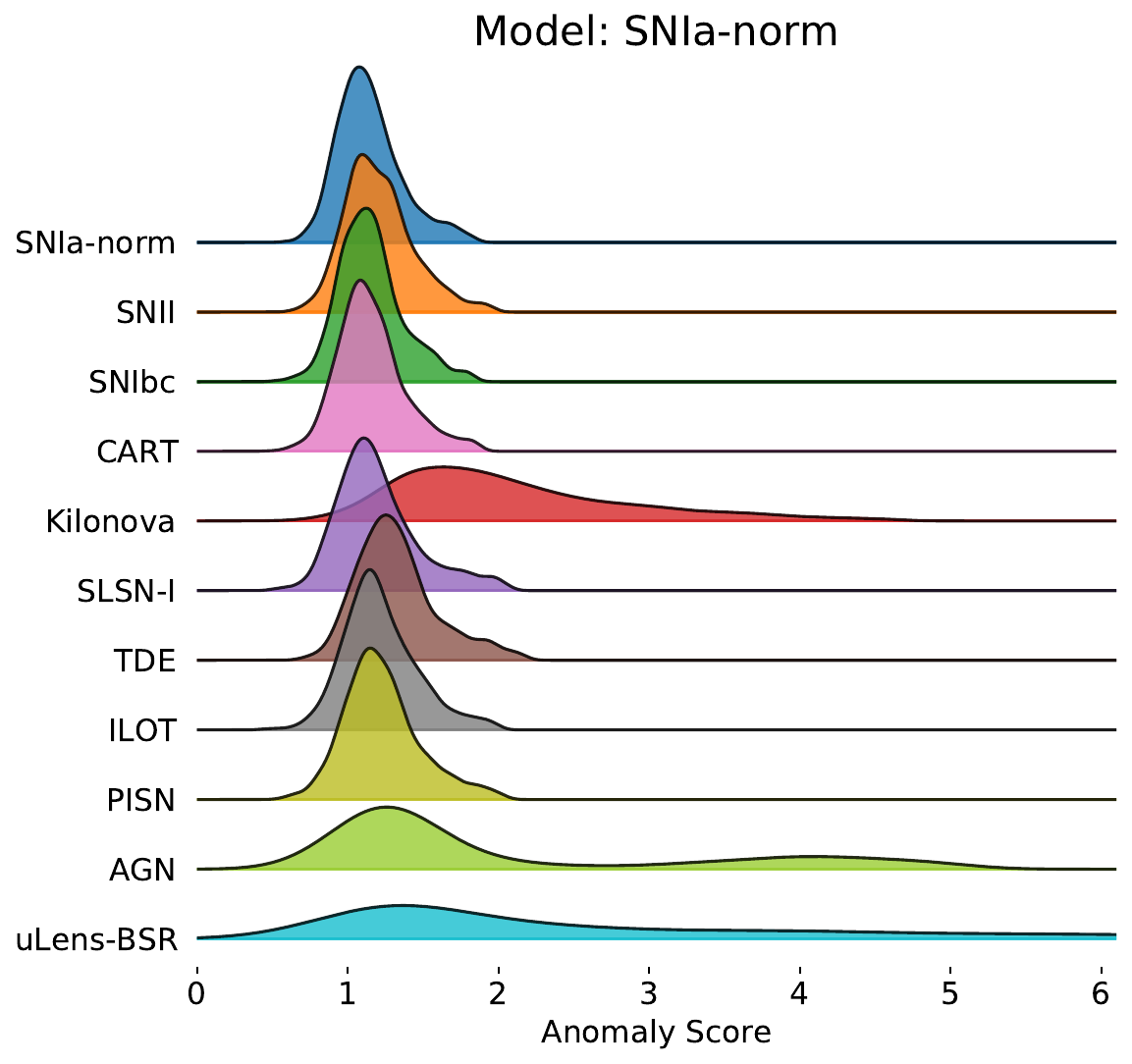}
    \includegraphics[width=0.495\linewidth]{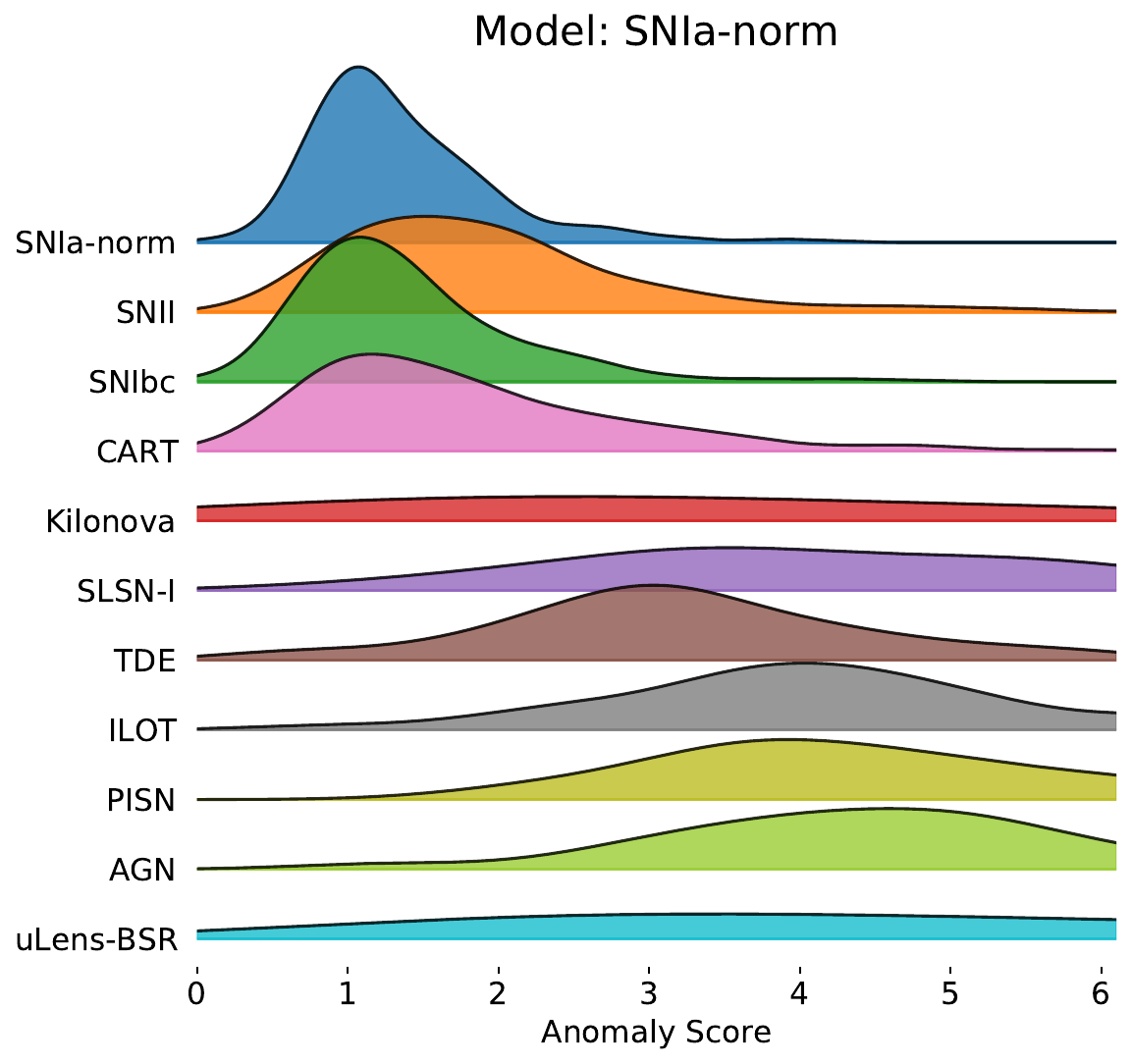}
    \caption{Anomaly score distribution recorded over the full light curve for the SNIa model tested on the simulated transient population of ten different classes. Classes that are dissimilar to SNIa have higher anomaly scores, while similar classes have lower anomaly score distributions. The Bazin plot (right) shows a larger separation of the distributions of the SNIa and anomalous classes than the DNN (left). Similar plots can be made for the other five trained models, but are not shown for brevity.}
    \label{fig:Anomaly_score_distribution}
\end{figure*}

\begin{figure*}
    \centering
    \textbf{DNN}\par\medskip
    \includegraphics[width=0.43\linewidth]{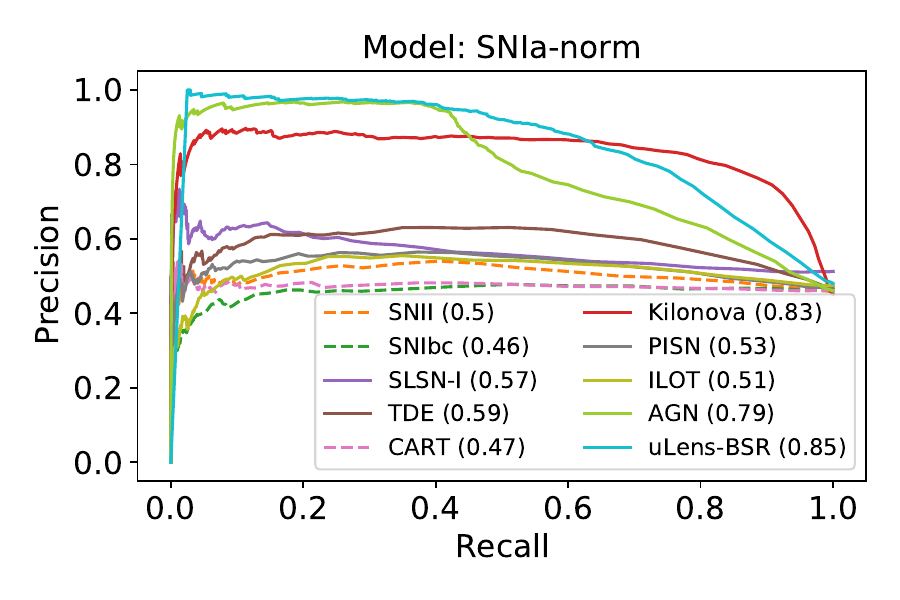}
    \includegraphics[width=0.43\linewidth]{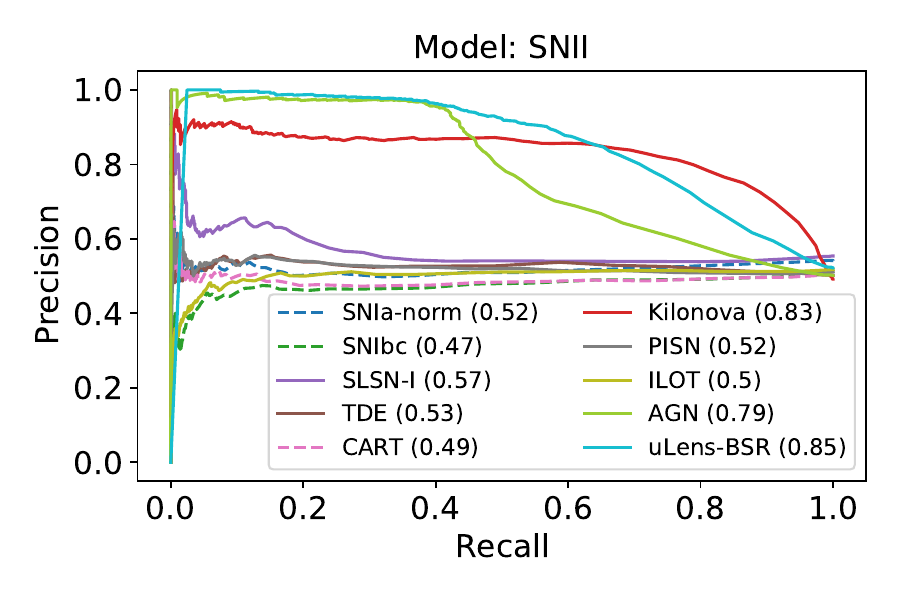}
    \includegraphics[width=0.43\linewidth]{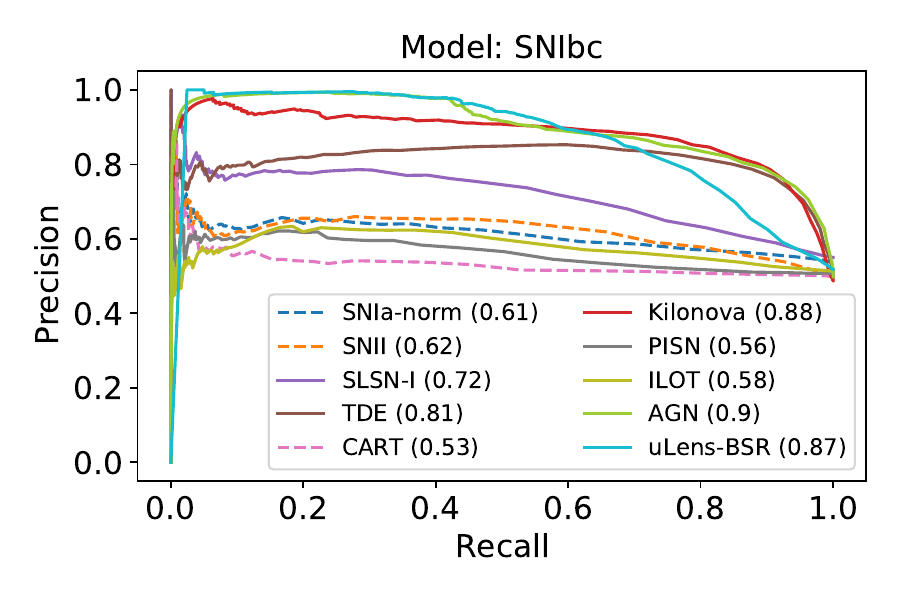}
    \includegraphics[width=0.43\linewidth]{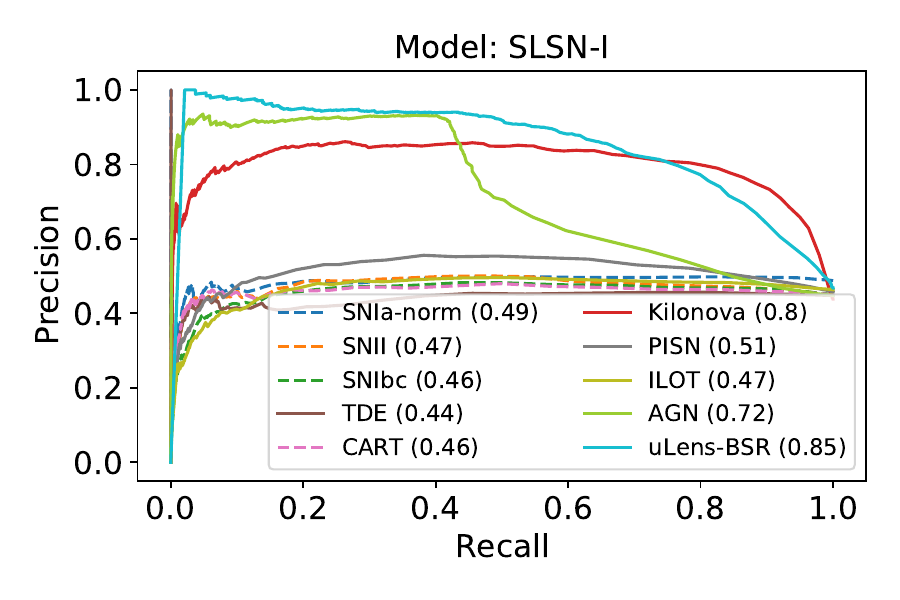}
    \includegraphics[width=0.43\linewidth]{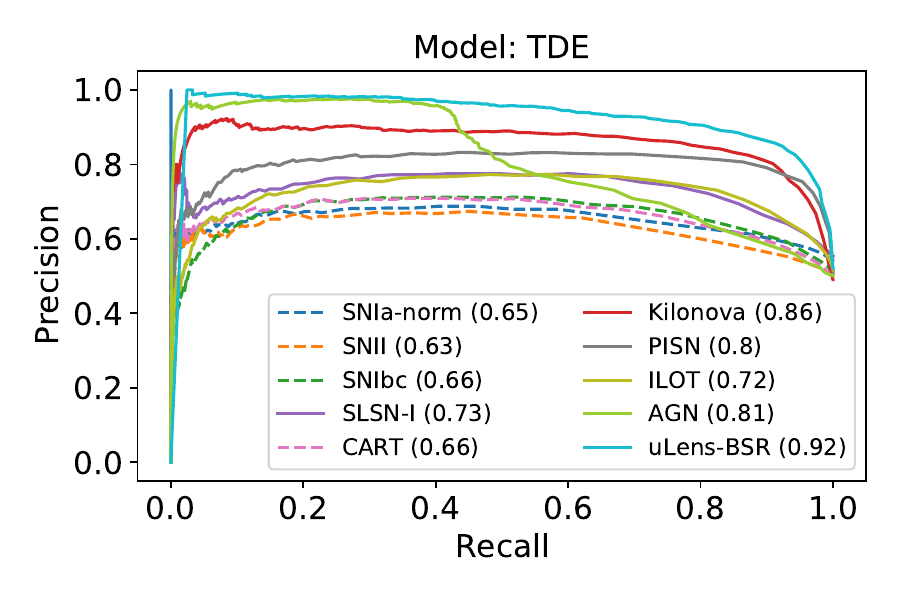}
    \caption{Precision-Recall (PR) curves for each trained DNN model against ten other transient classes. In each subfigure, we use the Model class as the reference class and the anomalous classes as the ones denoted in the legend. The solid lines are used to highlight the classes that are most different to common supernovae, and we expect to be be identified as anomalous. The dashed lines indicate the classes that often look similar to common supernovae. The area under the precision-recall curves (AUCPR) are shown in the brackets in the legends. The plots are made by plotting the precision against the recall for a range of different threshold anomaly scores. We use the anomaly scores over the full light curves of all transients in the simulated testing set to make these plots.}
    \label{fig:DNN_PR_curves}
\end{figure*}

\begin{figure*}
    \centering
    \textbf{Bazin}\par\medskip
    \includegraphics[width=0.43\linewidth]{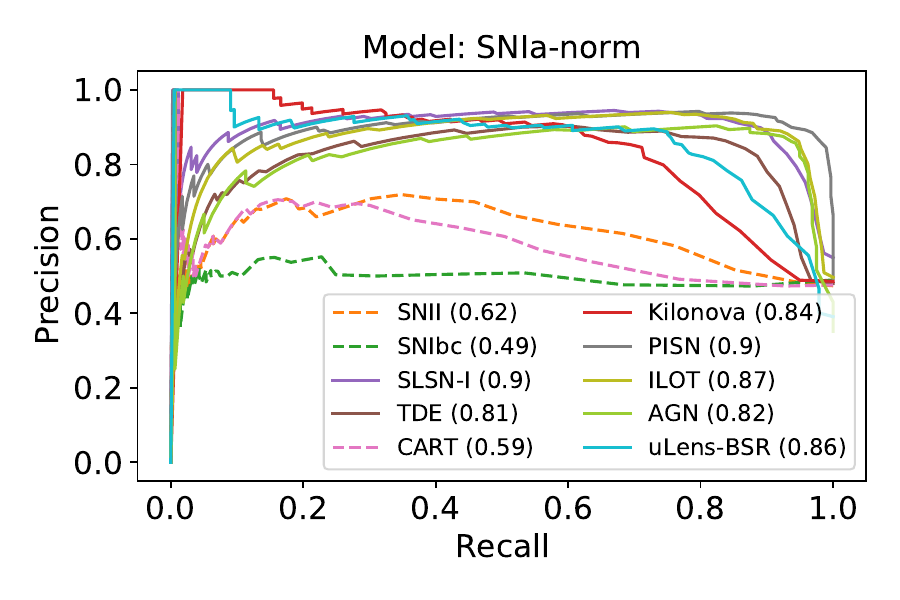}
    \includegraphics[width=0.43\linewidth]{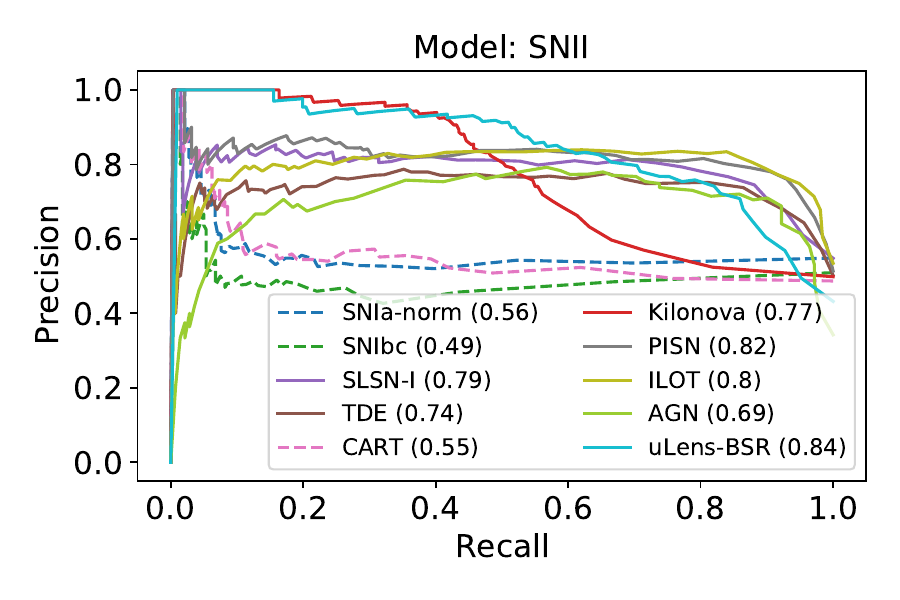}
    \includegraphics[width=0.43\linewidth]{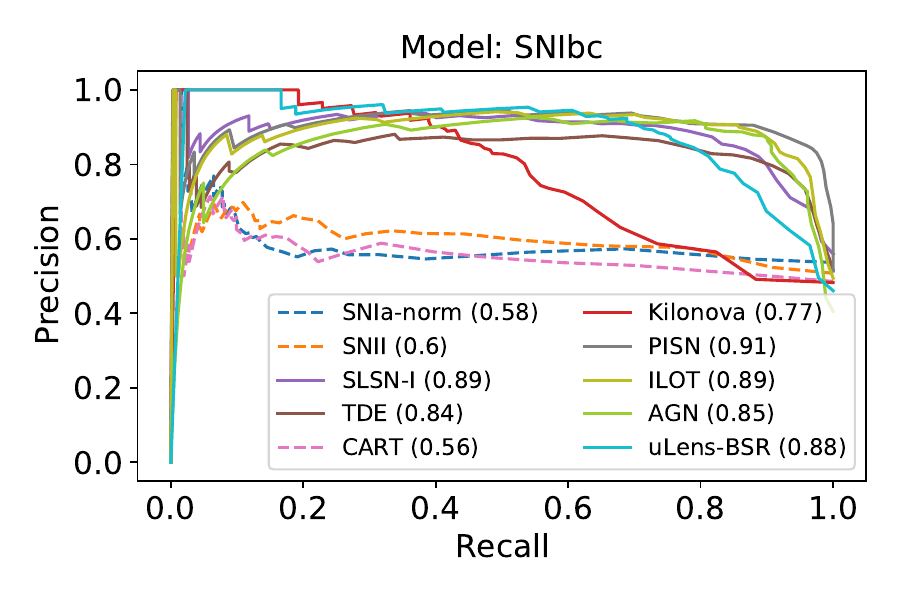}
    \includegraphics[width=0.43\linewidth]{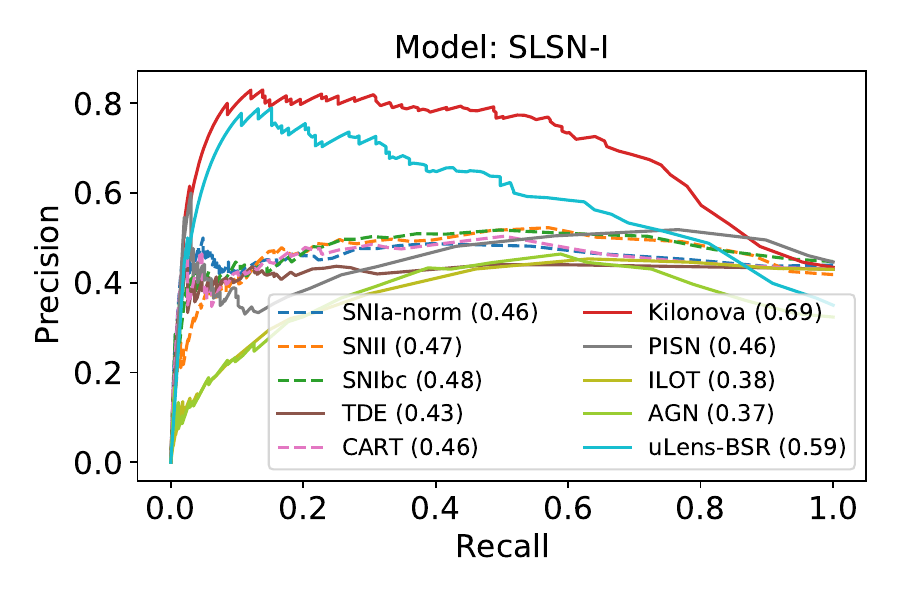}
    \includegraphics[width=0.43\linewidth]{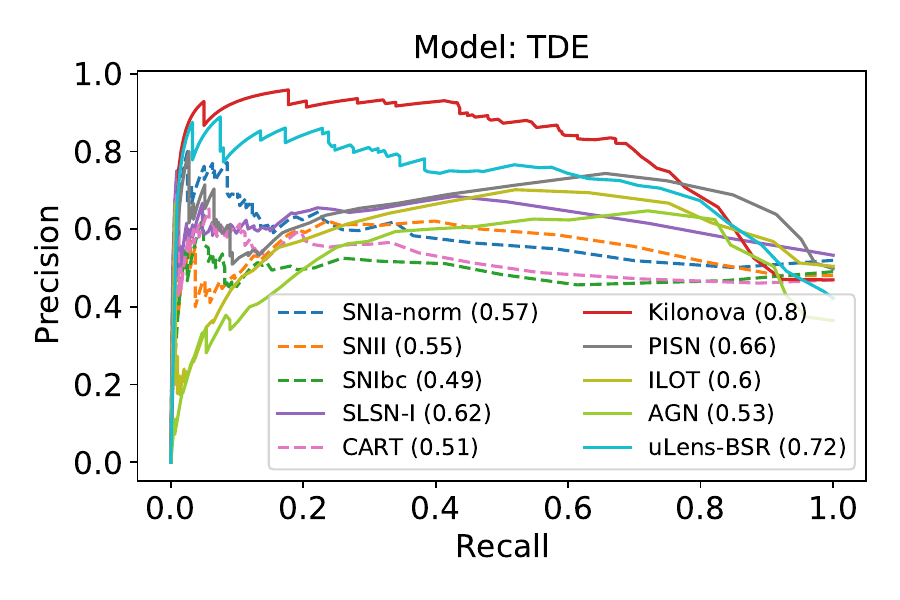}
    \caption{Precision-recall curve for each Bazin model against ten other transient classes. In each subfigure, we use the Model class as the reference class and the anomalous classes as the ones denoted in the legend. The solid lines are used to highlight the classes that are most different to common supernovae, and we expect to be be identified as anomalous. The dashed lines indicate the classes that often look similar to common supernovae. The area under the precision-recall curves (AUCPR) are shown in the brackets in the legends. The plots are made by plotting the precision against the recall for a range of different threshold anomaly scores. We use the anomaly scores over the full light curves of all transients in the simulated testing set to make these plots.}
    \label{fig:Bazin_PR_curves}
\end{figure*}

\begin{figure*}
    \begin{flushleft}
        \hspace{0.25\linewidth} \textbf{DNN} \hspace{0.45\linewidth} \textbf{Bazin}\par\medskip
        \vspace{-1.5em}
    \end{flushleft}
    \centering
    \includegraphics[width=0.49\linewidth]{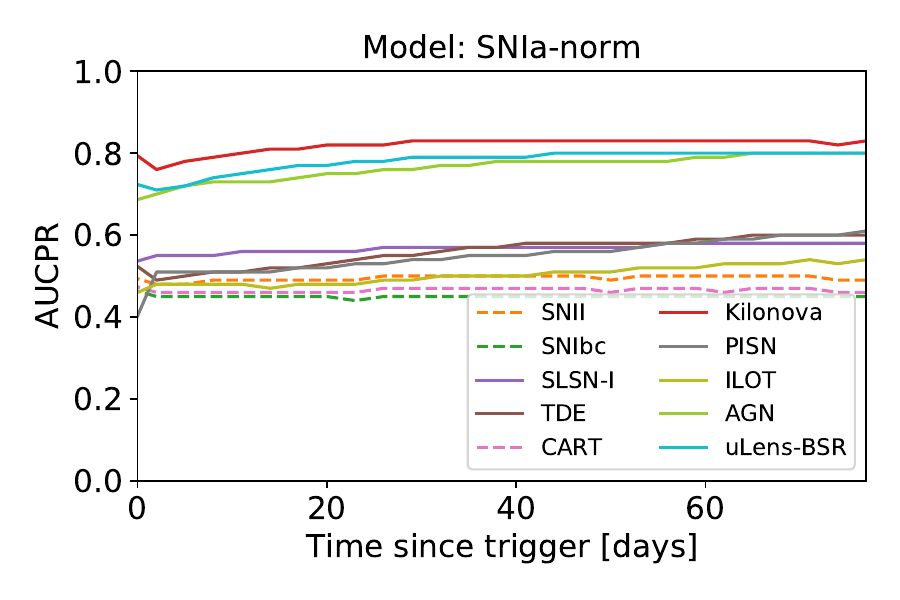}
    \includegraphics[width=0.49\linewidth]{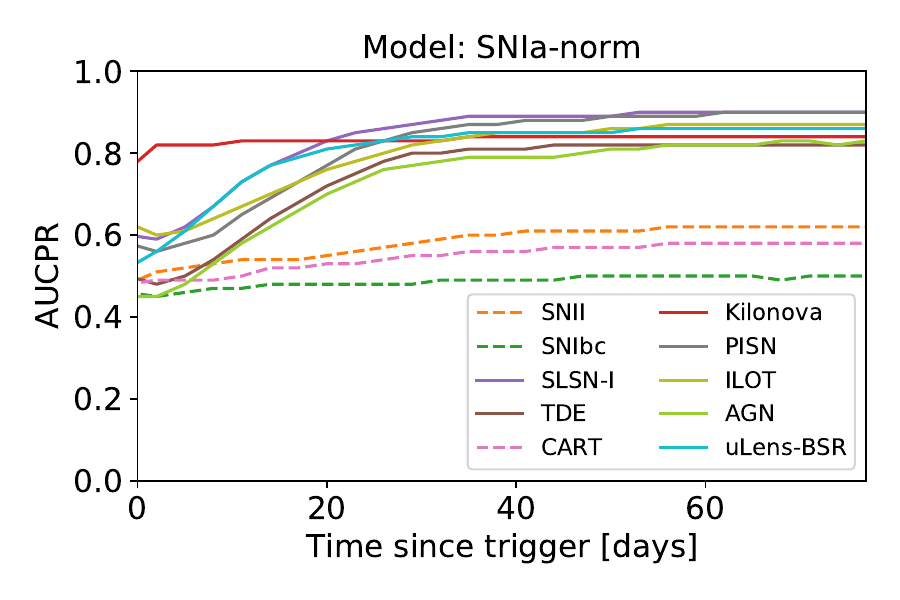}
    \caption{Area under the precision-recall curves (AUCPR) vs time since trigger assuming the SNIa model as the reference class and the anomalous classes denoted in the legend. These are made by reproducing the SNIa-norm precision-recall curves in Figures \ref{fig:DNN_PR_curves} and \ref{fig:Bazin_PR_curves} at all time steps since trigger (instead of only over the full light curve) and recording the AUCPRs. The solid lines are used to highlight the classes that are most different to common supernovae, and we expect to be be identified as anomalous. The dashed lines indicate the classes that often look similar to common supernovae. The plots make use of the simulated transient dataset.}
    \label{fig:AUCPR_vs_time}
\end{figure*}

\section{Results}
\label{sec:Results}
In this section, we explore the performance of the DNN and the Bazin parametric approach at predicting light curves and identifying anomalies on a simulated ZTF dataset. For the DNN, we trained five autoregressive models one for each transient class: SNIa, SNII, SNIbc, SLSN, TDE. And similarly, for the Bazin function, we defined a prior distribution for each class. Each training set consisted of $\sim$8000 light curves and we tested the performance of the models on $\sim$2000 light curves from each transient class.

\subsection{Generating light curves}
The DNN was designed to predict one time-step in the future given a light curve up to a specified time. On the other hand, the Bazin model fits an entire past light curve and can make predictions for any set of future times. However, the primary purpose of this paper is to predict just one time-step in the future to compare that prediction to observed fluxes (and hence evaluate an anomaly score). Before delving into the anomaly detection results, we first show the power of these two methods at building a generative model of a transient class. 

In Figure \ref{fig:Bazin_generative_plots}, we illustrate the use of our Bazin parametric approach as a generative model of an example SNIa. Each panel fits only a partial light curve (shown in the grey region) and generates the rest of the light curve from this information. In the first panel, where no observations are being used, we are effectively plotting the prior distribution. As more observations are included in the fit, predictions improve, and once the peak of the light curve has been observed the predictions are much more accurate. 

In Figure \ref{fig:DNN_generative_plots}, we show the power of our DNN as a generative model of a full light curve to compare it against the parametric approach. As the DNN was specifically designed to only predict one time-step (3 days) in the future and was not designed to generate an entire light curve, we can only obtain a sequence of predictions by feeding in the predicted values back into the DNN and iteratively predicting each consecutive time-step in the light curve. While the predictions in the first few time-steps are accurate, the small inaccuracies quickly compound before the predictions reduce down to the zero-flux background prediction. We emphasise that our architecture is not suited to generating full light curves, and that an autoencoder (specifically designed to fit an entire light curve) would perform much better at this task. Nonetheless, it is interesting to see how the network's predictions evolve over time.

These two plots (Figures \ref{fig:Bazin_generative_plots} and \ref{fig:DNN_generative_plots}) illustrate how the two models differ in their approach to the problem - one being a fitting function regressing fluxes over time and the other being a predicting algorithm regressing future fluxes over past data. The Bazin model which was designed as a generative model of a light curve, obviously produces much more realistic light curves than the DNN. Forcing the DNN to generate a full light curve by iteratively inputting predicted values back into the model produces poor predictions.

In the rest of the plots in this paper, we only use the predictions one time-step (3 days) in the future of a partial light curve. We emphasise that a key difference between Figures \ref{fig:Bazin_generative_plots} and \ref{fig:DNN_generative_plots} and Figures \ref{fig:Bazin_predictive_plots} and \ref{fig:DNN_predictive_plots}, is that the former two figures generate an entire light curve given some partial light curve, while the latter figures iteratively predict only the next time-step given a partial light curve. Hence, to make the first panels of Figures \ref{fig:Bazin_predictive_plots} and \ref{fig:DNN_predictive_plots}, we use all the subplots of Figures \ref{fig:Bazin_generative_plots} and \ref{fig:DNN_generative_plots}, respectively, by recording the predictions one time-step after each panel's grey shaded region.

\subsection{Using model predictions to identify anomalies}
Following the methods outlined in \S\ref{sec:Models}, we trained a separate DNN and defined a separate Bazin prior for each of the five transient classes. In Figures \ref{fig:DNN_predictive_plots} and \ref{fig:Bazin_predictive_plots}, we illustrate the performance of these five models on example light curves in the testing sets of each class. The top panel of each subfigure shows the example light curve with uncertainties and posterior draws of the predictions, with the bold line showing the median of these draws. Each prediction is causal and hence only uses data from the previous time-steps. The bottom panel of each subfigure plots the anomaly score defined in equation \ref{eq:Anomaly_score}. In most figures the final anomaly score is close to $\chi=1$ indicating that the models are effective at predicting future transients fluxes from their own class. 

The DNN and Bazin model's different approach to the regression problem causes slight differences in the anomaly score plots. The Bazin model regresses fluxes over time to learn the shape of each light curve as a function of time since trigger; while the DNN, on the other hand, regresses future fluxes over past data in a light curve and does not learn anything about time since trigger. This causes the DNN plots to have larger anomaly scores near the explosion time of each transient because the DNN's expectation of observing background flux is abruptly disrupted by the transient phase of an event. Furthermore, the Bazin model tends to have prediction light curves that are less smooth than the DNN, with some posterior samples having large deviations from the mean. This behaviour is because the Bazin model plots are produced from several independent fits to partial light curves. 

In Figure \ref{fig:anomaly_example}, we illustrate an example simulated kilonova with observations predicted using the SNIa models. The poor predictions and high anomaly scores indicate that this transient is flagged as anomalous with respect to the SNIa model - showing a first-order success in our method. Most kilonovae in the dataset were similarly flagged as highly anomalous at a similar epoch. The short timescale of the simulated kilonovae, which typically rise and fall within several days, requires much lower values for  $\tau_{\mathrm{fall}}$ and $\tau_{\mathrm{rise}}$ than the SNIa prior distributions (in Figure \ref{fig:Bazin_parameter_distribution}) permit.  Thus, the SNIa Bazin model struggles to model kilonova light curves, and in Figure \ref{fig:anomaly_example} it is dominated by this prior whereby the predicted light curves have a much larger $\tau_{\mathrm{fall}}$ than the data suggests, which causes a very high anomaly score. The DNN model appears to be more flexible, and is better at predicting the example kilonova light curve, but still has a very high anomaly score.

To compare the anomaly scores of all transients in our data set against our trained models, we have modelled every transient in the testing sets with each of the regressive models. Since we obtain anomaly scores as a function of time, we record the anomaly score of each transient over a full light curve, and report the median of all scores in Figure \ref{fig:similarity_matrix}. 

The plot highlights the similarity of each trained class to every other class, and acts as a similarity matrix for the shown transient classes. Higher numbers indicate classes that are more dissimilar, and lower number indicate classes that are more similar. For each trained model, the lowest number generally corresponds to the same class, which confirms effective training of our models, showing that each model can predict the future fluxes of transient light curves from its own class well.  The model trained on SNe Ia also has low scores for core-collapse SNe (SNIbc, SNII) and CARTs which highlights their similarity to SNe Ia. Kilonovae stand out as very anomalous for every trained model, indicating that the short lifetime and low luminosity of these classes cannot be well-predicted with the trained models. The Bazin matrix shows starker differences between the model class and the other testing classes, hinting that it may be better at identifying anomalies than the DNN. Overall, Figure \ref{fig:similarity_matrix} highlights some interesting similarities between transient classes, and confirms what may already be known about their general behaviour. It highlights the overall performance of our method on the testing sets, and shows that we are able to identify anomalous classes with this method.

However, Figure \ref{fig:similarity_matrix} only represents the median of the anomaly scores across the testing sets. In Figure \ref{fig:Anomaly_score_distribution}, we plot the histograms of the full light curve anomaly scores of the SNIa model predicting the light curves from ten different testing classes. The plot shows that the DNN SNIa model cannot easily differentiate classes, other than the kilonova, uLens-BSR, and AGN classes. The Bazin plot, on the other hand, shows that the anomalous classes (kilonova, SLSN, TDE, ILOT, PISN, AGN, and uLens-BSR) all have histograms that separate well from the SNIa class, while the classes that are known to look similar to SNIa (SNII, SNIbc, and CARTs) do not separate as well, consistent with expectations.

We refer to anomalies as all classes that are not from the reference class. To identify anomalies, a threshold anomaly score would need to be chosen such that the reference class is not often flagged as anomalous but all other transient classes are flagged as anomalous. This threshold score would need to be chosen to have a high \textit{precision} (also known as purity) and a high \textit{recall} (also known as completeness) of anomalies. That is, we would choose a threshold anomaly score that correctly identifies most of the transients from non-reference classes as anomalous, while not identifying many of the transients from the reference class as anomalous. The precision-recall metric is a good way to measure this trade-off. It is also a particularly good metric for imbalanced datasets \citep{Saito2015PRvsROC} which are inherent to anomaly detection.

The true positive rate (TPR) and false positive rate (FPR) are also commonly used metrics in machine learning applications. However, they are only useful for balanced datasets when the relative rate of the positive and negative classes in the test set are similar. Anomalies by definition are rare, and in such cases, the precision is a much more appropriate metric. Precision is a measure of how pure our anomaly predictions are, and recall is a measure of how many anomalies we can expect to find. Precision and recall together provide a well-rounded metric of the performance of anomaly detection methods.

For the purposes of explanation in this paragraph, we use a SNIa as an example of a reference class transient and a kilonova as an example of a non-reference class transient. We define a True Anomaly (TA) as a transient from a non-reference class that was predicted as being an anomaly (e.g. correctly identifying a kilonova as anomalous), a False Anomaly (FA) as a transient from the reference class that was predicted as being an anomaly (e.g. incorrectly identifying a SNIa as anomalous), a True Not Anomaly (TNA) as a transient from the reference class that was predicted as being non-anomalous (e.g. correctly identifying a SNIa as non-anomalous), and a False Not Anomaly (FNA) as a transient from the non-reference class that was predicted as being non-anomalous (e.g. incorrectly identifying a kilonova as non-anomalous). These definitions are summarised in Table \ref{tab:True_anomaly_table}.

\begin{table}
    \centering
    \begin{tabular}{c|c|c|}
          & Actual Anomaly & Actual Not Anomaly \\ \hline
         Predict Anomaly & TA & FA \\\hline 
         Predict Not Anomaly & FNA & TNA \\ \hline 
    \end{tabular}
    \caption{Definitions of True Anomalies (TA), False Anomalies (FA), False Not Anomalies (FNA), and True Not Anomalies (TNA).}
    \label{tab:True_anomaly_table}
\end{table}

The \textit{precision} is the fraction of predicted anomalies that are indeed actual anomalies and the \textit{recall} is the fraction of actual anomalies that were correctly predicted as anomalies, as follows,
\begin{equation}
    \mathrm{precision} = \frac{\mathrm{TA}}{\mathrm{TA}+\mathrm{FA}}.
\end{equation}
\begin{equation}
    \mathrm{recall} = \frac{\mathrm{TA}}{\mathrm{TA}+\mathrm{FNA}},
\end{equation}

We construct precision-recall curves to evaluate the performance of our different models. Each point on a precision-recall curve corresponds to a different threshold anomaly score. A good model will have both a high precision and high recall, and hence the area under precision-recall curve will be close to one.

We plot the precision-recall curves for each of the five trained models compared to each other class in Figures \ref{fig:DNN_PR_curves} and \ref{fig:Bazin_PR_curves}. The first subfigure of Figure \ref{fig:Bazin_PR_curves} illustrates that the Bazin model of a SNIa is very effective at identifying all anomalous classes assuming SNIa as the reference class with a high precision and recall (except for core-collapse SNe and CARTs which are known to look broadly similar to SNe Ia). The area under the curves of these anomalous classes are above 0.8. Choosing a threshold score along these curves near the inflection on the top right point such that the precision and the recall is high will be a good choice for identifying anomalies with respect to SNe Ia. The performance of the Bazin SNIbc model is similar and the Bazin SNII model is only slightly worse, but similarly predicts most classes except for common SN types as anomalous. The other Bazin models (SLSN and TDE models) are much poorer at identifying anomalies (assuming the model as the reference class). Given that SNIa, SNII, and SNIbc are the most common types, developing an algorithm that identifies all classes that are not these common types would be an effective anomaly detection algorithm for most astronomers.

The plots so far show that our method is able to identify anomalies relative to \textit{common} SN classes when using full light curves. However, what is often more important for large scale surveys is identifying anomalies in real-time so that we can prioritise which transients should receive follow-up observations. We have made similar precision-recall curves to figures \ref{fig:DNN_PR_curves} and \ref{fig:Bazin_PR_curves} for every time-step since trigger instead of over the full light curve. We summarise these for the SNIa models as a plot of the area under the precision-recall curve (AUCPR) of each class as a function of time since trigger in Figure \ref{fig:AUCPR_vs_time}. The AUCPR increases with time since trigger and plateaus around 25 days since trigger, which is often close to the end of the transient phase of most SNe Ia. The Bazin model clearly performs much better than the DNN model of SNIa at identifying anomalies (except for kilonovae) at all times, where we assume SNIa as the reference class. 

We have decisively shown that the Bazin models of the common SN classes are significantly better at identifying anomalies than the DNN models. In Appendix \ref{sec:Appendix_DNN_vs_Bazin}, we highlight that the poor performance of the DNN compared to the Bazin model is because it is too flexible at predicting light curves; and after being trained on one class, it is still able to accurately predict fluxes in a different class of transients. The DNN model is actually better at predicting the future fluxes of transients within a trained class, but is also able to predict the future fluxes of transients from different classes well. While this flexibility allows for good flux predictions, it is not good for anomaly detection. Future work should look at developing a better DNN model that penalises flux predictions from anomalous transients while rewarding flux predictions from the trained class. The remaining plots in this section and \S\ref{sec:application_to_real_ztf_data} use the Bazin framework instead of the DNN.

\begin{figure*}
    \centering
    \includegraphics[width=1.0\linewidth]{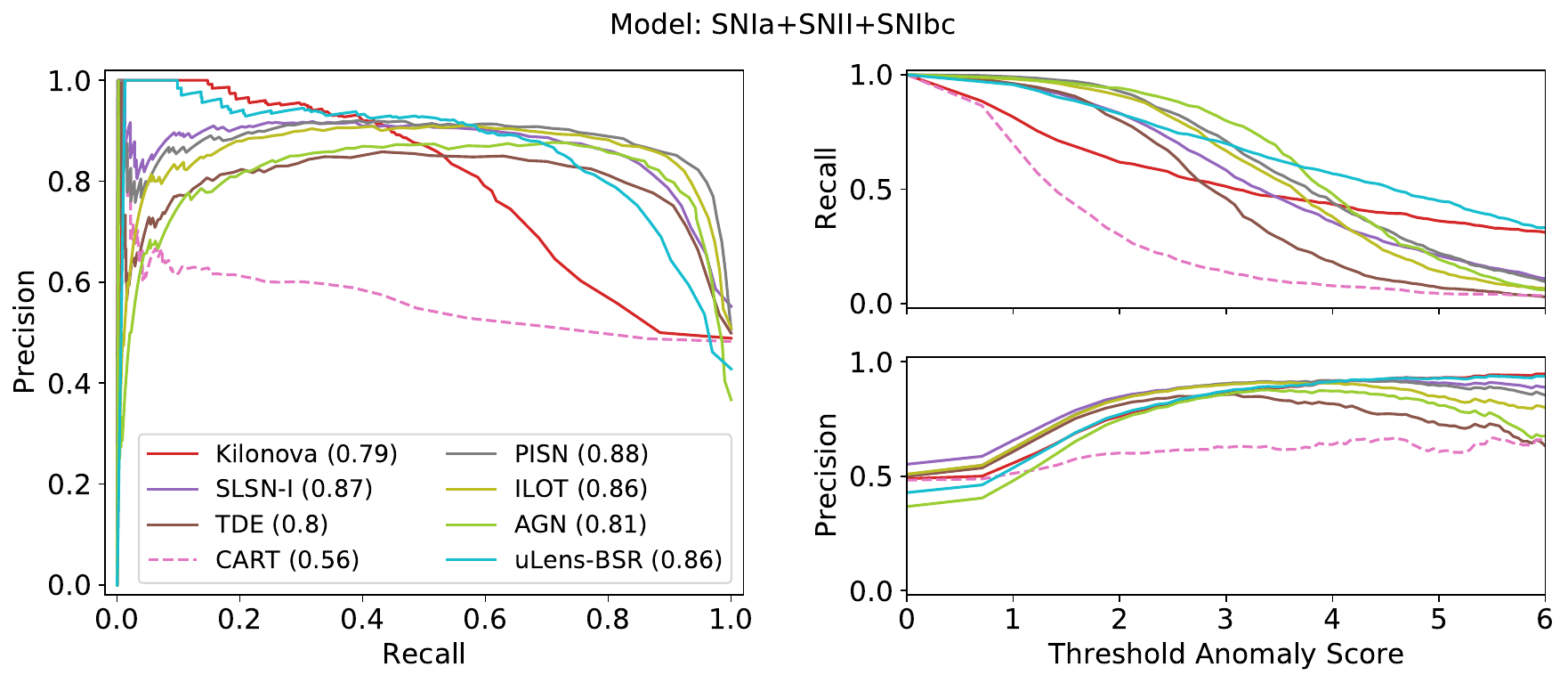}
    \caption{The precision and recall curves plotted at different threshold anomaly scores, assuming the combination of the Bazin SNIa, SNII, and SNIbc models are the reference classes and the anomalous classes are denoted in the legend.  The recall and precision are plotted against threshold anomaly scores in the right panels and the precision-recall curve is plotted at different thresholds in the left panel. The area under the precision-recall curves are shown in brackets in the legend.  We use the anomaly scores over the full light curves of all transients in the simulated testing set to make these precision-recall curves.}
    \label{fig:PR_curve_SNIaSNIISNIbc}
\end{figure*}

\begin{figure}
    \centering
    \includegraphics[width=1.0\linewidth]{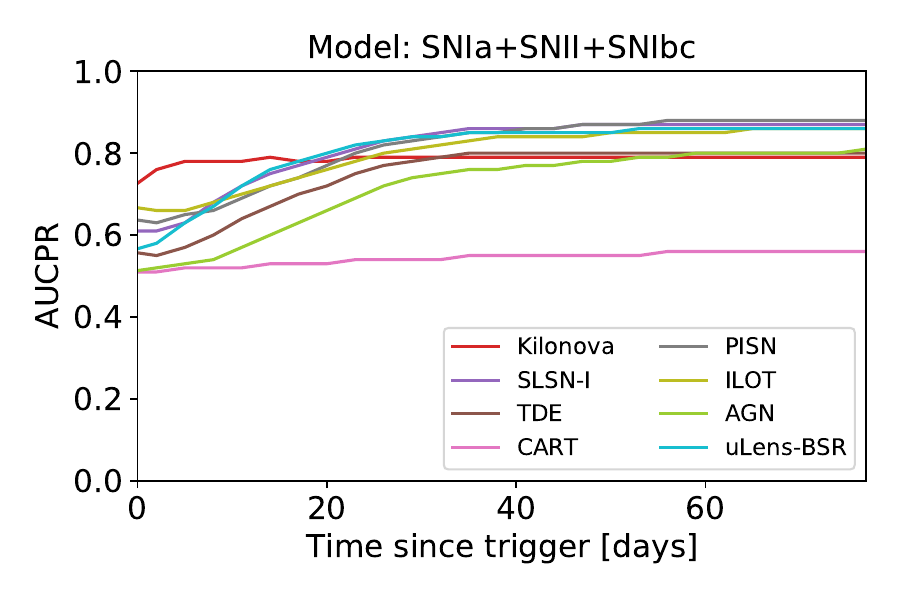}
    \caption{The area under the precision-recall curve (AUCPR) vs time since trigger assuming the combination of the Bazin SNIa, SNII and SNIbc models are the reference classes and the anomalous classes are denoted in the legend. These are made by reproducing the precision-recall curves in Figure \ref{fig:PR_curve_SNIaSNIISNIbc} at all time steps since trigger and recording the AUCPRs. The plot makes use of the simulated transient dataset.}
    \label{fig:AUCPR_curve_SNIaSNIISNIbc}
\end{figure}

\subsection{Identifying anomalies against common classes}
With this many models of each transient, the question of which transient model should ideally be used to identify anomalies remains. We have developed a framework for identifying anomalies with respect to particular model classes. However, often astronomers are interested in identifying anomalies with respect to common classes. For this case, we suggest a simple addition of the anomaly scores from the models that would be considered ``common''. For example, a good choice, might be to define the total anomaly score as the mean of the common SNIa, SNII, and SNIbc classes, as follows,
\begin{equation}
    \chi_{\mathrm{total}} = \frac{1}{3} \left( \chi_{\mathrm{SNIa}} + \chi_{\mathrm{SNII}} + \chi_{\mathrm{SNIbc}} \right).
\end{equation}
This enables us to identify anomalies with respect to all three of these common classes. 

We combine the results of the Bazin models of the SNIa, SNII, and SNIbc classes and plot the resultant precision-recall curve in Figure \ref{fig:PR_curve_SNIaSNIISNIbc}. We also plot the precision and recall against varying thresholds in the right panels of Figure \ref{fig:PR_curve_SNIaSNIISNIbc}. These plots makes it clear that as we increase the threshold anomaly score, the recall decreases, because more and more transients from the anomalous classes are predicted as non-anomalous. In contrast, the precision increases as we increase the threshold anomaly score before reaching a maximum and decreasing after passing a certain score. Examining the anomaly score distributions in Figure \ref{fig:Anomaly_score_distribution} helps to explain this decrease. The peak of the Bazin anomaly score distribution for TDEs occurs at a score of $\tilde{\chi} \approx 3$. Once we sweep above this score, more TDEs are predicted as being non-anomalous, and thus the precision begins to decrease. Therefore, to maximise the precision, the anomaly score threshold should be carefully chosen to ensure most of the predicted anomalous transients are true anomalies.  We emphasise that this is different from most classification applications, where the class decision boundary is such that selecting an arbitrarily high threshold score will lead to a higher precision.

In Figure \ref{fig:AUCPR_curve_SNIaSNIISNIbc} we plot the AUCPR vs time by recording the AUCPR for precision-recall curves made at every time-step. The combined model performs very well at distinguishing all classes except for CARTs with high AUCPRs. As noted in \citet{Muthukrishna19RAPID}, CARTs are difficult to distinguish from common SNe based only on the light curves. At 25 days after trigger, the AUCPR curves plateau at scores above 0.75 for most classes. While this is a good metric to measure the performance of our model, it is not obvious at what time after a transient's detection, and at what anomaly score threshold we should ideally follow up a transient.

\begin{figure}
    \centering
    \includegraphics[width=1.0\linewidth]{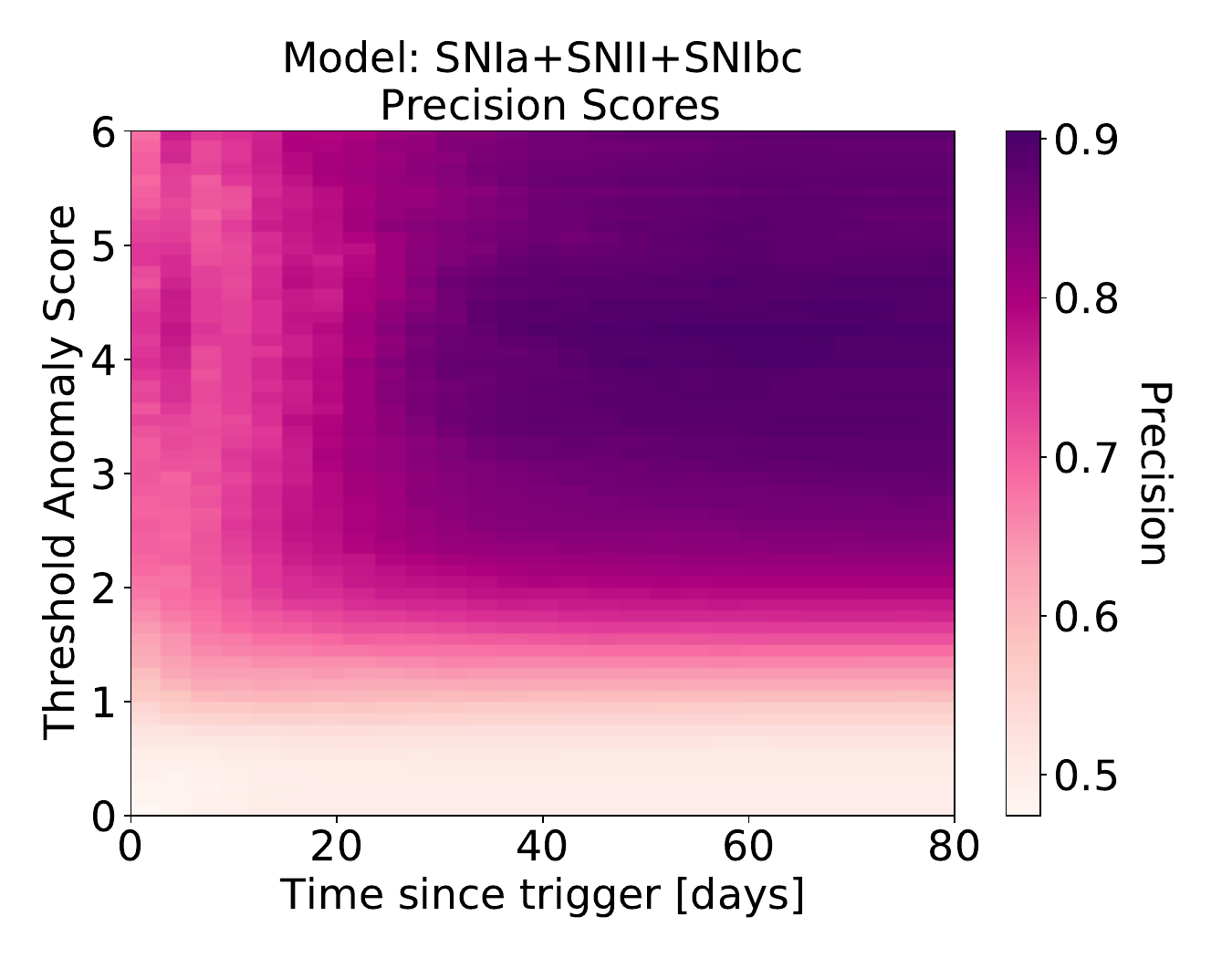}
    \caption{The precision score (indicated by the colour bar) plotted at varying threshold anomaly scores and times since trigger assuming the combination of the Bazin SNIa, SNII and SNIbc models are the reference classes and that all other classes (Kilonova, SLSN-I, TDE, CART, PISN, ILOT, AGN, uLens-BSR) are the anomalous classes. The plot makes use of the simulated transient dataset.}
    \label{fig:precision_vs_time_vs_threshold}
\end{figure}

In Figure \ref{fig:precision_vs_time_vs_threshold}, we plot the precision score at different times and different threshold anomaly scores. 
We have aggregated all the test classes (shown in the legends of Figures \ref{fig:PR_curve_SNIaSNIISNIbc} and \ref{fig:AUCPR_curve_SNIaSNIISNIbc}) and used these as the anomalous class, and set the Bazin models of the SNIa, SNII, and SNIbc classes as the model class. To obtain a precision around 0.9 on our test population, which indicates that 90\% of predicted anomalies will be true anomalies, we would need to wait until more than 25 days after a transient's detection, and only follow up transients with an anomaly score above 3. However, as we often want to perform spectroscopic followup early, we can see that even at detection, we can be 70\% confident that a transient that has an anomaly score between 3 to 5 will be a true anomaly given the transient population in our test set. These precision scores are dependent on the number of actual anomalies in the test set, which in this plot, is around 50\% of the test set. Figure \ref{fig:precision_vs_time_vs_threshold} illustrates that the precision increases as the light curves evolve, and that we can achieve good confidence in a predicted anomaly being a true anomaly from within a few days after detection.

\section{Application to ZTF Observational data}
\label{sec:application_to_real_ztf_data}
In this section we illustrate our method being applied to real observations from the public ZTF MSIP survey instead of simulations. 
\begin{figure*}
\centering
    {\includegraphics[width=0.2473\linewidth]{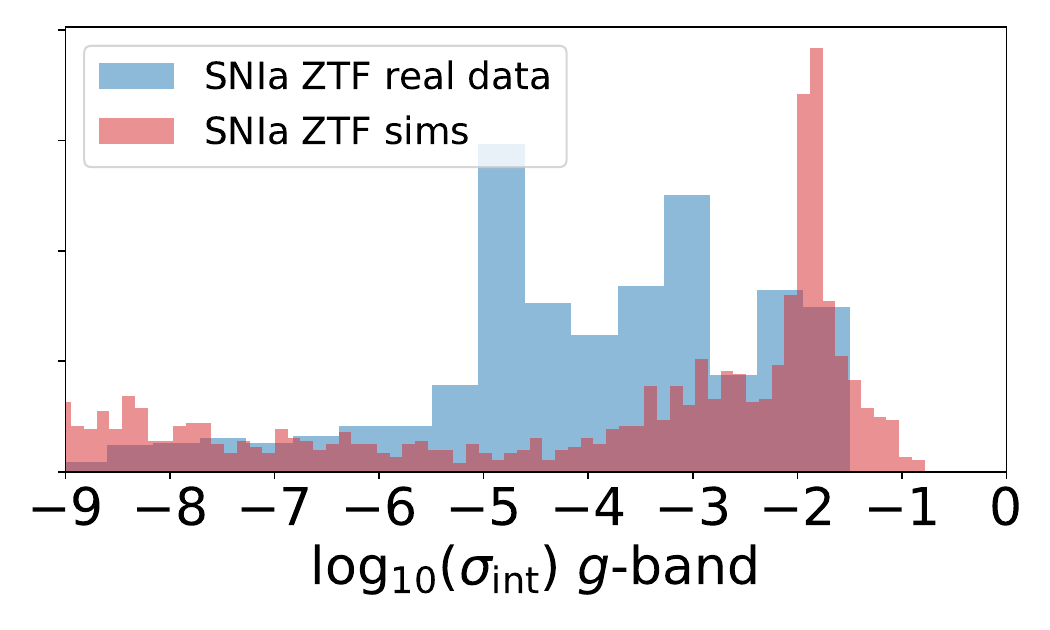}}
    {\includegraphics[width=0.2473\linewidth]{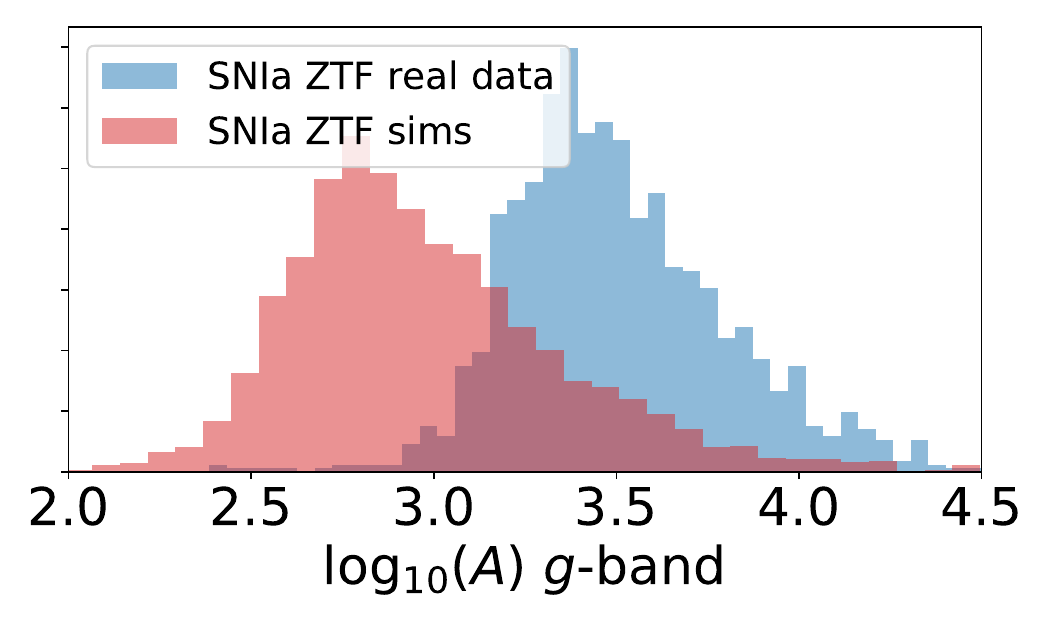}}
    {\includegraphics[width=0.2473\linewidth]{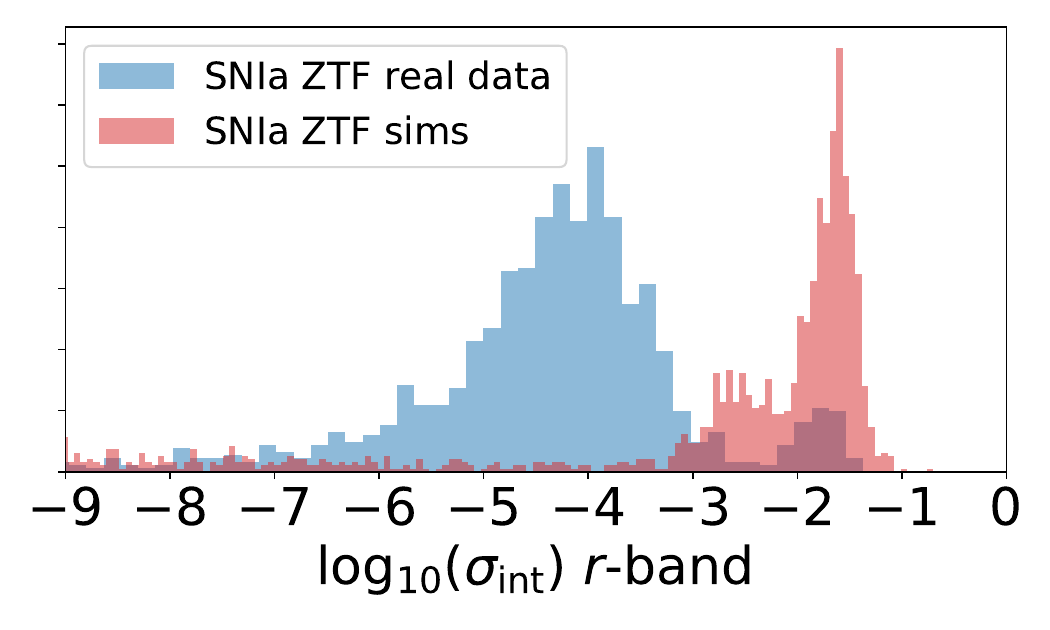}}
    {\includegraphics[width=0.2473\linewidth]{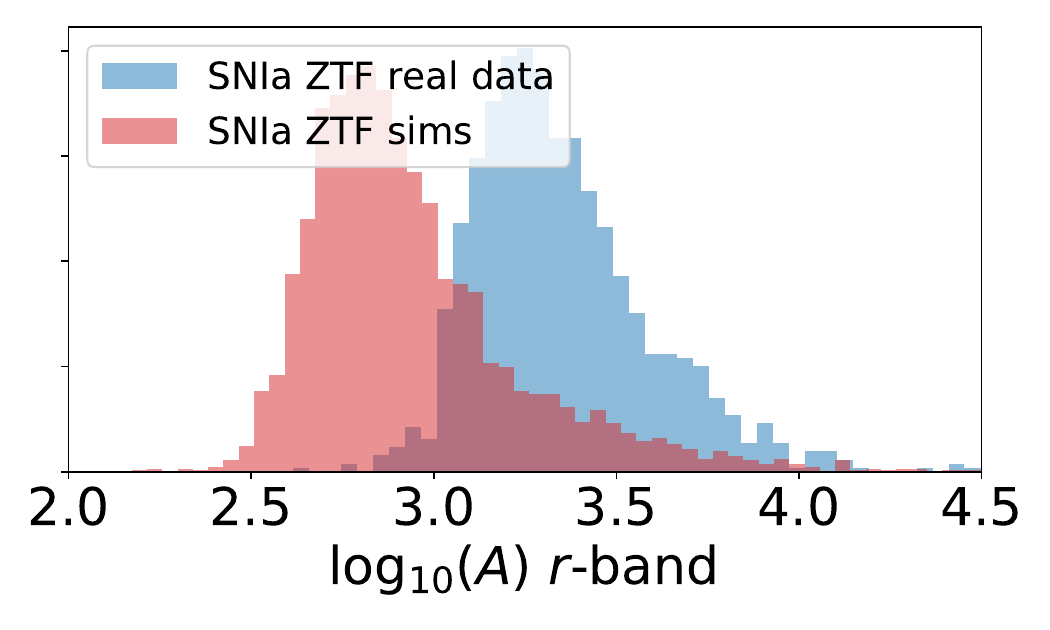}}
    {\includegraphics[width=0.2473\linewidth]{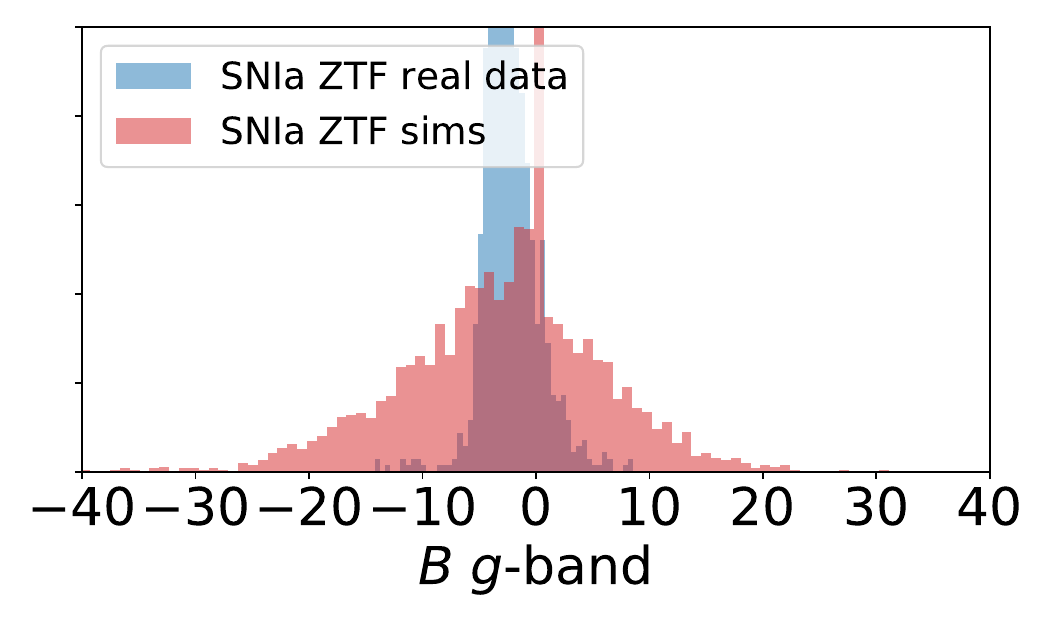}}
    {\includegraphics[width=0.2473\linewidth]{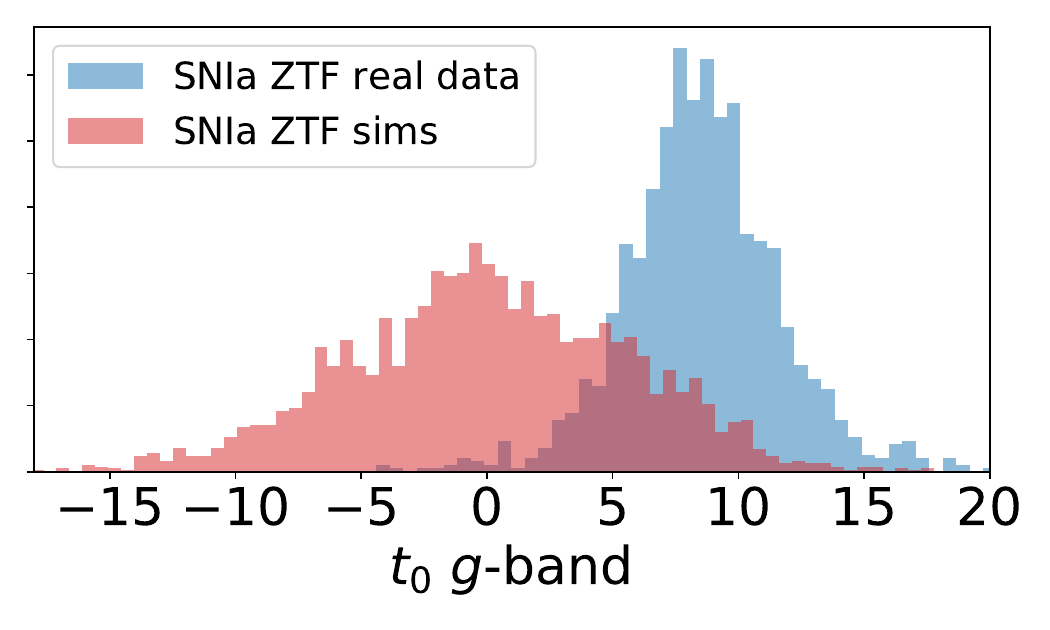}}
    {\includegraphics[width=0.2473\linewidth]{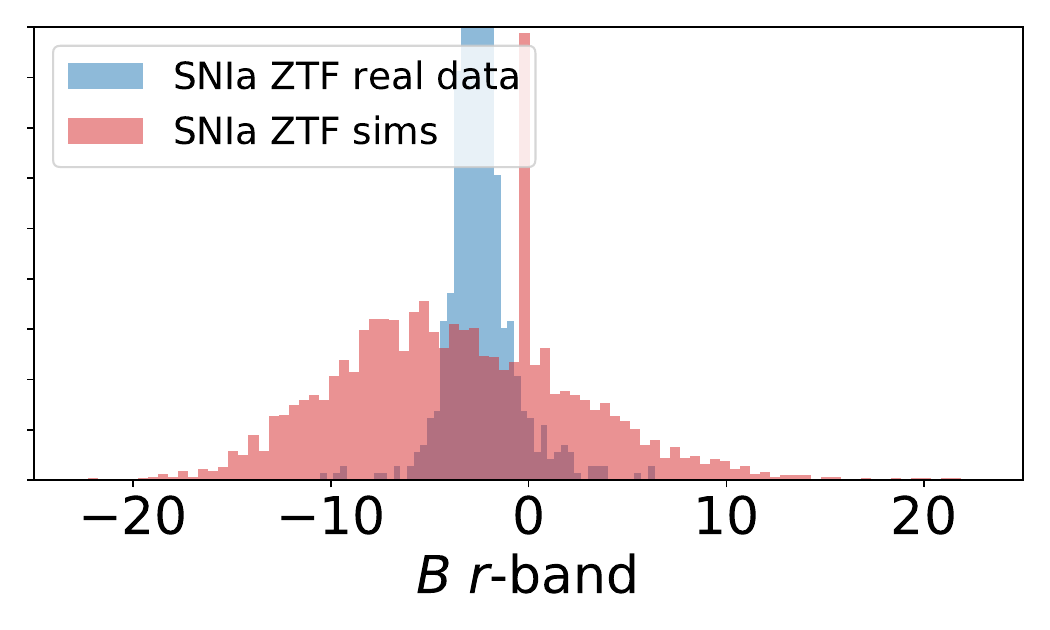}}
    {\includegraphics[width=0.2473\linewidth]{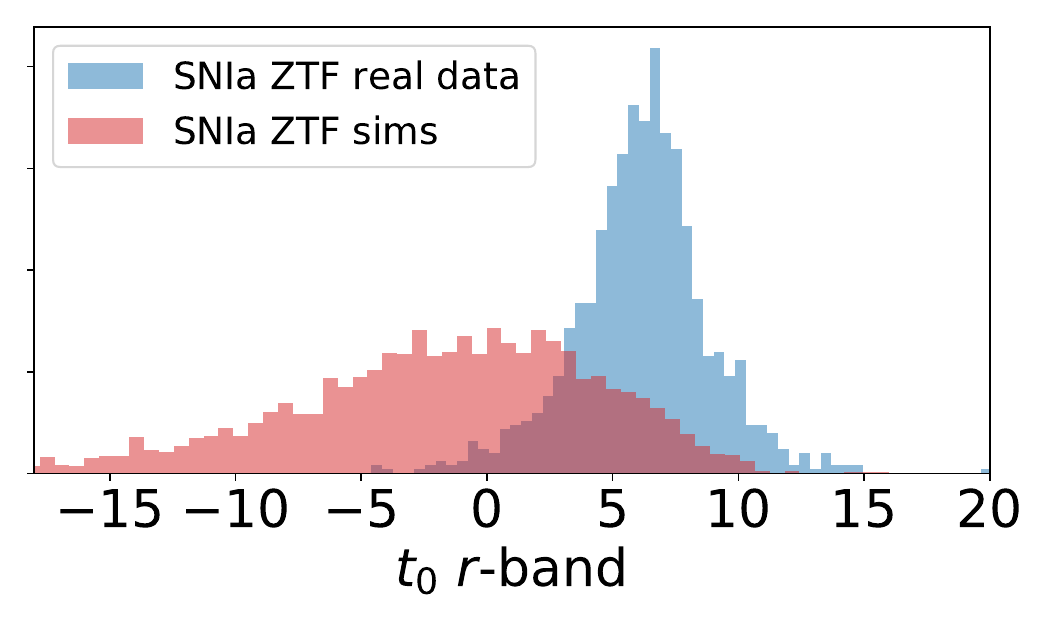}}
    {\includegraphics[width=0.2473\linewidth]{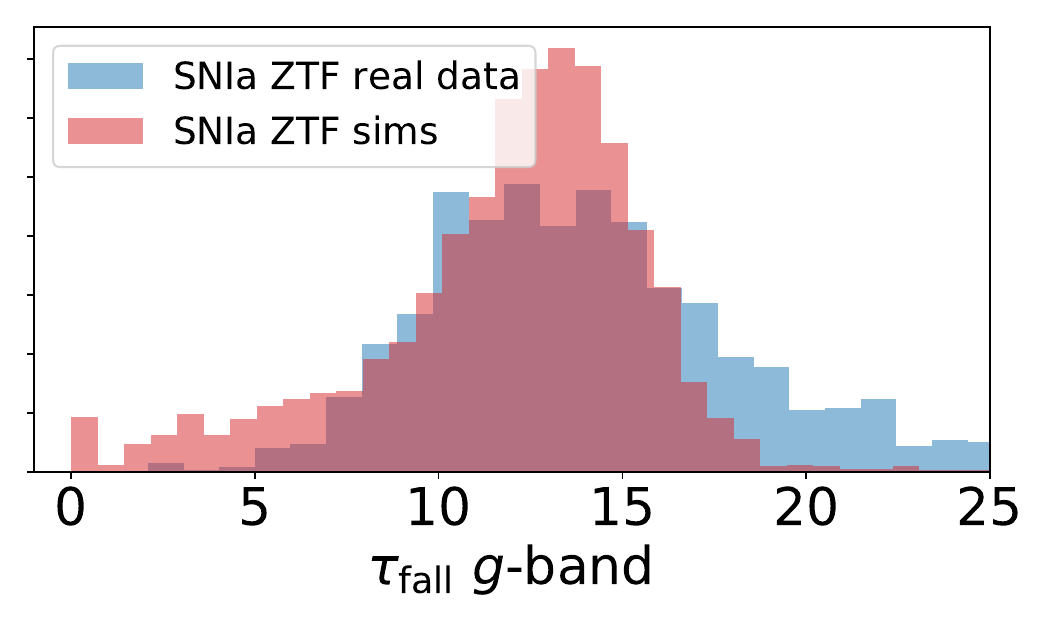}}
    {\includegraphics[width=0.2473\linewidth]{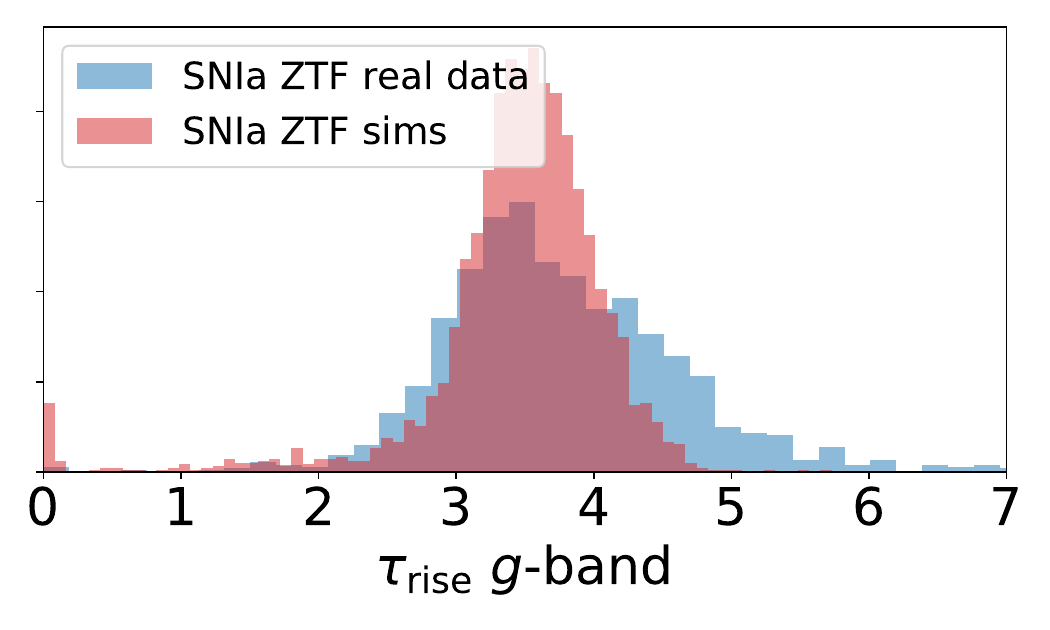}}
    {\includegraphics[width=0.2473\linewidth]{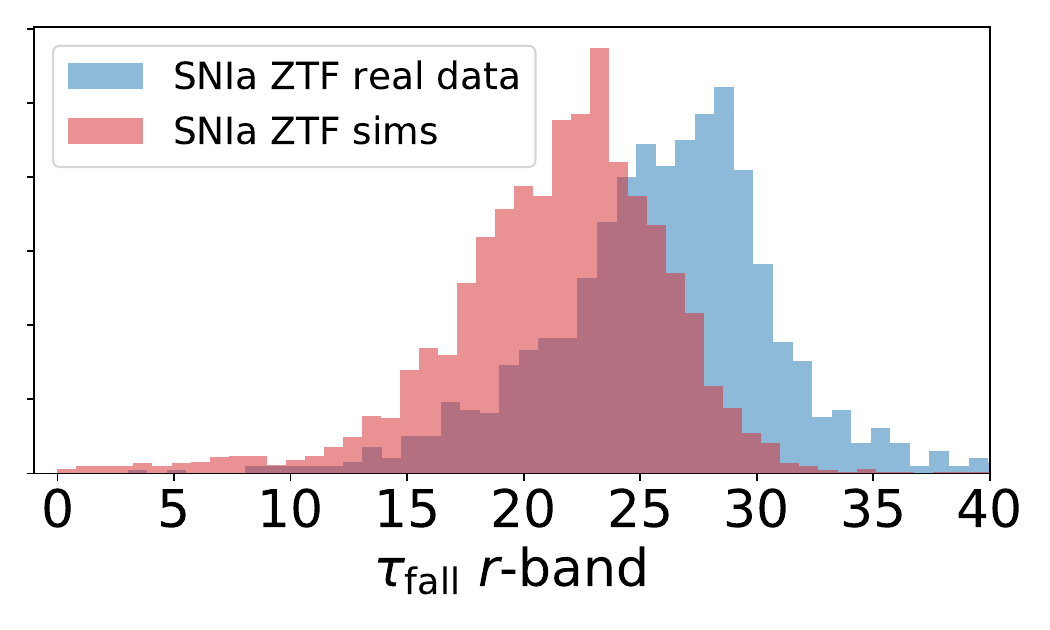}}
    {\includegraphics[width=0.2473\linewidth]{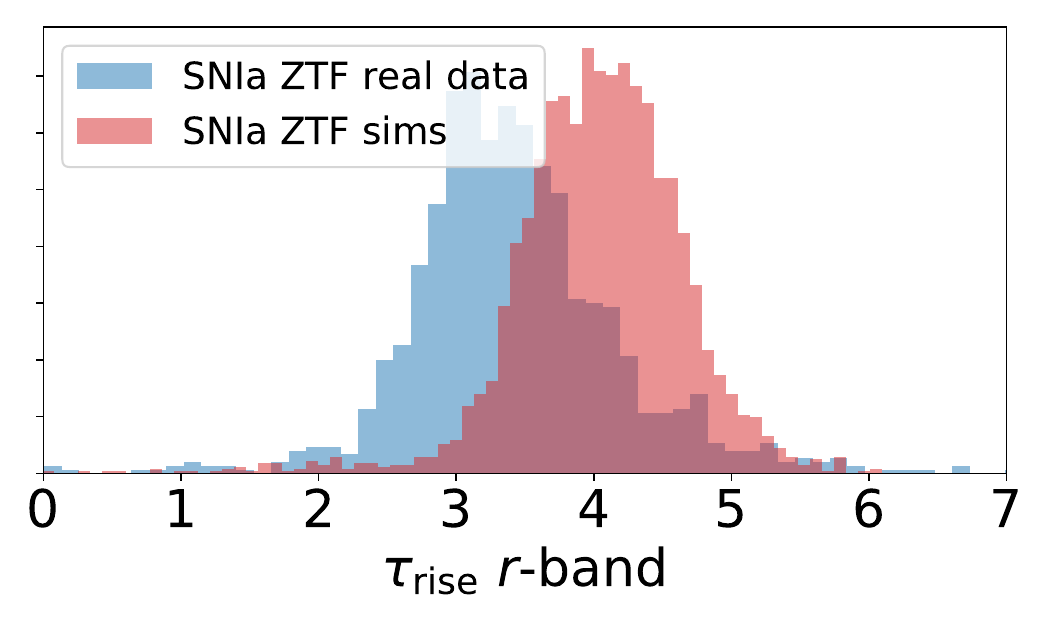}}

\caption[Distribution of Bazin function parameters for SNIa for the ZTF sims from \texttt{SNANA} and the real observations taken from the collection from the OSC and ZTF data stream.]{Comparison of the distributions of the best fit Bazin parameters for the population of simulated ZTF light curves and the collection of real observations taken from the ZTF data stream. We only show the parameter fits of light curves that have at least ten observations and at least one observation before peak.}
    \label{fig:real_data_Bazin_parameters}
\end{figure*}

The worldwide collection of extragalactic transients is dominated by SNe Ia, and for ZTF, there are not enough observations of other transient types to build an effective training set for the non-SNIa classes. Thus, in this section, we only build a model of SNe Ia, and identify anomaly scores relative to this class. To obtain a labelled dataset of ZTF transients, we searched the Transient Name Server\footnote{We searched the catalog, \url{https://www.wis-tns.org/}, on 22 Feb 2022} for objects with ZTF aliases. The number of collected transients under each broad classification label (after grouping similar labels) are listed in Table \ref{tab:real_data_population}.
\begin{table}
    \centering
    \begin{tabular}{c|c}
         Type & Number of Transients \\ \hline
         SNIa   &   3146    \\
         SNII   &   735     \\
         SNIbc  &   179     \\
         CV   &   111     \\
         SLSN  &   84     \\
         TDE   &   27     \\
         AGN   &   29     \\
    \end{tabular}
    \caption{Class distribution of real ZTF transients with labels taken from the Transient Name Server.}
    \label{tab:real_data_population}
\end{table}
We used SNIa as the reference class, and used Cataclysmic Variables (CVs), SLSNe, TDEs, and AGN as the anomalous classes. While AGN and CVs are relatively common, we have not included them as a model class because the wide variability of AGN make them difficult to model with the Bazin function, and because our collection of CVs is not large enough and of high enough quality to train an effective model. We also expect that because AGN and CVs are recurring phenomena, most of these will be quickly identified by LSST. However, because of the massive number of alerts from LSST, a non-insignificant number of CVs and AGN will still likely continue to be discovered over the course of the survey and thus we have included them as anomalous classes of interest to understand how our method will respond to these transients.

Unlike the simulations, the real data from the ZTF MSIP data stream are distributed in alert packets that contain magnitudes instead of flux units. We convert these magnitudes and magnitude uncertainties to flux counts and uncertainties as follows,
\begin{align}
    &F = 10^{-0.4(\mathrm{mag} - \mathrm{26.2})} \\
    &\sigma_F = |F \sigma_{\mathrm{mag}} \times 0.4\log{10}|
\end{align}
where $F$ is the flux, $\mathrm{mag}$ is the magnitude in the ZTF alert packet, $\sigma_F$ is the uncertainty in the flux, and $\sigma_\mathrm{mag}$ is the magnitude uncertainty. We have selected a zeropoint of 26.2 to scale the observations such that the flux and flux uncertainty distributions match the simulations.

We performed the same processing methods detailed in Sections \S\ref{sec:Data} and \S\ref{sec:Models}. To compare our simulations to the real observations, we plot the Bazin parameter distributions when fit to the simulated SNIa light curves and our collection of real SNIa light curves in Figure \ref{fig:real_data_Bazin_parameters}. The SNIa simulation distributions are the same as that shown in Figure \ref{fig:Bazin_parameter_distribution}, and the real data distributions were made by optimising the likelihood for all real SNIa light curves that had at least nine data points and at least one point before peak. The $\tau_{\mathrm{fall}}$ and $\tau_{\mathrm{rise}}$ distributions match the simulations reasonably well. However, there are a large fraction of real light curves that have sparse data that are not well observed before peak (to constrain $\tau_{\mathrm{rise}}$) and well after peak (to constrain $\tau_{\mathrm{fall}}$), and hence cause $\tau_{\mathrm{fall}}$ to be slightly overestimated for the real data. As in \S\ref{sec:Models}, we choose to use the population parameter distributions as the prior for the Bazin model. However, we have used the $\tau_{\mathrm{fall}}$ and $\tau_{\mathrm{rise}}$ prior from the simulations because the missing observations in the real light curves could lead to slight overestimations of these parameters and because we visually confirm that the shape of the distributions match. The real data also has much brighter peak fluxes as indicated by the larger values of $\log_{10}{(A})$. This is most probably due to selection effects where brighter SNe Ia are more likely to be discovered and classified. The distribution on $t_0$ is also slightly offset, mainly because the real data did not have any pre-trigger observations as there was no available forced photometry to get reliable non-detection fluxes when this work was conducted\footnote{ZTF recently released a forced photometry service that provides fixed-position PSF photometry on all publicly available ZTF images upon request. This service was not available at the start of this study, but future work should examine the improvement in performance when trained on data with forced photometry.}. 

\begin{figure}
    \centering
    \includegraphics[width=1.0\linewidth]{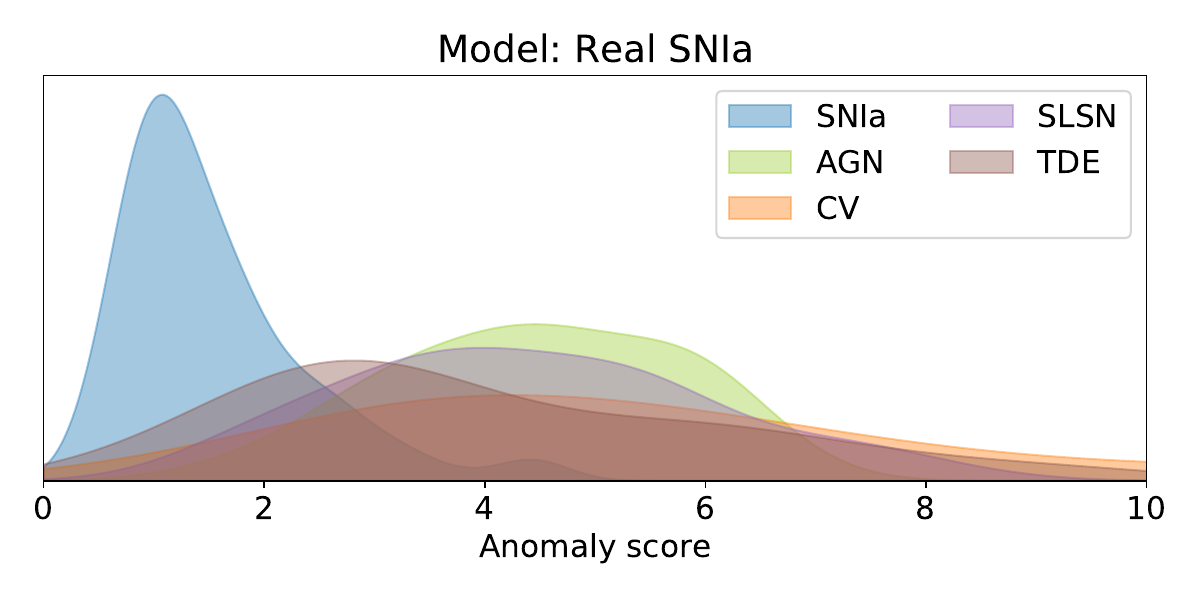}
    \caption{Anomaly score distribution recorded over the full light curve for the Bazin model for real ZTF SNIa observations tested on the real ZTF transient population of five different classes. For clarity, we estimate the probability density function of the anomaly scores using a Kernel Density Estimation (KDE) with a smooth Gaussian kernel. Classes that are dissimilar to SNIa have higher anomaly scores, while similar classes have lower anomaly score distributions.}
    \label{fig:real_data_Anomaly_score_distribution}
\end{figure}

Using the Bazin model trained on the SNIa class, we applied the model to our collection of observed AGN, CVs, SLSNe, TDEs, and SNe Ia, and plot the resulting distribution of anomaly scores for each class in Figure \ref{fig:real_data_Anomaly_score_distribution}. We define these non-SNIa classes shown in the plot as the anomalous classes of interest.

\begin{figure*}
    \centering
    \includegraphics[width=1.0\linewidth]{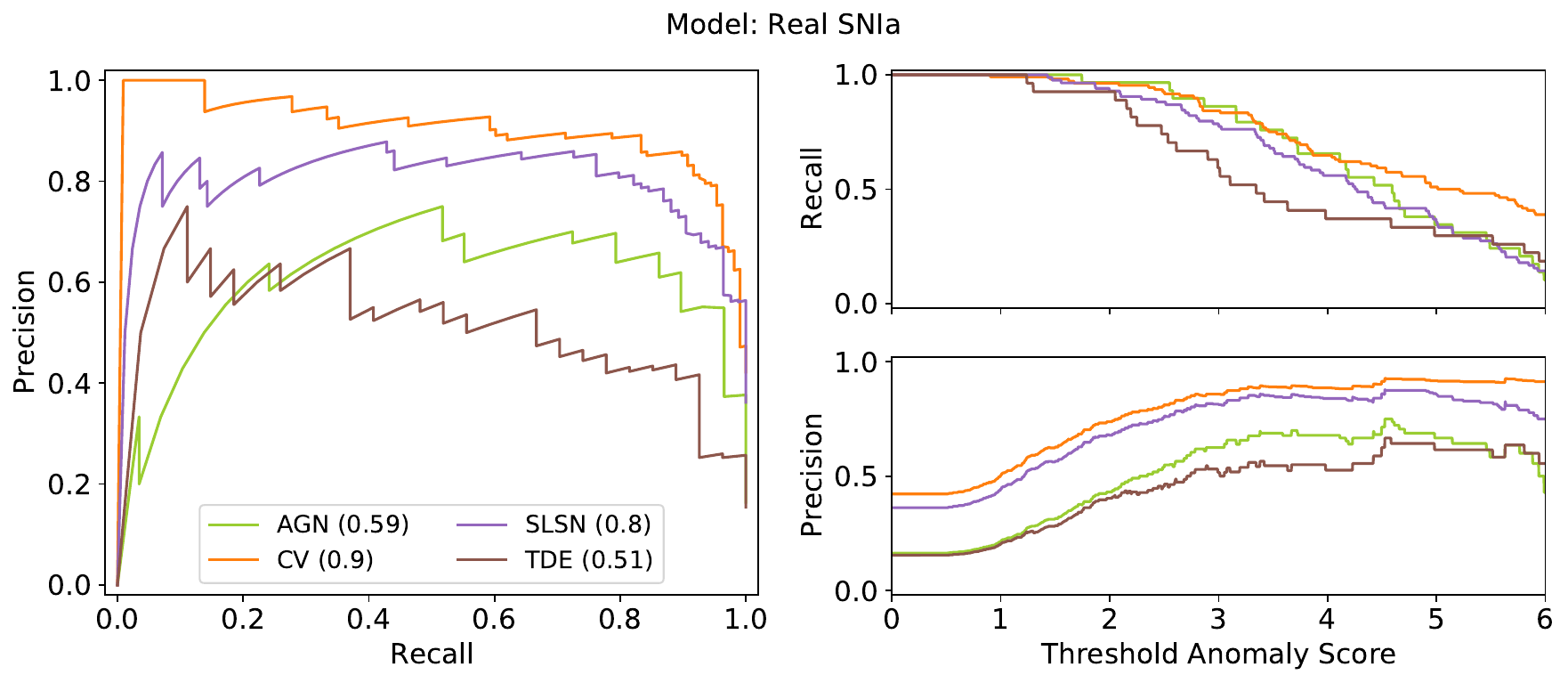}
    \caption{Precision-recall curve based on a Bazin model for real ZTF SNIa observations where we assume the SNIa class as the reference class and the anomalous classes are the ones denoted in the legend. We test the performance of this model on real observations of AGN, SLSN, CV, and TDE classes. We use the anomaly scores over the full light curves of all transients in the real data testing set to make the precision-recall curves.}
    \label{fig:real_data_PR_curve}
\end{figure*}

In Figure \ref{fig:real_data_PR_curve}, we compute the precision and recall at different threshold anomaly scores. Examining the right panels of the figure, we can see that to obtain a recall > 90\% for the anomalous transients, we need to select a threshold anomaly score $\tilde{\chi}_{\mathrm{thresh}} \lesssim 2$. However, to obtain the highest precision for anomalous transients, we need to select a threshold anomaly score $\tilde{\chi}_{\mathrm{thresh}} \approx 4.5$. The best precision is above 80\% for CVs and SLSNe (indicating that 80\% of all predicted anomalies are true anomalies) but is only above 65\% for the AGN and TDE classes. We note that we only have a very small collection of AGN and TDEs (29 and 27 transients, respectively), and thus these percentages are not necessarily a good representation of how the metrics will perform on new transients from this class. The TDEs and AGN in this real collection have much lower anomaly scores than the TDEs and AGN in our simulated population, and further analysis of the difference between these observed and simulated populations is required.

Nonetheless, this section has highlighted that our framework for identifying anomalies might be effective when applied to a collection of real data without the use of simulations. To better evaluate our method on interesting and real anomalous transients, a larger collection of rare and anomalous transients is needed. Future work in transient anomaly detection should collect a set of rare or anomalous transients that anomaly detection algorithms can be evaluated against.

\section{Conclusions}
\label{sec:Conclusions}
Upcoming wide-field surveys of the transient universe will probe deeper, wider, and faster than ever before, providing an opportunity for the discovery of entirely new classes of transient phenomena. However, discovery in astronomy has often been driven by serendipity, whereby identifying new phenomena has fortuitously occurred after human eyes sifted through data. With the huge amounts of data from surveys such as the LSST (expected to observe over 10 million transient alerts each night), a methodology aimed at automating the discovery of new transients through dedicated anomaly detection algorithms has become necessary.

Standard supervised classification approaches are unable to deal with the scope for new discovery offered by the wealth of data from upcoming surveys because they can only identify transients that they have been specifically trained on. Anomaly detection algorithms enable an opportunity to automatically flag unusual and interesting transients for further followup. In this paper, we have detailed the development of a real-time anomaly detection framework for identifying unusual transients in large-scale transient surveys. We have built two separate frameworks. The first is a probabilistic deep neural network (DNN) built using Temporal Convolutional Networks aimed at predicting future observations in a light curve. And the second is based on a parametric fit to a partial light curve using the Bazin function \citep{Bazin_function}, where we extrapolate a prediction 3 days after each partial light curve fit to compare it to the DNN approach. Each of the approaches can be well optimised to deal with the millions of alerts that ongoing and upcoming wide-field surveys such as ZTF and LSST will produce. 

Our two methods allow us to identify anomalies as a function of time, and we have demonstrated its performances on both ZTF-like simulations and real ZTF light curves from the public MSIP survey. In particular, we have demonstrated that we are able to identify rare transients as anomalous with respect to common supernova classes (SNIa, SNII, SNIbc). We obtain a high recall and precision of anomalous transients culminating in area under the precision-recall curve (AUCPR) scores above 0.79 for most rare classes by the end of the light curve with the Bazin approach. Our ability to identify anomalies improves over the lifetime of the light curves. Based on the anomaly scores and the epoch of discovery, our framework enables a prioritised followup of unusual transients.

Our method is very effective at identifying rare transient classes, including kilonovae, SLSNe, TDEs, PISNe, ILOTs, and uLens-BSR as anomalous with respect to common supernovae. AGN outbursts, while not rare, are also identified as anomalous very effectively by our methods. We expect that most AGN, with their long-lived variability, will be quickly identified by surveys such as LSST. However, in the first year of LSST, AGN will be a large contaminant of anomalies detected. But, using additional data such as the distance of the transient from the nucleus of a galaxy, and cross-matches with other surveys, can help to eliminate AGN. As our method is not suited to classifying AGN or CVs and will identify them as anomalous, we recommend that classifiers trained on CVs and AGN be used in conjunction with our anomaly detection framework.

CARTs, on the other hand, proved too difficult to identify with our approach. The reason for this difficulty is most likely because their fast rise times are very similar to many core-collapse SNe. As noted in \citet{Muthukrishna19RAPID}, their light curve shapes in the $g$ and $r$ passbands is very similar to common supernovae and identifying them will always be difficult from light curve shape alone.

Both the DNN and Bazin approaches are very fast and will be easily scaleable to surveys as large as LSST, but the DNN is considerably faster at inference time. The DNN also has the advantage over the Bazin function of being completely data-driven, and thus makes it easier to apply to transient types that are not well fit by the Bazin function.
However, while we have shown that our DNN approach is very good at the prediction of fluxes, we have also noted that it is too flexible to act as a good anomaly detector when compared with our Bazin approach. The DNN method trained on a particular supernova class is able to accurately predict supernova within that class, but is so flexible, that it makes reasonable predictions of transients in other classes too. This flexibility means that it is not good at detecting anomalies. On the other hand, the Bazin approach is very effective at identifying transients outside the modeled supernova class, making it an effective anomaly detector. 

In future work, we hope to apply our method on a ZTF transient broker to gauge our success at identifying real anomalous transients. We think that applying an anomaly detection framework in conjunction with a transient classifier will provide more valuable information on whether a newly discovered transient is interesting enough for further follow-up observations. An issue with this work, is that there has been no distinction between anomalies and \textit{interesting} anomalies. It is possible that without good \textit{real-bogus} cuts on a data stream, our approach may flag unusual transient phenomena that don't align with our trained supernova classes but are uninteresting to most astronomers. To deal with this, future work should consider Active Learning frameworks that use methods such as \textit{Human-in-the-loop learning} that focus on specifically targeting what users define as interesting phenomena (recent work by \citealt{Ishida2021_Timeseries,Lochner2020Astronomaly} have begun working on Active Learning for anomaly detection).

Overall, this paper presents a novel and effective method at identifying anomalous transients in real-time surveys. Anomaly detection coupled with other classification approaches enables astronomers to prioritise follow-up candidates. This work and other recent approaches to anomaly detection are going to be critical for discovery in the new era of large-scale astronomical surveys.

\section*{Acknowledgements}
DM would like to thank the Cambridge Trust and the Cambridge Australia Poynton Scholarship for studentship funding. ML acknowledges support from South African Radio Astronomy Observatory and the National Research Foundation (NRF) towards this research. Opinions expressed and conclusions arrived at, are those of the authors and are not necessarily to be attributed to the NRF.  KSM acknowledges funding from the European Research Council under the European Union's Horizon 2020 research and innovation programme (ERC Grant Agreement No. 101002652). This project has been made possible through the ASTROSTAT-II collaboration, enabled by the Horizon 2020, EU Grant Agreement No. 873089.

We would also like to thank the organisers of the 2019 Kavli Summer Program in Astrophysics hosted at the University of California, Santa Cruz for fostering this collaboration. The program was funded by the Kavli Foundation, The National Science Foundation, UC Santa Cruz, and the Simons Foundation. We further thank Francois Lanusse, Stephen Smartt, Simon Hodgkin, Michael Muthukrishna, Thomas Espaas, and Calum Ashcroft for helpful comments and discussions that improved the paper.

\section*{Data availability}
The PLAsTiCC models used to create the simulations used in this work were released by \citet{KesslerPlasticcModels} and are available at \url{https://zenodo.org/record/2612896#.YYAz1NbMJhE} \citep{plasticc_modelers_2019_2612896}. We used the \texttt{SNANA} \citep{Kessler2010SNANA:Analysis} software developed for PLAsTiCC along with observing logs from the public ZTF MSIP survey to simulate light curves. These light curves were first used in \citet{Muthukrishna19RAPID}, and can be shared on reasonable request to the corresponding author. The real ZTF observational data used in \S\ref{sec:application_to_real_ztf_data} are publicly available from the ZTF MSIP survey. The list of supernovae used in \S\ref{sec:application_to_real_ztf_data} can be shared upon request to the corresponding author.




\bibliographystyle{mnras}
\bibliography{references}



\appendix



\section{Laplace approximation}
\label{sec:Appendix_Fisher_matrix_approximation}
In \S\ref{sec:Model_Bazin}, we detail how the light curves are fit using a Bayesian parametric function. We would ideally obtain distributions over the parameters for each light curve fit using MCMC. However, because of the how computationally intensive this would be for the many light curves in our training set (and in large scale surveys), we resort to optimising the fit, and approximating the posterior with the Laplace approximation.

\begin{figure*}
    \centering
    \includegraphics[width=0.75\linewidth]{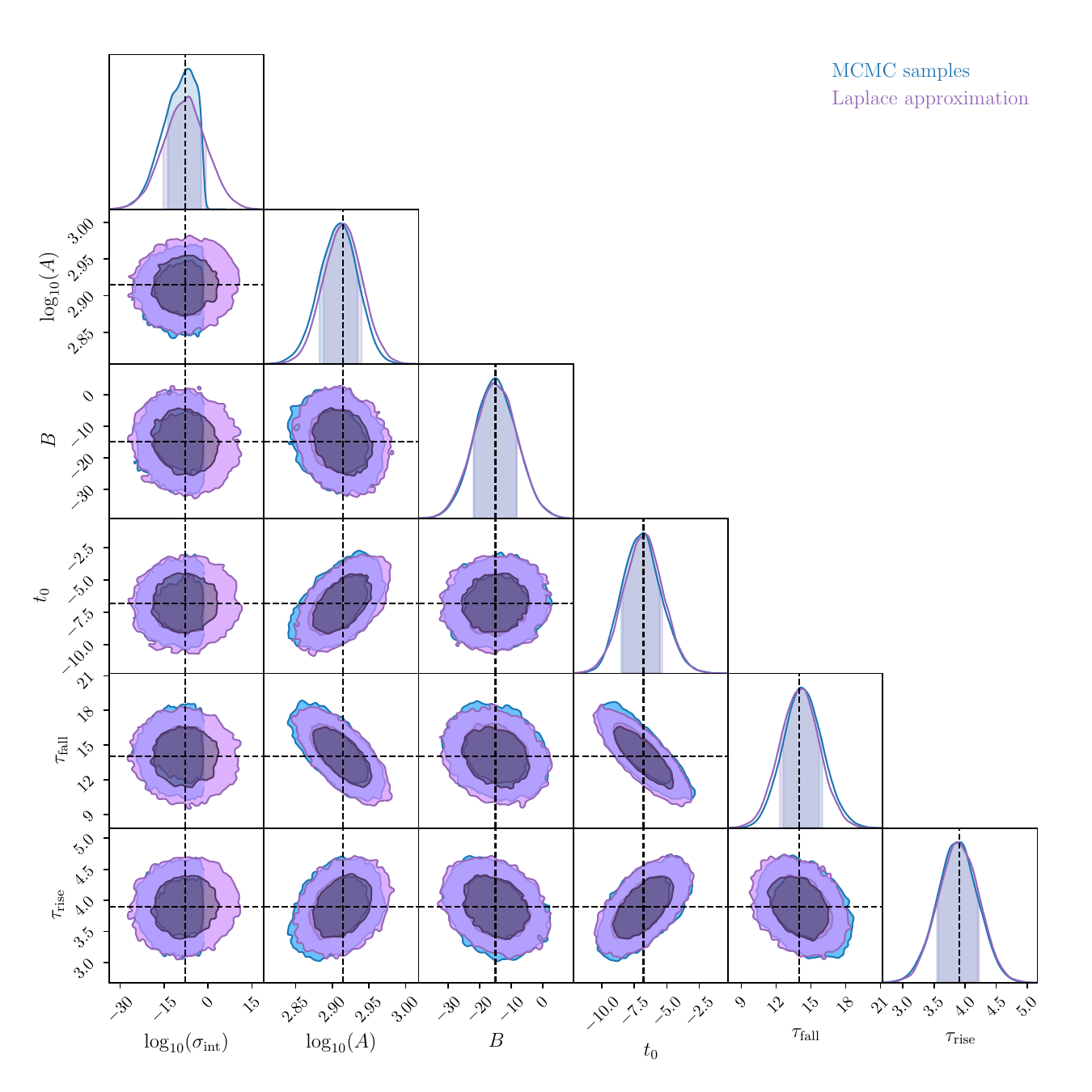}
    \caption[Example Bazin fit parameter distributions of example SNIa light curve using MCMC samples and the Laplace approximation.]{Example Bazin fit parameter distributions of example simulated SNIa light curve using MCMC samples and the Laplace approximation. The dashed lines shows the optimal fit from the Nelder-Mead optimisation routine.}
    \label{fig:example_MCMC_Bazin_contours}
\end{figure*}

In Figure \ref{fig:example_MCMC_Bazin_contours}, we compare our fits to an example SNIa light curve using MCMC and our Laplace approximation. The parameter distributions appear well-approximated by the Laplace approximation for all parameters except for $\log_{10}{(\sigma_{\mathrm{int}})}$. The MCMC samples disfavour high values of $\log_{10}{(\sigma_{\mathrm{int}})}$, and have a sharp cut-off near zero. Since the Laplace method approximates the posterior as a Gaussian, it does not approximate this behaviour well. The very large values of $\log_{10}{(\sigma_{\mathrm{int}})}$ lead to unrealistically large estimates of the predicted flux. Thus, we use the mode of this parameter instead of sampling over it for all Bazin models in this paper.

\section{Analysis of predictive uncertainty}
\label{sec:Appendix_analysis_of_predictive_uncertainty}
\begin{figure*}
    \centering
    \includegraphics[width=1.0\linewidth]{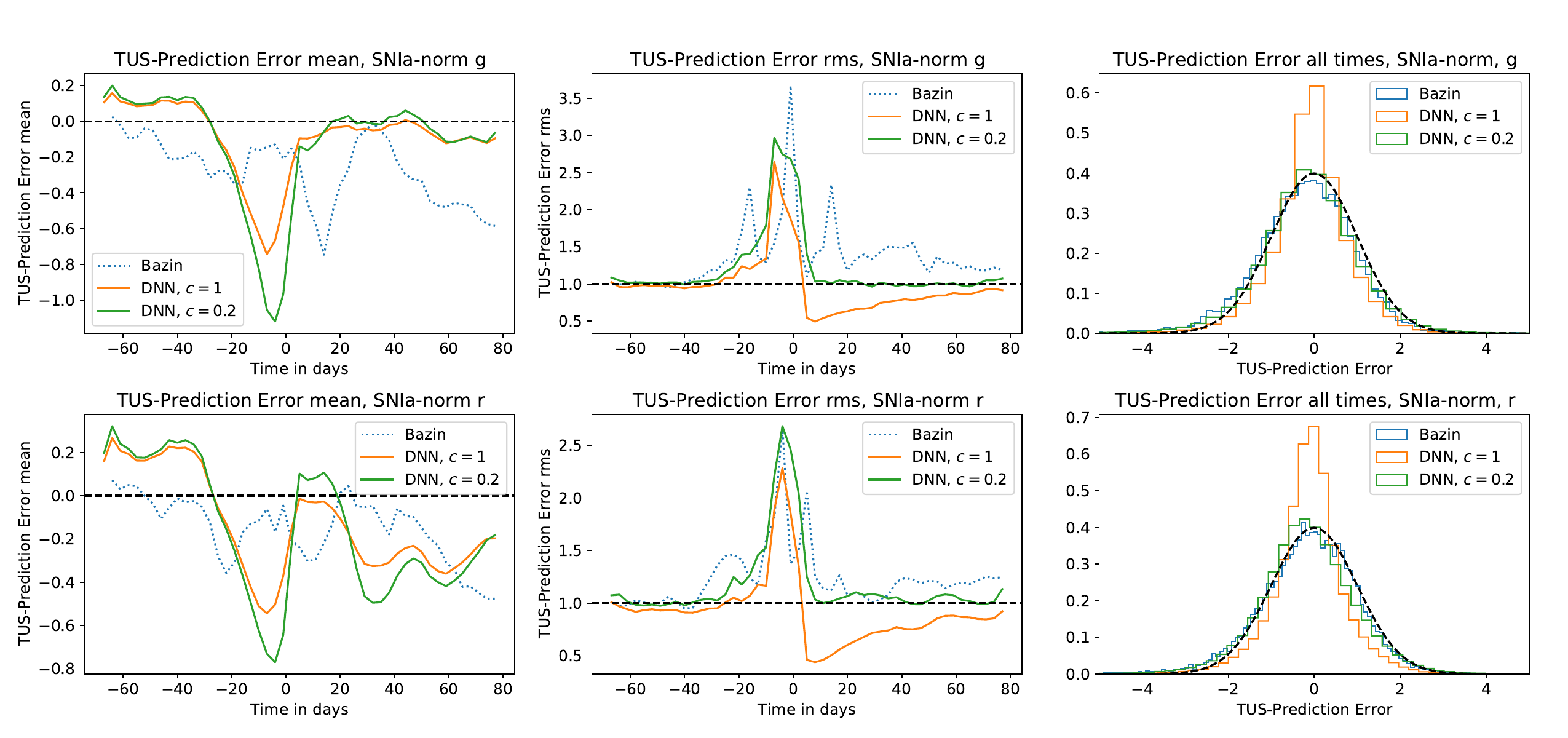}
    \caption{Distribution of the Total-Uncertainty-Scaled Prediction Error (TUSPE) for the SNIa model at different times. Equation \ref{eq:scaled_error} is computed at each time-step for all SNIa in the testing set. We show the mean and root mean square (rms) for each scaled error distribution at each time-step in the first three panels, with the $g$ band shown in the top row of panels, and the $r$ band shown in the bottom row of panels. In the last column, we plot the distribution of scaled errors across all times. After trigger, a scaling factor of $c=0.2$ on the predictive uncertainty for the DNN improves the scaled error. Similar plots were made for the SNIbc, SNII, SLSN and TDE models, but are not shown here for brevity. We plot a unit Gaussian (which is an ideal Scaled Error distribution) as a black dashed line to help guide the eye. The plots were made using the simulated dataset.}
    \label{fig:Scaled_errors}
\end{figure*}

We performed the following analysis to assess the computed predictive uncertainties. We first defined the Total-Uncertainty-Scaled Prediction Error as follows,
\begin{equation}
        \mathrm{TUSPE}_{spt} = \frac{(y_{spt} - D_{spt})} {\sqrt{c^2\sigma_{y,{spt}}^2 + \sigma_{D,{spt}}^2}}.
\label{eq:scaled_error}
\end{equation}
The instantaneous anomaly score in equation \ref{eq:Anomaly_score} is just the average over passbands of the squared scaled error. We plotted the distribution of the scaled error for the SNIa model and recorded the mean and root mean square (rms) at each time-step. For an unbiased model, the mean of the scaled error should be close to $0$, and for a model that correctly estimates the predictive uncertainty, the rms should be close to $1$. We have plotted the scaled error as a function of time since trigger for the SNIa DNN and Bazin model in Figure \ref{fig:Scaled_errors}. Around the early phase of the transient (before trigger) the rms and bias is large because the models are not effective until more of the light curve has been observed. In this work, we are mainly interested in observations after trigger, and so, we recognise that a good model would have a rms near $1$ after trigger. While the rms is good for the Bazin model, it is too small for the DNN, indicating that the predictive uncertainty is overestimated. Our DNN poorly estimates the predictive uncertainty, and to calibrate it, we scale it by a factor of $c=0.2$ to obtain the green plot in Figure \ref{fig:Scaled_errors}. We calibrated this scaling factor on the training set, not the test set. We made similar plots for the other five transient models (not shown for brevity), and determined that the optimal factor to correct all the DNN models was close to $c=0.2$. We refer the reader to \citet{Caldeira_Nord2020} for further examples of how uncertainty quantification with deep neural networks can be inaccurate. 

\section{Comparison of DNN and Bazin predictive power}
\label{sec:Appendix_Comparison_of_DNN_and_Bazin}
    We performed the following analysis to evaluate why the DNN model was less effective at identifying anomalies than the Bazin model. To compare the models on an even scale, we defined the Measurement-Uncertainty-Scaled Prediction Error as follows,
        \begin{equation}
                \mathrm{MUSPE}_{spt} = \frac{(y_{spt} - D_{spt})} {\sigma_{D,{spt}}}.
        \label{eq:physical_flux_error}
        \end{equation}
    This differs from equation \ref{eq:scaled_error} because we are normalising the flux prediction error by the square-root of the measurement variance instead of the square-root of the total variance. The scaled error in equation \ref{eq:scaled_error} would not let us easily compare between the methods because the predictive variance differs for the DNN and Bazin model, and thus the denominators would be different for each model. Equation \ref{eq:physical_flux_error}, on the other hand, is just the flux prediction error in units of the measurement error, and thus allows us to compare the DNN and Bazin models on an even scale.
    
    \subsection{DNN overfitting evaluation}
\begin{figure*}
        \centering
        \includegraphics[width=1.0\linewidth]{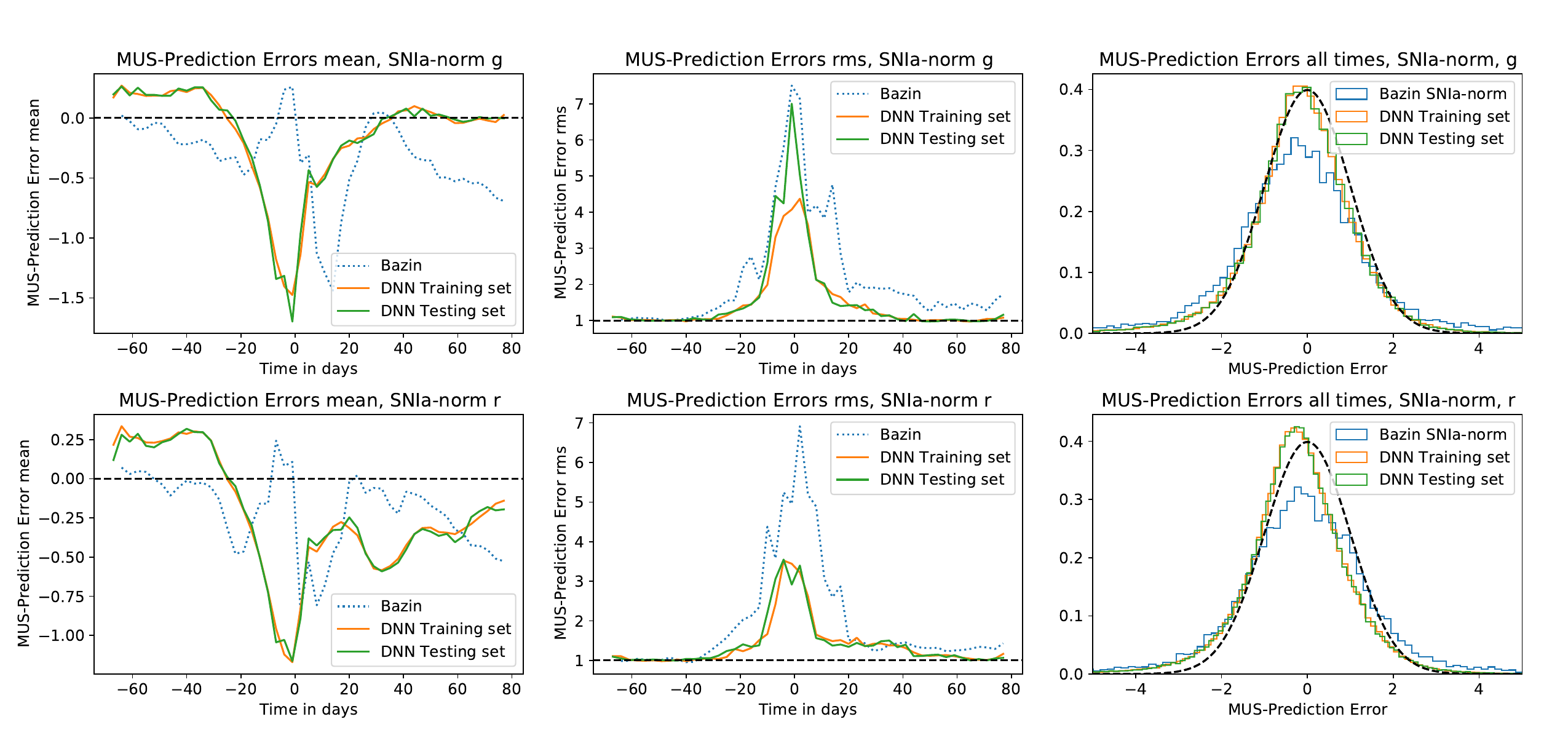}
        \caption[Distribution of the Measurement-Uncertainty-Scaled Prediction Errors (MUSPE) for the SNIa model at different times.]{Distribution of the Measurement-Uncertainty-Scaled Prediction Errors (MUSPE) for the SNIa model at different times. Equation \ref{eq:physical_flux_error} is computed at each time-step for all SNIa in DNN training set (orange lines), testing sets (green lines) and for the Bazin training set (blue dotted line). We show the mean, and root mean square (rms) for each prediction error distribution at each time-step in the first three panels, with the $g$ band shown in the top row of panels, and the $r$ band shown in the bottom row of panels. In the last column, we plot the distribution of scaled errors across all times. Similar plots were made for the SNIbc, SNII, SLSN and TDE models, but are not shown here for brevity. We plot a unit Gaussian as a black dashed line to help guide the eye. The plots were made using the simulated dataset.}
        \label{fig:DNN_train_vs_test_flux_errors}
        \end{figure*}

        In Figure \ref{fig:DNN_train_vs_test_flux_errors}, we plot the distribution of prediction errors on all light curves in the DNN SNIa training set (orange lines) and testing set (green lines). The prediction error distributions are very similar, and add further confirmation that our DNN SNIa model has not overfit the training set. We made similar plots and conclusions for the SNIbc, SNII, SLSN and TDE models, but have not shown them here for brevity.
        
        The blue line in Figure \ref{fig:DNN_train_vs_test_flux_errors} shows the prediction error distributions for the Bazin model, and it appears that they perform slightly worse than the DNN, with a slghtly wider prediction error distribution. Hence, we conclude that the DNN is slightly better at modelling SNIa than the Bazin model. However, this does not explain why the DNN is worse at identifying anomalies.
    
    \subsection{DNN vs Bazin}
        \label{sec:Appendix_DNN_vs_Bazin}
\begin{figure*}
        \centering
        \textbf{Bazin}\par\medskip
        \vspace{-0.8em}
        \includegraphics[width=0.99\linewidth]{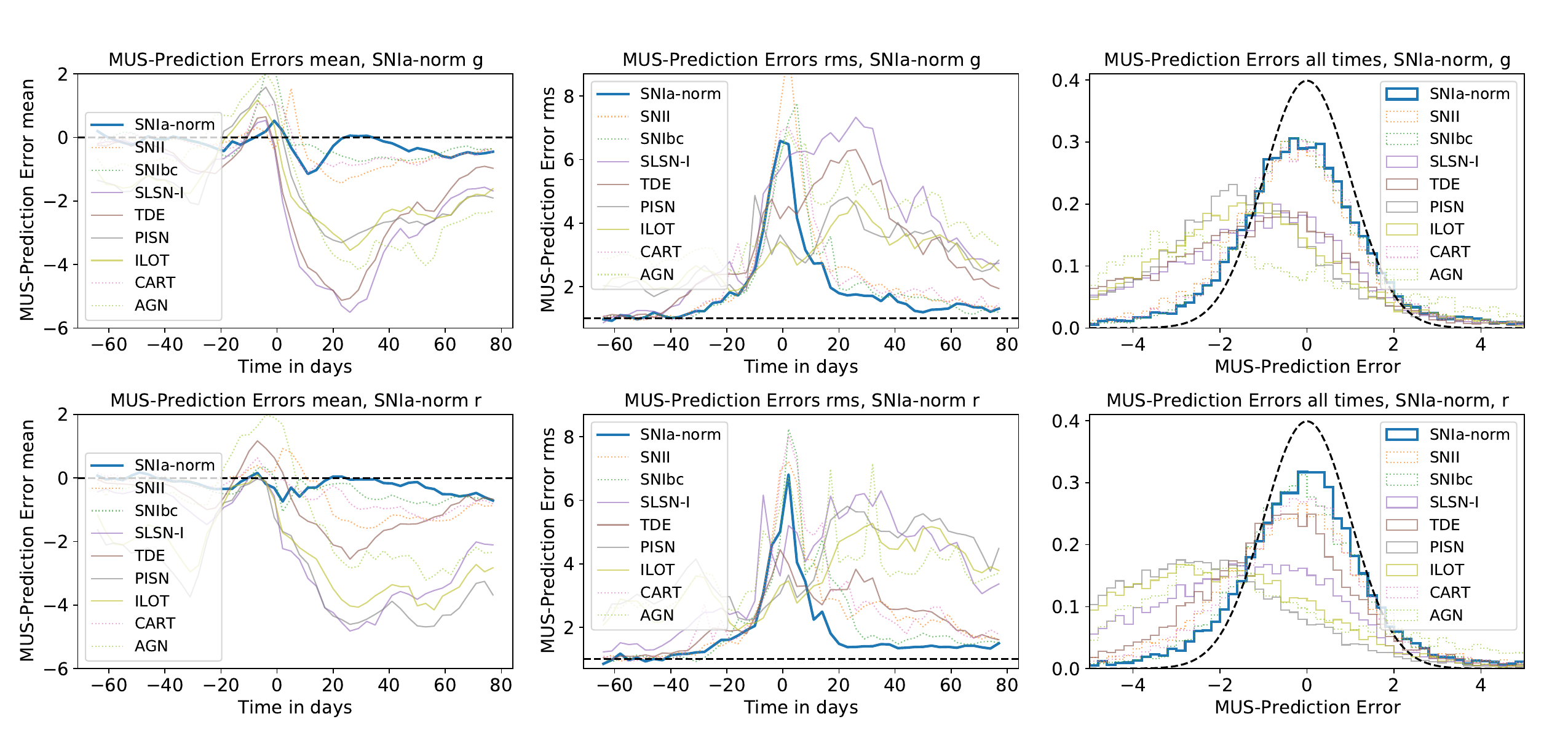}
        \caption[Distribution of the Measurement-Uncertainty-Scaled Prediction Errors (MUSPE) for the Bazin SNIa model at different times.]{Distribution of the Measurement-Uncertainty-Scaled Prediction Errors (MUSPE) for the Bazin SNIa model at different times. We apply the Bazin SNIa model to each other class' datasets and compute Equation \ref{eq:physical_flux_error} at each time-step. We have not plotted the kilonova or uLens-BSR classes here because they show significant deviations, indicating that the SNIa model is very bad at predicting these classes, and hence easily identifies them as anomalies. We show the mean and root mean square (rms) for each prediction error distribution at each time-step in the first three panels, with the $g$ band shown in the top row of panels, and the $r$ band shown in the bottom row of panels. In the last column, we plot the distribution of scaled errors across all times. Similar plots were made for the SNIbc, SNII, kilonova, SLSN and TDE Bazin models, but are not shown here for brevity. We plot a unit Gaussian as a black dashed line to help guide the eye. The plots were made using the simulated dataset.}
        \label{fig:Bazin_flux_errors}
        \end{figure*}

\begin{figure*}
        \centering
        \textbf{DNN}\par\medskip
        \vspace{-0.7em}
        \includegraphics[width=0.99\linewidth]{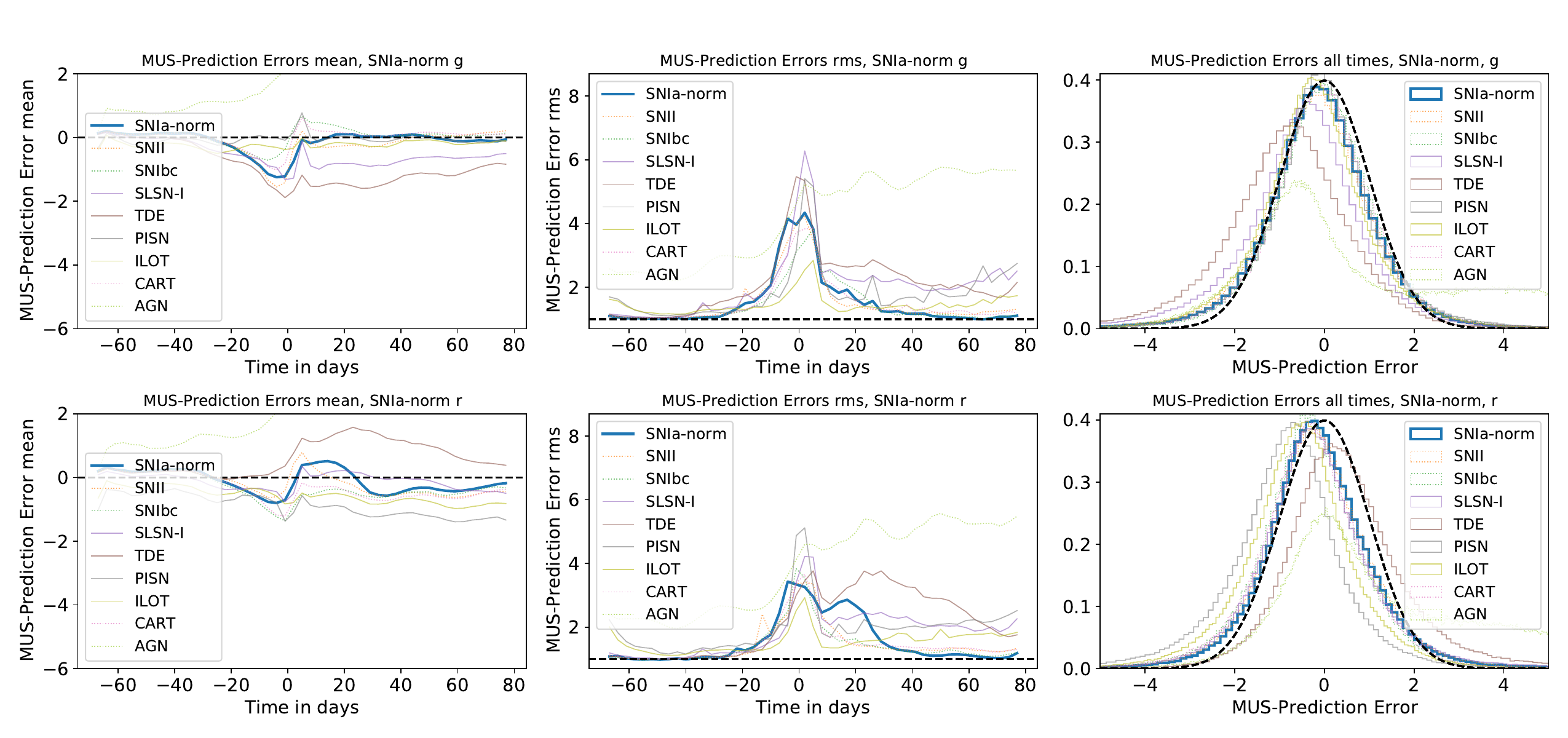}
        \caption[Distribution of the Measurement-Uncertainty-Scaled Prediction Errors (MUSPE) for the DNN SNIa model at different times.]{Distribution of the Measurement-Uncertainty-Scaled Prediction Errors (MUSPE) for the DNN SNIa model at different times. We apply the DNN SNIa model to each other class' datasets and compute Equation \ref{eq:physical_flux_error} at each time-step. We have not plotted the kilonova or uLens-BSR classes here because they show significant deviations, indicating that the SNIa model is very bad at predicting these classes, and hence easily identifies them as anomalies. We show the mean and root mean square (rms) for each prediction error distribution at each time-step in the first three panels, with the $g$ band shown in the top row of panels, and the $r$ band shown in the bottom row of panels. In the last column, we plot the distribution of scaled errors across all times. Similar plots were made for the SNIbc, SNII, kilonova, SLSN and TDE Bazin models, but are not shown here for brevity. We plot a unit Gaussian as a black dashed line to help guide the eye. The plots were made using the simulated dataset.}
        \label{fig:DNN_flux_errors}
        \end{figure*}

        To analyse why the SNIa Bazin model is better than than the DNN at identifying anomalies, we plot the prediction errors for the SNIa models applied to each of the other transient classes for Bazin and DNN in Figures \ref{fig:Bazin_flux_errors} and \ref{fig:DNN_flux_errors}, respectively. We expect that the SNIa models should predict the SNIa light curves best, and indeed, we see that these blue lines for the SNIa have nearly the best prediction error distributions. In Figure \ref{fig:Bazin_flux_errors}, the prediction errors are significantly worse for the more anomalous classes (SLSNe, TDEs, PISNe, ILOTs) with deviations ranging up to 5 sigma. However, the prediction errors for these classes in the DNN are much smaller, not much more than 1 sigma deviations. This indicates that the Bazin model is much worse at predicting the observations from these anomalous classes than the DNN, and hence is better at identifying them as anomalies, despite Figure \ref{fig:DNN_train_vs_test_flux_errors} highlighting that the SNIa DNN model is better able to predict SNIa observations.
        
\bsp	
\label{lastpage}
\end{document}